\documentstyle[12pt,epsf,epsfig]{article}
\newcount\timecount
\newcount\hours \newcount\minutes  \newcount\temp \newcount\pmhours
\hours = \time
\divide\hours by 60
\temp = \hours
\multiply\temp by 60
\minutes = \time
\advance\minutes by -\temp
\def\hour{\the\hours}
\def\minute{\ifnum\minutes<10 0\the\minutes
            \else\the\minutes\fi}
\def\clock{
\ifnum\hours=0 12:\minute\ AM
\else\ifnum\hours<12 \hour:\minute\ AM
      \else\ifnum\hours=12 12:\minute\ PM
            \else\ifnum\hours>12
                 \pmhours=\hours
                 \advance\pmhours by -12
                 \the\pmhours:\minute\ PM
                 \fi
            \fi
      \fi
\fi
}

\def\monthname{\relax\ifcase\month 0/\or January\or February\or
   March\or April\or May\or June\or July\or August\or September\or
   October\or November\or December\else\number\month/\fi}

\def\bold#1{\setbox0=\hbox{$#1$}%
     \kern-.025em\copy0\kern-\wd0
     \kern.05em\copy0\kern-\wd0
     \kern-.025em\raise.0433em\box0 }

\def\beq{\begin{equation}}
\def\eeq{\end{equation}}

\def\ga{\mathrel{\raise.3ex\hbox{$>$\kern-.75em\lower1ex\hbox{$\sim$}}}}
\def\la{\mathrel{\raise.3ex\hbox{$<$\kern-.75em\lower1ex\hbox{$\sim$}}}}
\def\gev{{\rm \, Ge\kern-0.125em V}}
\def\tev{{\rm \, Te\kern-0.125em V}}
\def\gyr{{\rm \, G\kern-0.125em yr}}
\def\ohsq{\Omega_{\chi} h^2}

\def\gappeq{\mathrel{\rlap {\raise.5ex\hbox{$>$}}
{\lower.5ex\hbox{$\sim$}}}}
\def\lappeq{\mathrel{\rlap{\raise.5ex\hbox{$<$}}
{\lower.5ex\hbox{$\sim$}}}}
\def\Toprel#1\over#2{\mathrel{\mathop{#2}\limits^{#1}}}

\def\PL{{\it Phys. Lett.} }
\def\PR{{\it Phys. Rev.} }

\def\NP{{\it Nucl. Phys.} }

\def\m12{m_{1\!/2}}

\begin{document}
\begin{titlepage}
\pagestyle{empty}
\baselineskip=21pt
\rightline{\tt hep-ph/0508198}
\rightline{CERN-PH-TH/2005-111, UMN--TH--2404/05, FTPI--MINN--05/19}
\vskip 0.2in
\begin{center}
{\large {\bf Supersymmetric Benchmarks with Non-Universal Scalar Masses 
or Gravitino Dark Matter
}} \\
\end{center}
\begin{center}
\vskip 0.2in
{\bf A.~De~Roeck}$^1$,
{\bf J.~Ellis}$^1$, {\bf F.~Gianotti}$^1$, 
{\bf F.~Moortgat}$^1$, {\bf K.~A.~Olive}$^{2}$
and {\bf L.~Pape}$^{1,3}$
\vskip 0.1in
{\it
$^1${Physics Department, CERN, CH-1211 Geneva 23, Switzerland}\\
$^2${William I. Fine Theoretical Physics Institute,
University of Minnesota, Minneapolis, MN 55455, USA}\\
$^3${Department of Physics, ETHZ, CH-8093 Zurich, Switzerland}}\\
\vskip 0.2in
{\bf Abstract}
\end{center}
\baselineskip=18pt \noindent

We propose and examine a new set of benchmark supersymmetric scenarios,
some of which have non-universal Higgs scalar masses (NUHM) and others
have gravitino dark matter (GDM). The scalar masses in these models are
either considerably larger or smaller than the narrow range allowed for
the same gaugino mass $m_{1/2}$ in the constrained MSSM (CMSSM) with
universal scalar masses $m_0$ and neutralino dark matter. The NUHM and GDM
models with larger $m_0$ may have large branching ratios for Higgs and/or
$Z$ production in the cascade decays of heavier sparticles, whose
detection we discuss. The phenomenology of the GDM models depends on the
nature of the next-to-lightest supersymmetric particle (NLSP), which has a
lifetime exceeding $10^4$~seconds in the proposed benchmark scenarios. In
one GDM scenario the NLSP is the lightest neutralino $\chi$, and the
supersymmetric collider signatures are similar to those in previous CMSSM
benchmarks, but with a distinctive spectrum.  In the other GDM scenarios
based on minimal supergravity (mSUGRA), the NLSP is the lighter stau
slepton ${\tilde \tau}_1$, with a lifetime between $\sim 10^4$ and $3
\times 10^6$~seconds. Every supersymmetric cascade would end in a ${\tilde
\tau}_1$, which would have a distinctive time-of-flight signature.  
Slow-moving ${\tilde \tau}_1$'s might be trapped in a collider detector or
outside it, and the preferred detection strategy would depend on the
${\tilde \tau}_1$ lifetime. We discuss the extent to which these mSUGRA
GDM scenarios could be distinguished from gauge-mediated models.

\vfill
\leftline{CERN-PH-TH/2005-111}
\leftline{August 2005}
\end{titlepage}
\baselineskip=18pt

\section{Introduction}

Among the most promising prospects for new physics that might be
discovered at the LHC is supersymmetry. Its appearance at the TeV energy
scale is motivated by the stability of the mass hierarchy
\cite{hierarchy}, the lightness of the Higgs boson as inferred from
precision electroweak measurements~\cite{LEPEWWG}, the possibility of
unifying the gauge interactions~\cite{GUTs}, and (assuming that $R$-parity
is conserved) the existence of a natural candidate for cold dark matter
(CDM)~\cite{EHNOS}. However, despite these phenomenological hints and
the intrinsic beauty of supersymmetry, there is no direct evidence for its
existence. Evidence for new physics at the TeV scale might be provided by
the anomalous magnetic moment of the muon, $g_\mu - 2$, if it is confirmed
that measurements are significantly different from the Standard Model
prediction \cite{g-2}. Supersymmetry could be one of the possible
interpretations, but not the only one.

In advance of experiments at the LHC and other accelerators intended to
explore the TeV energy range in more detail, it is valuable to understand
the variety of possible signatures of supersymmetry, taking into account
the constraints imposed by present accelerator experiments as well as
astrophysics and cosmology. Experimental signatures of supersymmetry
depend quite sensitively on the possible sparticle masses, which in turn
depend on the unknown mechanism of supersymmetry breaking. There have
already been a number of surveys of the supersymmetric parameter space
\cite{oldBench,Bench,SPS,Bench2}, some focusing on specific benchmark
points that exemplify distinct possibilities, and others tracking the
phenomenological variations along lines in this space that are consistent
with the cosmological and other constraints.

Most such studies have focused on the CMSSM, in which the
supersymmetry-breaking gaugino, scalar and trilinear parameters $m_{1/2},
m_0$ and $A_0$, respectively, are each assumed to be universal at some
input GUT scale $\sim 10^{16}$~GeV, and the gravitino is assumed not to be
the lightest supersymmetric particle (LSP). However, this is not the only
possibility: universality is not strongly favoured within our current
understanding of possible underlying theoretical frameworks such as string
theory, and the gravitino might be the LSP. Therefore, in this paper we
investigate benchmark scenarios that sample models with patterns of
supersymmetry breaking different from that in the CMSSM.

The most dubious of the CMSSM universality assumptions may be that for the
soft supersymmetry-breaking scalar masses $m_0$. Universality between
sparticles in the same gauge multiplet is inevitable, and universality
between sparticles in different generations that share the same quantum
numbers is motivated by constraints from flavour-changing neutral
interactions~\cite{FCNC}.  Moreover, sparticles with different quantum
numbers may originate from common GUT multiplets, in which case their soft
supersymmetry-breaking scalar masses should also be universal at the GUT
scale, apart from possible non-universal GUT $D$ terms. However, none of
these arguments give any reason why the soft supersymmetry-breaking scalar
masses of the electroweak Higgs multiplets should be universal, and this
may be the Achilles heel of the CMSSM. Accordingly, some of the non-CMSSM
scenarios we study in this paper have non-universal Higgs masses (NUHM)
\cite{nonu,eos3}.

We assume in this paper that $R$-parity is conserved, so that the LSP is
stable and a candidate for the CDM required by astrophysics and cosmology
\cite{EHNOS}. However, even if one accepts the CMSSM universality
assumptions, it is not evident that the LSP is necessarily the lightest
neutralino $\chi$, as usually assumed in CMSSM analyses. Another plausible
candidate is the gravitino~\cite{ekn} - \cite{feng}, the supersymmetric
partner of the graviton, in which case options for the next-to-lightest
supersymmetric particle (NLSP) include the lightest neutralino and the
lighter stau slepton ${\tilde \tau}_1$. The gravitino mass $m_{3/2}$ is
poorly constrained by accelerator experiments, and the astrophysical and
cosmological constraints on the gravitino differ from those on the
neutralino. One approach to gravitino dark matter (GDM) models is to
retain the standard CMSSM universality assumptions and simply assume that
the gravitino is lighter than the lightest neutralino $\chi$. One of the
GDM benchmarks we explore is of this type, with a neutralino NLSP.  
However, allowing for flexibility in the gravitino mass introduces an
additional arbitrary parameter, and it is difficult to scan and
characterize the larger-dimensional parameter space opened up in this way.

As an alternative scenario with a lower-dimensional parameter space, we
consider a sample of GDM models in which the gravitino mass is fixed to
equal the universal soft supersymmetry-breaking masses of observable
scalar sparticles: $m_{3/2} = m_0$ at the GUT scale. This assumption is
motivated by minimal supergravity (mSUGRA) models of supersymmetry
breaking, which also impose a relation between the soft trilinear and
bilinear supersymmetry-breaking parameters: $A_0 = B_0 +
m_0$~\cite{mSUGRA}. This latter assumption may be used to fix the ratio of
MSSM Higgs vacuum expectation values $\tan \beta$, further reducing the
GDM parameter space \cite{VCMSSM}. For definiteness, we use the value $A_0
= (3 - \sqrt{3})m_0$ found in the specific Polonyi model of supersymmetry
breaking in a hidden sector~\cite{pol}. Even with this extra assumption,
there is still a two-dimensional region of the $(m_{1/2}, m_0)$ parameter
space allowed by the accelerator, astrophysical and cosmological
constraints. Below we propose GDM benchmark scenarios that explore some of
phenomenological possibilities in this case, in which the NLSP is the
${\tilde \tau}_1$~\footnote{One could also discuss GDM models with
non-universal Higgs masses, but this would increase the dimensionality of
the parameter space still further.}.

The structure of this paper is as follows. In Section~2 we give an
overview of the CMSSM, NUHM and GDM parameter spaces, and in Section~3 we
discuss in more detail the proposed new benchmark scenarios, specifying
the parameter choices and presenting their spectra as computed in the {\tt
SSARD} \cite{SSARD} and {\tt ISASUGRA} \cite{ISASUGRA} programs. We also
present calculations of the LSP relic density \cite{us,them}, the $b \to s
\gamma$ decay branching ratio \cite{bsg} and the supersymmetric
contribution to $g_\mu - 2$ \cite{g-2} in each of these scenarios, as well
as the value of $\chi^2$ found in a global fit to laboratory observables
including $m_W$ and $\sin^2 \theta_W$~\cite{ehow3}.  Next, in Section~4 we
present and discuss several of the most important sparticle decay
branching ratios in the various new benchmark scenarios. Subsequently, in
Section~5 we discuss the detectability of these benchmark scenarios at the
LHC~\cite{AHiggs,CMSTP}, ILC~\cite{ILC} and CLIC~\cite{CLIC}.  Section~6
contains a dedicated discussion of possible measurements of the metastable
${\tilde \tau}_1$ NLSP mass and its decays in the various mSUGRA-motivated
scenarios. Finally, Section~7 discusses our main conclusions and possible
directions for future work.

\section{Overview of the CMSSM, NUHM and GDM Parameter Spaces}

Before discussing specific benchmark scenarios, we first give an overview
of the CMSSM, NUHM and GDM parameter spaces, which helps to motivate the
specific parameter choices we make. In both the NUHM and GDM models, we
implement the LEP limits on $m_{\chi^\pm}$ \cite{LEPSUSY} and $m_h$
\cite{LEPHWG} and the $b \to s \gamma$ constraint \cite{bsg} in the same
way as already documented for the CMSSM \cite{Bench,Bench2,us}. We
restrict our attention to models with $\mu > 0$, as favoured by $g_\mu -
2$ \cite{g-2}. We do not impose a numerical range on the supersymmetric
contribution to this quantity, but we do refer later to a likelihood
analysis that incorporates it as well as $m_W, \sin^2 \theta_W$ and $b \to
s \gamma$ in the global $\chi^2$ function \cite{ehow3}.

\begin{figure}
\begin{center}
\begin{tabular}{c c}
\mbox{\epsfig{file=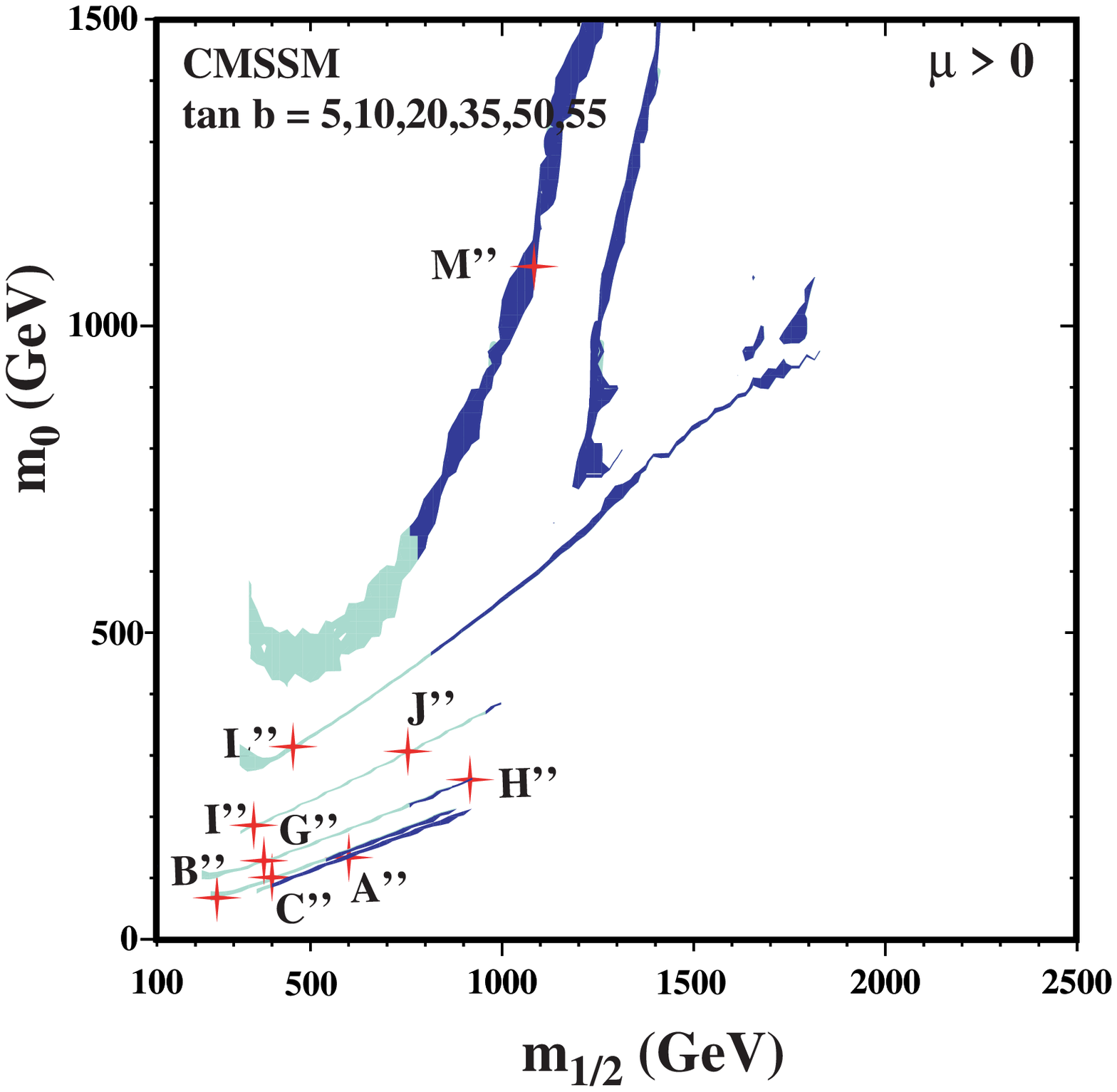,height=7cm}} &
\mbox{\epsfig{file=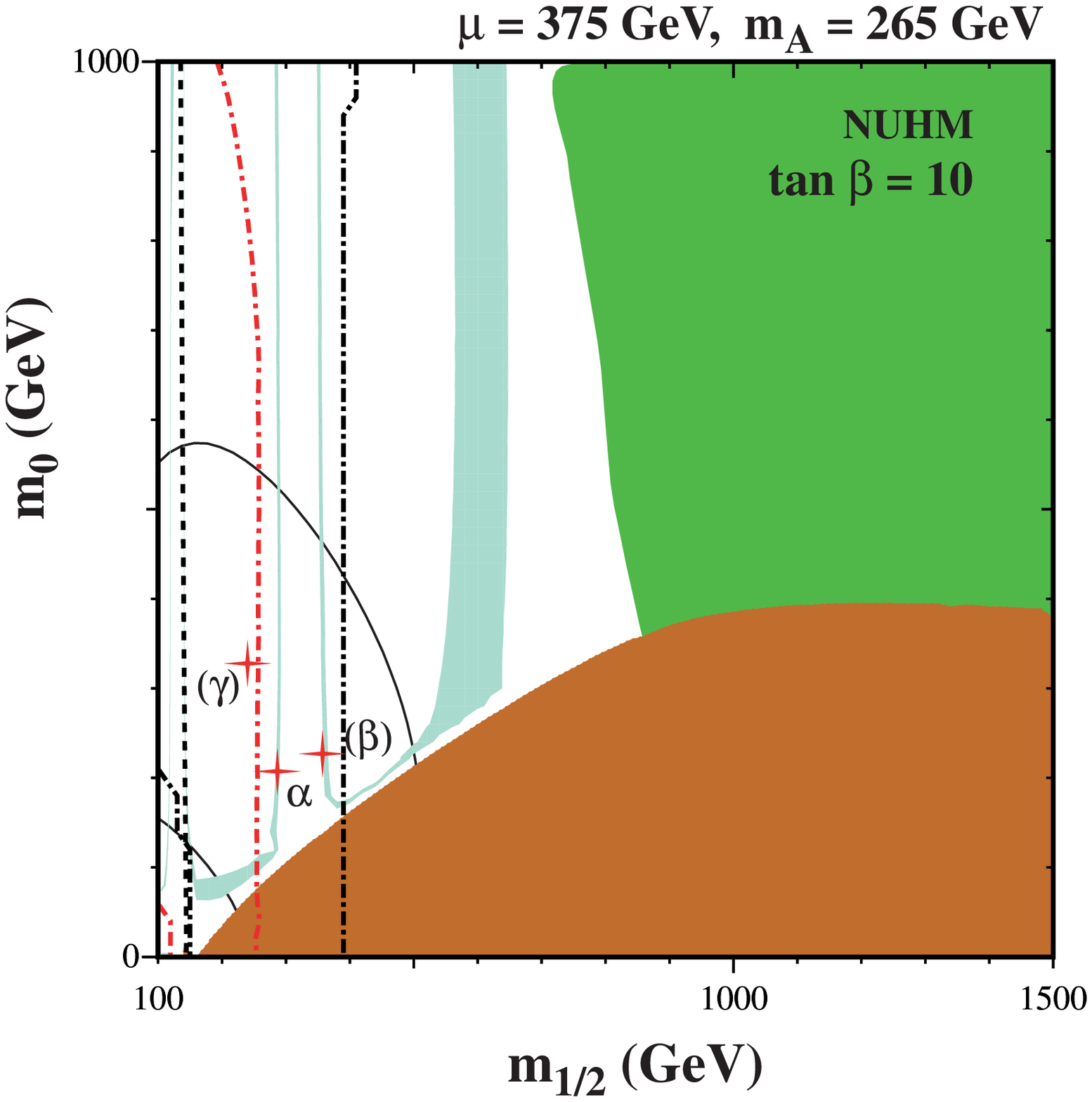,height=7cm}} \\
\end{tabular}
\end{center}   
\begin{center}
\begin{tabular}{c c}
\mbox{\epsfig{file=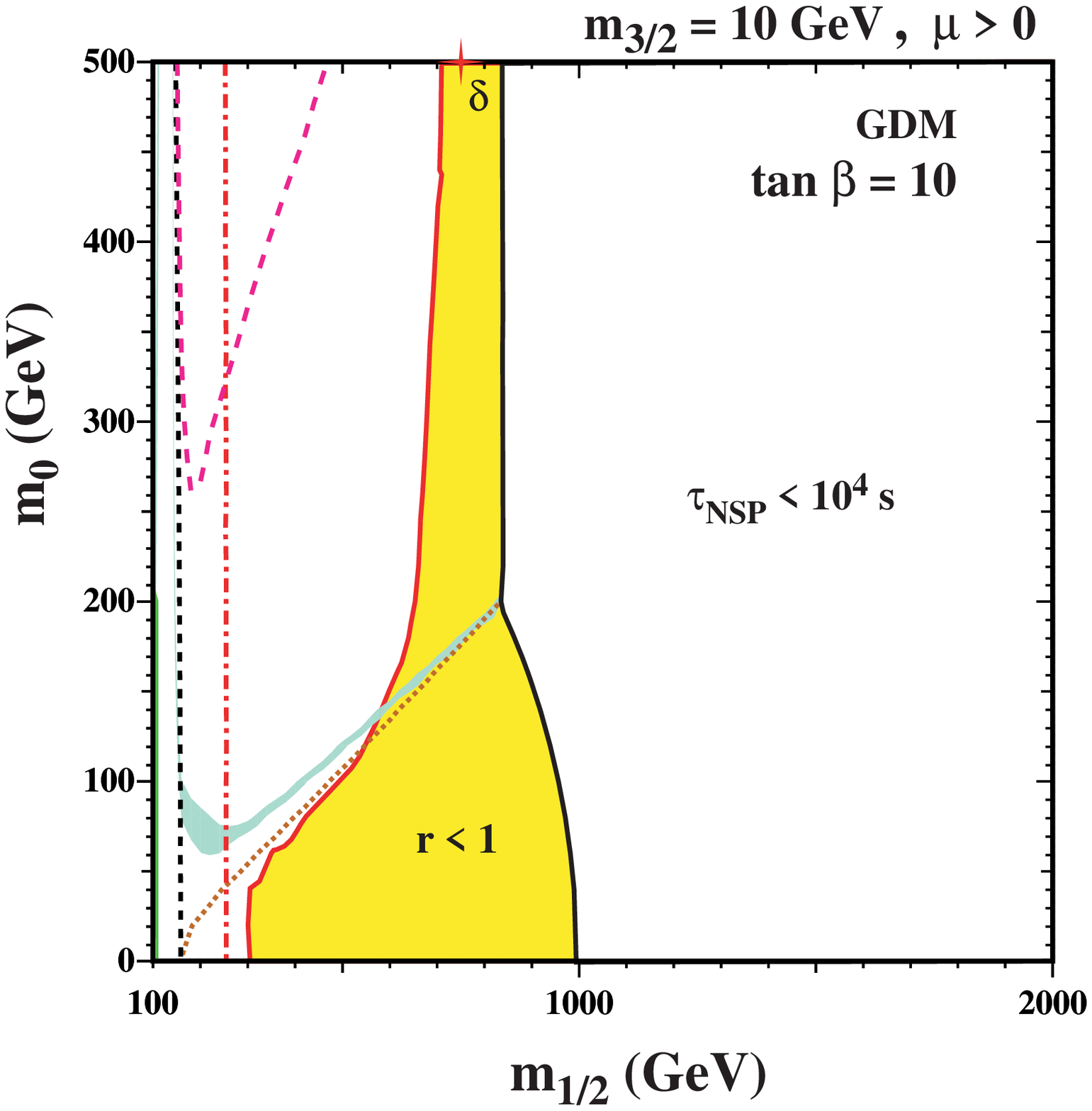,height=7cm}} &
\mbox{\epsfig{file=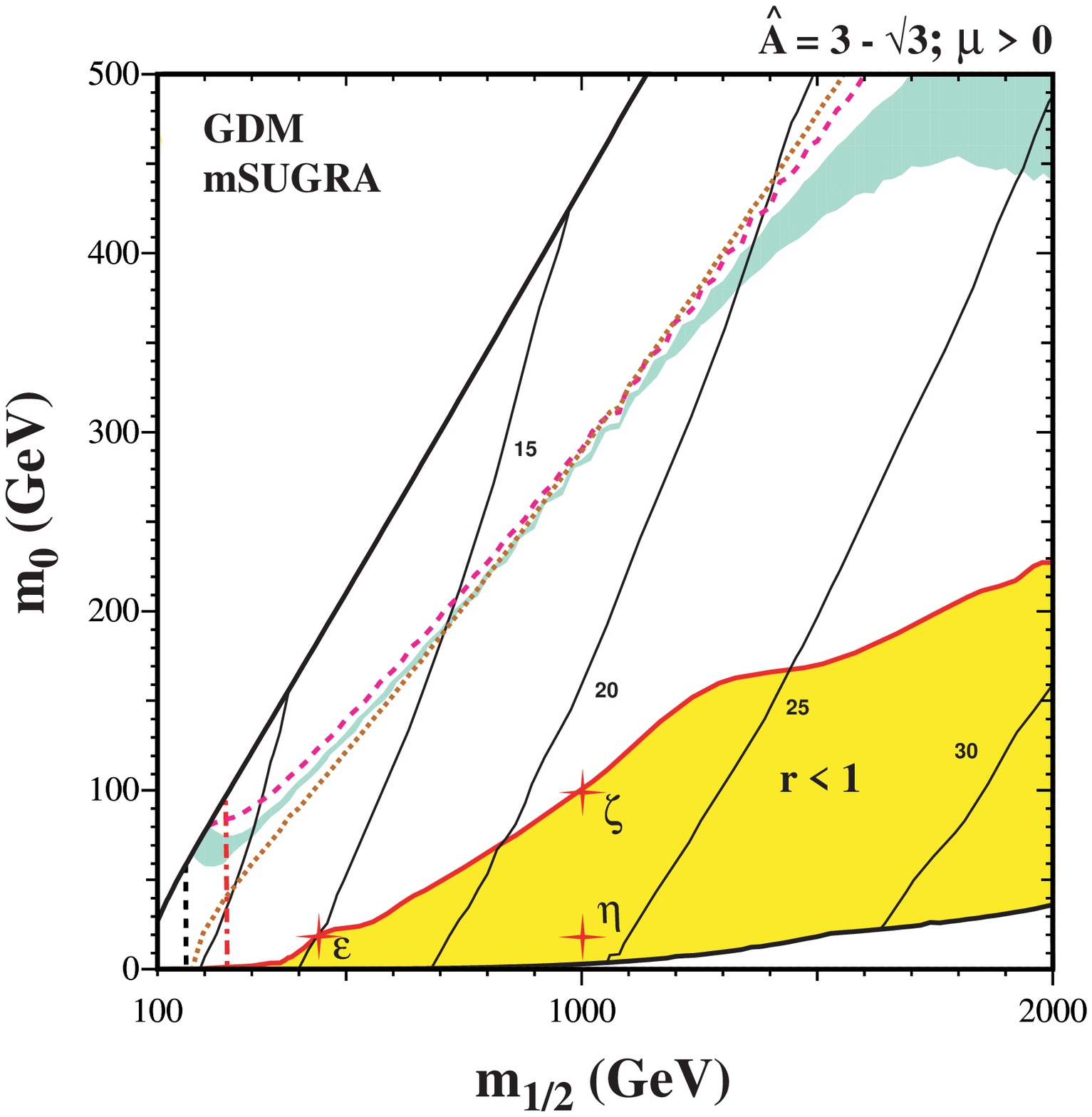,height=7cm}} \\
\end{tabular}
\end{center}   
\caption{\label{fig:planes}\it
The allowed regions in the $(m_{1/2}, m_0)$ planes for (a) the CMSSM with 
a neutralino LSP [light/dark (blue) shaded strips], (b) the NUHM with a 
neutralino LSP for $\mu = 375$~GeV and $m_A = 265$~GeV [light (blue) 
shaded strips], 
(c) the CMSSM with a 
gravitino LSP weighing 10~GeV, and (d) mSUGRA with a 
gravitino LSP [light (yellow) shaded 
regions labelled $r < 1$], for $A \equiv {\hat A} m_0: {\hat A} = 3 - 
\sqrt{3}$. In each 
case, the proposed benchmark scenarios are indicated by 
(red) crosses. Panel (a) displays WMAP strips for $\tan \beta = 
5, 10, 20, 35, 50, 55$, panels (b, c) have $\tan \beta = 10$, and in panel 
(d) $\tan \beta$ is 
fixed by the electroweak vacuum conditions with the values indicated 
along the (solid black) contours.
}
\end{figure}

We first consider the CMSSM, assuming that the gravitino is not the LSP.
We recall that the cosmological CDM density constraint:
\begin{equation}
0.094 < \Omega_{CDM} h^2 < 0.129
\label{WMAP}
\end{equation}
provided by WMAP \cite{WMAP} and earlier data
restricts the CMSSM parameter space to narrow strips in the $(m_{1/2},
m_0)$ planes for specific choices of $\tan \beta$ and $A_0$, if
one assumes that all the CDM is provided by the neutralino LSP. The WMAP
strips foliate the $(m_{1/2}, m_0)$ plane in the CMSSM, as seen in panel
(a) of Fig.~\ref{fig:planes}, where we display updated WMAP strips for
$\mu > 0, A_0 = 0$ and $\tan \beta = 5, 10, 20, 35, 50$ and 55, as 
calculated assuming $m_t = 172.7$~GeV, the current central 
value~\cite{newmt}. The lighter
(darker) parts of these strips are (in)compatible with $g_\mu - 2$ at the
2-$\sigma$ level, if one uses $e^+ e^-$ data to calculate the Standard 
Model contribution~\cite{g-2}:
\begin{equation}
\delta a_\mu = (25.2 \pm 9.2) \times 10^{-10} .
\label{g-2}
\end{equation}
Also shown are updated versions $A^{\prime \prime}, B^{\prime \prime}, 
...$
of the previously-proposed CMSSM benchmark scenarios \cite{Bench2} that lie on these
WMAP strips. The update from primed to double-primed points
is largely due to the change in our adopted top quark mass (from 175 to 
172.7~GeV~\cite{newmt})
and improvements to {\tt SSARD} which now include full two-loop running of 
the RGEs.
We recall that most of the regions below the WMAP strips are
forbidden, because there the LSP would be charged, namely the ${\tilde
\tau}_1$, which would be stable in this case.

As noted above, the latest version of {\tt SSARD} has been used to
calculate these WMAP strips. It has an improved treatment of higher-order
corrections, which are significant at large $\tan \beta$ and particularly
in the very sensitive rapid-annihilation funnel region, which is now
barely visible for $\tan \beta = 50$ and $m_t = 172.7$~GeV, 
but present for $\tan \beta = 55$, as seen in
Fig.~\ref{fig:planes}a. We have adjusted minimally the parameters of
previous CMSSM benchmark points with $\mu > 0$, so as to leave unchanged
the previous collider phenomenology and keep the relic density
$\Omega_\chi h^2$ within the WMAP range. The corresponding changes in
(mainly) $m_0$ and (occasionally) $m_{1/2}$ and $\tan \beta$ are tabulated
in Table~\ref{tab:msugra.spectr_Olive} where we indicate the values of
$m_{1/2}$ and $m_0$ for the previous (single-primed) points \cite{Bench2}. 
Note
that we have not updated the previous `focus-point' benchmark scenarios
(these are extremely sensitive to $m_t$, whose measurements at the 
Tevatron
collider are still evolving significantly), nor the points with $\mu < 0$ 
that are disfavoured by $g_\mu - 2$.

\begin{table}
\centering
\renewcommand{\arraystretch}{0.95}
{\bf Updated CMSSM benchmark scenarios}\\
{~}\\
\begin{tabular}{|c||r|r|r|r|r|r|r|r|r|}
\hline
Model          
& A$^{\prime\prime}$   &  B$^{\prime\prime}$  &  C$^{\prime\prime}$  &  G$^{\prime\prime}$  &  H$^{\prime\prime}$  &   I$^{\prime\prime}$  &  J$^{\prime\prime}$  & L$^{\prime\prime}$   & M$^{\prime\prime}$ \\
\hline
$m_{1/2}$ 
& 600 & 250 & 400 & 375 &  910 & 350 & 750 & 450 & 1075 \\
&     &     &     &     & (935)&     &     &     &(1840)\\
\hline
$m_0$          
& 135 &  65 &  95 & 125 &  260 & 180 & 300 & 310 & 1100 \\ 
&(120)& (60)& (85)&(115)& (245)&(175)&(285)&(300)&      \\ 
\hline
$\tan{\beta}$  
& 5   & 10  & 10  & 20  &  20  & 35  & 35  & 50  &  55  \\
&     &     &     &     &      &     &     &     & (50) \\
\hline
\end{tabular}
\caption[]{\it 
We list updated GUT-scale input parameters for CMSSM benchmark 
scenarios with $\mu > 0$ in the coannihilation and rapid-annihilation 
funnel regions. In scenarios where the values of $m_{1/2}, m_0$ and/or 
$\tan \beta$ have been changed so as to keep $\Omega_\chi h^2$ within the 
WMAP range, as calculated with the latest version of {\tt SSARD}, the 
previous input values are listed below in brackets. These points are 
valid for $m_t = 172.7$~GeV.} 
\label{tab:msugra.spectr_Olive} 
\end{table}

For any given value of $m_{1/2}$ and $m_0$, the NUHM has two additional
parameters, reflecting the degrees of non-universality $\delta_{u,d}$ of 
the masses of the two MSSM Higgs doublets~\cite{nonu}: $m_{H_{u,d}}^2 = 
(1 + \delta_{u,d}) m_0^2$. One representative example of an NUHM
plane is shown in panel (b) of Fig.~\ref{fig:planes}, where we have
assumed for definiteness that $\mu = 375$~GeV and $m_A = 265$~GeV,
using $m_t = 178$~GeV~\footnote{At this value of $\tan \beta =10$, the 
NUHM plane for $m_t = 172.7$~GeV would be virtually identical.}. The
pale (turquoise) region allowed by cosmology includes a small `bulk'
region at low $m_{1/2}$ and $m_0$ that extends into a short coannihilation
strip. There is then a near-vertical rapid-annihilation funnel whose
location is determined by the choice of $m_A$, which is flanked by very
narrow allowed WMAP strips at both larger and smaller values of $m_{1/2}$.
There is then a continuation of the coannihilation strip and finally a
third, broader vertical band where the relic density falls within the
range allowed by WMAP. The latter band is a result of the fact that the 
LSP is becoming more Higgsino-like as $m_{1/2} > \mu$.
The dark (brick) shaded region at large $m_{1/2}$
and small $m_0$ is forbidden because here the LSP would the (stable)  
${\tilde \tau}_1$ and the medium (green) shaded region at large 
$m_{1/2}$ and $m_0$ is excluded by $b
\to s \gamma$. Only regions between the two near-vertical black dot-dashed
lines have effective potentials that are stable up to the GUT scale, and
are hence permissible theoretically. The near-vertical black dashed and
red dot-dashed lines represent the LEP constraints on $m_{\chi^\pm}, m_h$
respectively~\footnote{The elliptical solid black lines bound the
preferred range of $g_\mu - 2$, if the Standard Model contribution is
calculated using $e^+ e^-$ data alone.}.  Thus, only the two narrow WMAP
strips above the dark (brick) shaded region between the near-vertical red
and black dot-dashed lines are consistent with all the constraints.

We see that one of the options opened up by the NUHM is a range of values
of $m_0$ that are considerably larger than the very narrow range of values
allowed by the CDM constraint in the CMSSM coannihilation strip, for any
fixed values of $m_{1/2}, \tan \beta$ and $A_0$.  In the next Section, we 
exploit this freedom
to increase $m_0$ by proposing three larger-$m_0$ scenarios shown as red
crosses in the $(m_{1/2}, m_0)$ plane. One of these has the same values of
$\mu, m_A$ as those chosen in panel (b) of Fig.~\ref{fig:planes}, whereas
the other two (indicated in brackets) have different values of $\mu, m_A$. 
As discussed
in more detail below, the specific values of $\mu, m_A, m_{1/2}$ and $m_0$
were chosen so as to offer various different sparticle cascade decay 
signatures including
decays of the second-lightest neutralino $\chi_2 \to h \chi, Z \chi$ that 
are not
favoured in CMSSM scenarios with a neutralino LSP, as was discussed
in~\cite{Bench2}.

Similar large values of $m_0$ are also attainable in the CMSSM, if the LSP
is the gravitino. A representative sample $(m_{1/2}, m_0)$ plane in such a
GDM scenario is shown in panel (c) of Fig.~\ref{fig:planes}, where we made
the particular choice $m_{3/2} = 10$~GeV~\footnote{As in the case of panel 
b), we use here $m_t = 178$~GeV, but the plane for $m_t = 172.7$~GeV would
again be virtually identical.}. The light (yellow) shaded region labelled 
by $r<1$ is that allowed not only by accelerator constraints (the LEP
$m_{\chi^\pm}$ and $m_h$ limits are shown as near-vertical dashed black
and dot-dashed red lines, respectively) but also by astrophysical and
cosmological constraints~\cite{CEFO}. Only below the diagonal dashed 
purple line can
one satisfy the CDM constraint on the relic density of gravitinos produced
in decays of the NLSP. However,
the most stringent cosmological constraints in this GDM scenario come from 
comparing
the baryon-to-entropy ratio inferred from measurements of the Cosmic
Microwave Background (CMB) with that inferred from the measured Big-Bang
Nucleosynthesis (BBN) calculations and light-element abundances, which
might have been altered by NLSP decay products \cite{CEFO,gdm}~\footnote{Here and in panel
(d), we restrict our attention to regions of GDM parameter space where the
NLSP lifetime $\tau_{NLSP} > 10^4$~s, for which the only relevant decay
products are photons and electrons.}. The light elements whose abundances 
we include in
this analysis are $^4$He, $^3$He, Deuterium, $^6$Li and $^7$Li. Also shown
in panel (c) of Fig.~\ref{fig:planes} is the dotted line where $m_\chi =
m_{{\tilde \tau}_1}$, above which the NLSP is the lightest neutralino
$\chi$, and below which the NLSP is the lighter stau slepton ${{\tilde
\tau}_1}$, which decays predominantly into $\tau$ gravitino. For 
comparison, the
region with pale (turquoise)  shading is the strip in the $(m_{1/2}, m_0)$
plane that would have been allowed if the lightest neutralino were the 
LSP.

The GDM benchmark point $\delta$ shown in panel (c) of
Fig.~\ref{fig:planes} again has much larger $m_0$ and important $\chi_2
\to h \chi, Z \chi$ decays. In this case, since the NLSP is the $\chi$ 
which has a
lifetime $\sim 1.8 \times 10^4$~s, it has no collider signature apart from
missing transverse energy. This scenario therefore looks qualitatively
similar to the CMSSM scenarios discussed earlier~\cite{Bench,Bench2},
apart from its different and larger value of $m_0$.

In probing the possibilities for GDM with values of $m_0$ below the
$m_\chi = m_{{\tilde \tau}_1}$ line in the $(m_{1/2}, m_0)$ plane, we 
propose below to study more restricted mSUGRA scenarios with $m_{3/2} = 
m_0, A_0
= (B_0 + 1) m_0$ and the Polonyi choice $A_0 = (3 - \sqrt{3}) 
m_0$~\cite{pol}, as seen in panel
(d) of Fig.~\ref{fig:planes}, which was produced using $m_t = 
178$~GeV~\footnote{Lowering $m_t$ to 172.7~GeV would mostly affect the 
$\tan \beta$ contours, typically increasing
$\tan \beta$ by $\sim 4$.}.  The dashed, dot-dashed and dotted lines and
the pale (turquoise) strip have the same significances as in panel (c).
Also shown as solid lines are contours of $\tan \beta$, as fixed by the
electroweak vacuum conditions. In the low-$m_0$ region, the NLSP is the
${\tilde \tau}_1$, and its lifetime within the light (yellow) shaded
region varies between $\sim 3 \times 10^6$ and $10^4$~s.

We propose to survey this wedge-shaped GDM region by studying the three
indicated points $\epsilon, \zeta, \eta$ located at the vertex and along
the top and bottom sides of the wedge. In all these benchmark scenarios,
the metastable ${\tilde \tau}_1$ would be detectable as a charged particle
with an anomalously long time of flight. As pointed out
in~\cite{Nojiri,Fengslep}, one might hope to trap some of the 
slower-moving
charged NLSPs and detect their decays. However, the strategies and
prospects for detecting and trapping ${\tilde \tau}_1$'s would be rather
different for NLSPs with lifetimes measured in hours or months, as we
discuss later.

\section{Proposed New Benchmark Scenarios}

\begin{figure}
\begin{center}
\begin{tabular}{c c}
\mbox{ \epsfig{file=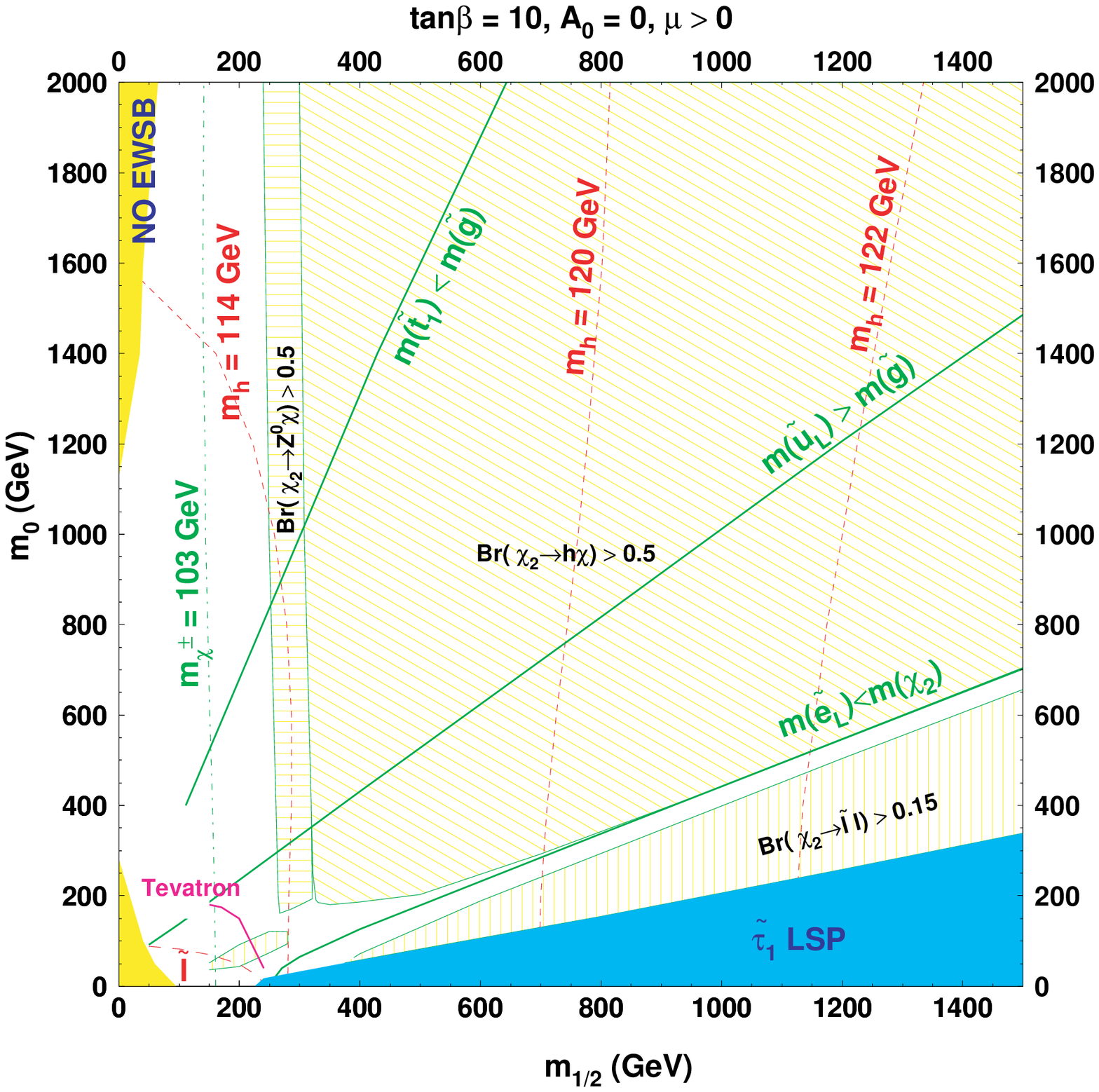,height=7.5cm}} &
\mbox{\epsfig{file=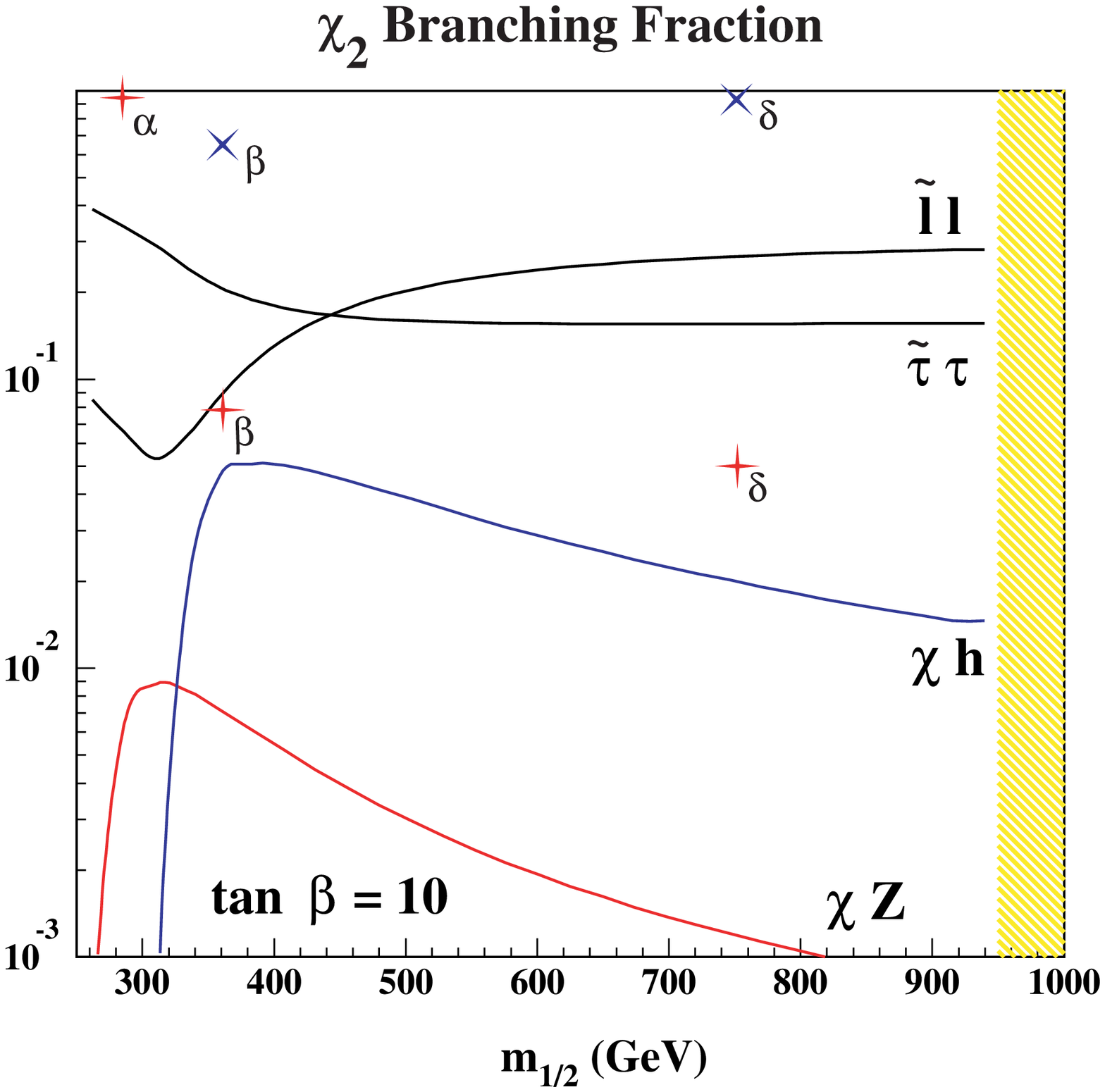,height=7.5cm}} \\
\end{tabular}
\end{center}   
\caption{\label{fig:Luc}\it
Left panel: Illustration of relevant aspects of the sparticle spectrum 
ordering and important decay modes in the $(m_{1/2}, m_0)$ plane of 
the CMSSM for $\tan \beta = 10, \mu 
> 0$ and $A_0 = 0$. Right panel: Decay modes of the second neutralino $\chi_2
\to \chi Z$ (plus signs) and $\chi h$ (crosses) in various NUHM and GDM 
benchmark scenarios,
compared with the corresponding branching ratios along the the WMAP line
for $\tan \beta = 10$ and $\mu > 0$ discussed
in~\protect\cite{Bench2}. These plots were obtained with {\tt PYTHIA
6.215}~\protect\cite{PYTHIA} interfaced to
{\tt ISASUGRA~7.69}~\protect\cite{ISASUGRA}.}
\end{figure}

We now describe in more detail the proposed new benchmark scenarios. As
shown in Fig.~\ref{fig:planes}(a), previous benchmarks were located along
the WMAP lines in the CMSSM parameter space where $0.094 < \Omega_\chi h^2
< 0.129$, which are very narrow in $m_0$ for any fixed values of $m_{1/2},
\tan \beta$ and $A_0$. As seen in Fig.~\ref{fig:Luc}, the orderings of
sparticle masses vary in important ways across the CMSSM $(m_{1/2}, m_0)$
plane, with important implications for the allowed and dominant sparticle
decay modes. However, within the CMSSM, points with larger $m_0$ generally
have larger values of $\Omega_\chi h^2$, and hence are cosmologically
unacceptable~\footnote{We recall, however, the allowed focus-point region
at very large $m_0$, whose location is very sensitive to $m_t$. The central value of
$m_t$ has changed significantly as new Tevatron Run~II
measurements have been taken into account~\cite{newmt}, and, awaiting its 
stabilization, we do not discuss the focus-point
region further in this paper. For reference, we note here also that we
assume $m_b^{\overline {MS}} (m_b) = 4.25$~GeV, and that
$A_0 = 0$, except for the mSUGRA GDM scenarios.}.

In particular, along the CMSSM WMAP strips and specifically in the
previous benchmark scenarios, the branching ratios $\chi_2 \to h \chi,
Z \chi$ are generally quite small. The decays $\chi_2 \to {\tilde \ell} 
\ell$
dominate in this region, as illustrated in both panels of 
Fig.~\ref{fig:Luc}.
This is rather exceptional, since, as seen in the left panel, there is a
large region of the CMSSM parameter space where $\chi_2 \to \chi h$ decays
dominate. These could play an important role in the discovery of the
lightest Higgs boson $h$, as well as in reconstructing the sparticle
cascade decays~\cite{AHiggs,CMSh}. There is also a band of low values of 
$m_{1/2}$ and
relatively large values of $m_0$ where $\chi_2 \to \chi Z$ decays are
important, which does not happen along the WMAP strips, as seen in the
right panel of Fig.~\ref{fig:Luc}. Even more strikingly, in models with
values of $m_0$ significantly smaller than the WMAP strip value for any
given choices of $m_{1/2}, \tan \beta$ and $A_0$, the LSP would be the
${\tilde \tau}_1$, but such a charged LSP would be forbidden by
astrophysics.  However, scenarios with values of $m_0$ very different from
those in the WMAP strips can be consistent with cosmology beyond the
specific CMSSM framework considered in~\cite{Bench,Bench2}, as we now
discuss.

\subsection{NUHM Benchmark Scenarios}

When one considers models with non-universal soft supersymmetry-breaking
contributions to the Higgs masses (NUHM), the electroweak vacuum
conditions no longer fix $|\mu|$ and the pseudoscalar Higgs mass $m_A$ as
they do in the CMSSM, so that both $m_0$ and the $\chi_2 - \chi$ mass
difference can be increased in such a way that the
decays $\chi_2 \to \chi h, Z$ are kinematically accessible but not
competing decay modes into sleptons. We have chosen to exemplify the new
phenomenology these decay modes open up by proposing consideration of
three NUHM benchmark scenarios with relatively small sparticle masses that
are beyond the reach of the Fermilab Tevatron collider, but would provide 
plenty of early physics opportunities for the LHC and
substantial follow-up opportunities for the ILC~\footnote{Their values of 
$m_{1/2}, m_0$ and $\tan \beta$ approximate those of benchmarks chosen 
for study by the CMS Collaboration~\cite{CMSh}.}.

As seen in Table~\ref{tab:points} and as a (red) cross in
Fig.~\ref{fig:planes}(b), NUHM benchmark $\alpha$ has $\tan \beta = 10$
and $m_{1/2} = 285$~GeV, which is close to the value $m_{1/2} = 250$~GeV
in the CMSSM benchmark B$^{\prime\prime}$ (or SPS1a~\cite{SPS}). However, 
as seen in panel
(b) of Fig.~\ref{fig:planes}, it has a larger value of $m_0 = 210$~GeV, as
compared to the CMSSM benchmark B$^{\prime\prime}$ value $m_0 = 65$~GeV.
This ensures that $\chi_2 \to {\tilde \ell} \ell$ decays are kinematically
forbidden, as well as $\chi_2 \to \chi h$, whereas $\chi_2 \to \chi Z$ is
possible and prominent, thanks to the mass difference $m_{\chi_2} - m_\chi
= 99$~GeV. As also seen in Table~\ref{tab:points}, this benchmark has
$\mu = 375$~GeV and $m_A = 265$~GeV, as for the $(m_{1/2}, m_0)$ plane
shown in panel (b) of Fig.~\ref{fig:planes}. The choice of $m_A$ has the
important consequence that rapid annihilation via the $A$ pole occurs at
lower $m_{1/2}$, reducing the $\chi$ LSP density to the allowed
cosmological range, as seen from the location of point $\alpha$ on one of
the thin WMAP strips in panel (b)  of
Fig.~\ref{fig:planes}~\footnote{There are similar effects for the other
NUHM benchmarks introduced below.}. We also note the degrees of
non-universality in the soft supersymmetry-breaking Higgs masses:
\begin{equation}
m^2_{H_u} = - (333 {\rm GeV})^2, \; m^2_{H_d} = + (294 {\rm GeV})^2,
\label{nonunivalpha}
\end{equation}
where $H_u$ ($H_d$) give masses to the $u$-type quarks ($d$-type 
quarks and charged leptons), respectively.
These correspond to violations of Higgs universality that are ${\cal 
O}(1)$~\footnote{Despite the negative value of 
$m^2_{H_u}$, the contribution of $|\mu|^2$ to the effective Higgs masses 
ensures that there is no vacuum instability below the GUT scale, apart 
from that induced at the electroweak scale by the top quark.}. 
Coincidentally, the value of $m_{1/2}$ at the benchmark
point $\alpha$ is close to the value of $m_{1/2}$ at the $\tan \beta
= 10$ CMSSM point that minimized $\chi^2$ in a fit to accelerator data
including $m_W, \sin^2 \theta_W, b \to s \gamma$ and $g_\mu  - 2$ \cite{ehow3}. The
value of $\chi^2$ is not very sensitive to $m_0$, so its value at
benchmark $\alpha$ is not very different from the minimum value in the 
CMSSM, as
seen in Table~\ref{tab:others}. This Table also lists this point's
values of $\Omega_\chi h^2, b \to s \gamma$ and the supersymmetric 
contribution to $g_\mu -  $2.

\begin{table}[p!]
\centering
\renewcommand{\arraystretch}{0.80}
{\bf Supersymmetric spectra in NUHM and GDM benchmark scenarios}\\
{~}\\
\begin{tabular}{|c||r|r|r||r|r|r|r|}
\hline
Model          & $\alpha$ & $\beta$ & $\gamma$ & $\delta$ & $\epsilon$ &  
$\zeta$ & $\eta$ \\ 
\hline
$m_{1/2}$      & 285 & 360 & 240 & 750 & 440 & 1000 & 1000 \\
$m_0$          & 210 & 230 & 330 & 500 &  20 &  100 &   20 \\
$\tan{\beta}$  & 10  & 10  & 20  & 10  & 15  & 21.5 & 23.7 \\
sign($\mu$)    & $+$ & $+$ & $+$ & $+$ & $+$ & $+$  & $+$ \\ 
$A_0$          & 0   & 0   &  0  &  0  & 25  & 127  & 25   \\
$m_t$          & 178 & 178 & 178 & 178 & 178 & 178  & 178 \\ \hline
Masses         &     &     &     &     &     &      &      \\ \hline
$|\mu |$       & 375 & 500 & 325 & 927 & 578 &1176  &1161  \\ \hline
h$^0$          & 115 & 117 & 114 & 122 & 119 & 124  & 124  \\
H$^0$          & 266 & 325 & 240 &1177 & 641 &1307  &1277  \\
A$^0$          & 265 & 325 & 240 &1177 & 641 &1307  &1277  \\
H$^{\pm}$      & 277 & 335 & 253 &1180 & 646 &1310  &1279  \\ \hline
$\chi^0_1$     & 113 & 146 &  95 & 323 & 183 & 436  & 436  \\
$\chi^0_2$     & 212 & 279 & 178 & 625 & 349 & 840  & 840  \\
$\chi^0_3$     & 388 & 515 & 341 & 954 & 578 &1176  &1165  \\
$\chi^0_4$     & 406 & 528 & 358 & 964 & 593 &1186  &1175  \\
$\chi^{\pm}_1$ & 212 & 279 & 177 & 625 & 349 & 840  & 840  \\
$\chi^{\pm}_2$ & 408 & 529 & 360 & 965 & 594 &1186  &1176  \\ \hline
$\tilde{g}$    & 674 & 835 & 575 &1610 & 986 &2097  &2097  \\ \hline
$e_L$, $\mu_L$ & 296 & 346 & 376 & 702 & 298 & 664  & 657  \\
$e_R$, $\mu_R$ & 216 & 241 & 328 & 571 & 169 & 383  & 370  \\
$\nu_e$, $\nu_{\mu}$
               & 285 & 337 & 367 & 697 & 287 & 660  & 652  \\
$\tau_1$       & 212 & 239 & 315 & 564 & 150 & 340  & 322  \\
$\tau_2$       & 298 & 348 & 377 & 700 & 302 & 661  & 655 \\
$\nu_{\tau}$   & 285 & 337 & 364 & 695 & 285 & 651  & 644  \\ \hline
$u_L$, $c_L$   & 648 & 793 & 612 &1532 & 897 &1892  &1889  \\
$u_R$, $c_R$   & 637 & 778 & 607 &1480 & 867 &1817  &1814  \\
$d_L$, $s_L$   & 653 & 797 & 617 &1534 & 901 &1893  &1891  \\
$d_R$, $s_R$   & 630 & 768 & 599 &1474 & 864 &1807  &1805  \\
$t_1$          & 471 & 596 & 434 &1159 & 682 &1465  &1472  \\
$t_2$          & 652 & 784 & 600 &1429 & 879 &1758  &1756  \\ 
$b_1$          & 590 & 727 & 540 &1395 & 824 &1726  &1723  \\
$b_2$          & 629 & 767 & 594 &1468 & 862 &1781  &1775  \\ \hline
\hline
\end{tabular}
\caption[]{\it
Proposed NUHM ($\alpha, \beta, \gamma$) and GDM ($\delta, \epsilon, \zeta, 
\eta$) benchmark points and mass spectra (in GeV), as 
calculated using {\tt SSARD}~\protect\cite{SSARD} and {\tt
FeynHiggs}~\protect\cite{Heinemeyer:2000yj}, using the
one-loop corrected effective potential computed at the electroweak scale
and one-loop corrections to the chargino and neutralino masses.
We use here $m_b(m_b)^{\overline {MS}} = 4.25$~GeV and
$m_t = 178$~GeV.}               
\label{tab:points}
\end{table}

\begin{table}
\centering
\renewcommand{\arraystretch}{0.90}
{\bf Relic density $\ohsq, b \to s \gamma$ and $g_\mu - 2$ in post-WMAP 
benchmark scenarios, also $\tau_{NLSP}$ in the GDM models}\\
{~}\\
\begin{tabular}{|c||c|c|c||c|c|c|c|}
\hline
~  & $\alpha$ & $\beta$ & $\gamma$ & $\delta$ & $\epsilon$ & $\zeta$ & 
$\eta$ \\
\hline
$\Omega_{LSP} h^2$   & 0.12 & 0.10 & 0.09 & 0.07 & $0.9 \times 10^{-3}$ & 
$0.9 \times 10^{-2}$ & $1.6 \times 10^{-3}$ \\
$\delta a_{\mu}$($10^{-9}$)
               & 1.5 & 1.0 & 2.6 & 0.2 & 1.8 & 0.5  & 0.5  \\
$B_{s \gamma}$($10^{-4}$)
               & 4.1 & 4.4 & 2.8 & 3.7 & 3.6 & 3.6  & 3.6  \\ 
$\tau_{NLSP}$(s) & $ --- $ & $ --- $ & $ --- $ & $1.8 \times 10^4$ &
$3.3 \times 10^6$ & $2.0 \times 10^6$ & $6.8 \times 10^4$ \\
$\chi^2$ & 1.93 & 3.67 & 1.98 & 6.81 & 1.15 & 6.25 & 5.99 \\
\hline
\end{tabular}
\caption[]{\it Comparison of $\Omega_{LSP} h^2$ for the benchmark points 
in Table~\ref{tab:points}, as
computed with the {\tt SSARD} code~\protect\cite{SSARD},
$\delta a_\mu$($10^{-9}$), the branching ratio for $b \to s \gamma$,
the NLSP lifetime for GDM scenarios  and the 
$\chi^2$ for a global fit to precision observables~\cite{ehow3}.}
\label{tab:others}
\end{table}

As also seen in Table~\ref{tab:points} and as a bracketed (red) cross 
in Fig.~\ref{fig:planes}(b), NUHM benchmark $\beta$ has $\tan
\beta = 10$ but the somewhat higher value of $m_{1/2} = 360$~GeV and
$m_0 = 230$~GeV.
However, it has $\mu = 500$~GeV and $m_A = 325$~GeV, for which choices 
the $(m_{1/2}, m_0)$ plane would look somewhat different. As for point 
$\alpha$, the choice of $m_A$ ensures a suitable relic density via 
rapid annihilation. The 
non-universal soft supersymmetry-breaking Higgs masses-squared are:
\begin{equation}
m^2_{H_u} = - (461 {\rm GeV})^2, \; m^2_{H_d} = + (229 {\rm GeV})^2.
\label{nonunivbeta}
\end{equation}
These again correspond to violations of Higgs universality that are ${\cal 
O}(1)$. With the larger value of $\mu$ as 
compared to point $\alpha$,
the $\chi_2 - \chi$ mass difference is increased
to 133~GeV, thereby opening up the decay mode $\chi_2 \to \chi h$ ($m_h = 
117$~GeV at this point). Benchmark $\beta$ lies somewhat further
from the CMSSM point with minimum $\chi^2$, but its overall value of
$\chi^2$ is still acceptable, as seen in Table~\ref{tab:others},
along with its values of $\Omega_\chi h^2, b \to s \gamma$ and $g_\mu - 
2$.

As also seen in Table~\ref{tab:points}, the final NUHM benchmark
$\gamma$ has $\tan \beta = 20$ and, as seen as a bracketed (red) cross in
Fig.~\ref{fig:planes}(b), the somewhat lower value of $m_{1/2} = 240$~GeV,
and the somewhat larger value of $m_0 = 330$~GeV. It also has $\mu =
325$~GeV and $m_A = 240$~GeV. The latter choice ensures an acceptable
relic density, although with different values of $\mu$ and $m_A$ from
those used in the rest of panel (b) of Fig.~\ref{fig:planes}. The
non-universal soft supersymmetry-breaking Higgs masses-squared are:
\begin{equation}
m^2_{H_u} = - (242 {\rm GeV})^2, \; m^2_{H_d} = + (373 {\rm GeV})^2.
\label{nonunivgamma}
\end{equation}
These also correspond to violations of Higgs universality that are ${\cal
O}(1)$, not orders of magnitude. With the choices of $\tan \beta, m_{1/2}$ 
and $\mu$, the $\chi_2 - \chi$ mass difference is {\it decreased} to 83~GeV,
closing off the $\chi_2 \to \chi Z, h$ decay modes. Moreover, with this
large value of $m_0$, the sleptons are far from the kinematic range for
$\chi_2$ decays, which are therefore dominated by non-specific three-body
decays mediated mainly by virtual $Z$ exchange. As seen in 
Table~\ref{tab:others}, the overall $\chi^2$ is again
quite small for this point, which has a value of $m_{1/2}$ rather similar
to CMSSM benchmark B$^{\prime\prime}$~\footnote{We note that the spectra
for the NUHM benchmark points were computed using $m_t = 178$~GeV.
The primary effect of lowering $m_t$ to 172.7~GeV would be to
lower $m_h$ by 2-4 GeV: squark and slepton masses would only change by 
0-2~\%.}.

\subsection{GDM Benchmark Scenarios}

As discussed in the Introduction, one may construct GDM scenarios with
values of $m_0$ that are either much larger or much smaller than that in
the CMSSM for the same values of $\tan \beta, m_{1/2}$ and $A_0$.

GDM benchmark $\delta$ described in Table~\ref{tab:points} has been chosen 
to
exemplify phenomenology at a significantly larger value of $m_0$,
specifically 500~GeV, as well as a moderately large value of $m_{1/2} =
750$~GeV. Thus, it samples the `panhandle' reaching to large $m_0$, as 
shown by the (red) cross in Fig.~\ref{fig:planes}(c). The $\chi_2 - \chi$ 
mass difference increases to 302~GeV in this scenario, and all the sparticles 
are significantly heavier than
in the previous NUHM benchmarks. We recall that the NLSP in this scenario
is the lightest neutralino $\chi$. However, the neutralino lifetime for
decays into a gravitino is $1.8 \times 10^4$~s, too long to be
detectable in the neighbourhood of a plausible collider experiment. 
Because of its larger values of $m_{1/2}$ and $m_0$, in particular, the 
overall $\chi^2$ for this model is somewhat further from the global 
minimum, but it cannot be excluded on this ground.

Here and in the following GDM models, we obtain a contribution to the LSP
relic density by first calculating the NLSP abundance following
thermalization, annihilation and freeze-out, and then reducing the density
by a factor $m_{3/2}/m_{NLSP}$ to allow for the subsequent NLSP decays to
the gravitino LSP. We see from Table~\ref{tab:others} that this
contribution to the relic gravitino density lies below the preferred WMAP
range, and this feature is even more marked for the other GDM models
discussed below. The total cold dark matter could be raised into the WMAP
range either by another gravitino production mechanism, such as thermal
production in the very early Universe~\cite{ekn}, or if there is another 
component in
the cold dark matter, such as an axion or superheavy relic.

Benchmark scenario $\delta$ is just one point in a three-dimensional space
parametrized by $(m_{1/2}, m_0, m_{3/2})$ for the specific choices $A_0 = 
0$ and $\tan \beta = 10$, which certainly includes
regions that would be more accessible in the early days of the LHC or at
the ILC. On the other hand, the high-$m_0$ panhandle extends to larger
values of $m_0$ that are not shown. This region presumably also extends to
larger values of $m_{1/2}$ than those shaded in Fig.~\ref{fig:planes}, but
these would have lower NLSP lifetimes, in which cases hadronic NLSP decays
would also have to be taken into account when evaluating the astrophysical
and cosmological constraints~\cite{CEFO}. We do not explore these options, 
since the
challenging nature of benchmark scenario $\delta$ already serves as an 
adequate
counterweight to the `easy' scenarios $\alpha, \beta$ and $\gamma$ with
their relatively low values of $m_{1/2}$ and $m_0$.

Since benchmark point $\delta$ is CMSSM-like, changing $m_t$ to 172.7 GeV
would produce very little change in the sfermion spectrum, the biggest
effect being 1.5~\% in $m_{\tilde t}$.  The Higgs mass would drop by 4
GeV, $m_A$ would drop by 3~\%, and $|\mu|$ would drop by 5~\%. The latter
two changes would also affect the heavy Higgses and heavy neutralinos and
charginos by approximately 5~\%.

In order to reduce the dimensionality of the parameter space to be
explored by the remaining GDM benchmark scenarios, we next assume an
mSUGRA framework in which $m_{3/2} = m_0$ and $A_0 = B_0 + m_0$. We
further assume that $A_0 = (3 - \sqrt{3}) m_0$~\cite{pol}. In this 
case, the
value of $\tan \beta$ is fixed by the electroweak vacuum conditions and
varies across the $(m_{1/2}, m_0)$ plane, as seen in panel (d) of
Fig.~\ref{fig:planes}. In addition to a (pale blue) WMAP strip with $\chi$ 
LSP at 
$m_{1/2} < 1$~TeV, we also see a (yellow) GDM region of parameter
space allowed by the astrophysical and cosmological constraints. It takes
the form of a wedge
that broadens as $m_{1/2}$ increases, throughout which the NLSP is the
${\tilde \tau}_1$. We choose as our next GDM benchmark $\epsilon$ the point
shown as a (red) cross that is close to the vertex of this wedge, with
$m_{1/2} = 440$~GeV and $m_0 = 20$~GeV. In this case, $\tan \beta = 15$,
the ${\tilde \tau}_1$ NLSP has a mass of 150~GeV and a lifetime of $3.3
\times 10^6$~s.  At this point, the global $\chi^2$ is not far from the
best fit in the CMSSM, as seen in Table~\ref{tab:points}.

As seen by two more (red) crosses in panel (d) of
Fig.~\ref{fig:planes}(d), we complement this LHC- and ILC-friendly point
with two points that are more challenging. {\it A priori}, values of
$m_{1/2}$ considerably beyond 2~TeV would be possible in this wedge:
they would be beyond the reach of either the LHC or the ILC, although the 
CLIC 
reach in ${\tilde \tau}_1$ pair production would extend beyond $m_{1/2} = 
4$~TeV. We 
consider two points with $m_{1/2} = 1$~TeV, which are already quite
challenging for the LHC. The upper edge of the wedge is defined by the
astrophysical and cosmological constraint on ${\tilde \tau}_1$ decays, and
corresponds to a ${\tilde \tau}_1$ lifetime $\sim 3.3 \times 10^6$~s. The
lower edge of the wedge that we consider corresponds to a lifetime of
$10^4$~s. Benchmark point $\zeta$ is close to the upper edge, with $m_0 =
100$~GeV and $\tan \beta \simeq 21.5$. Here the ${\tilde \tau}_1$ NLSP has 
a mass of
340~GeV and a lifetime of $2 \times 10^6$~s. Benchmark point $\eta$ is 
close
to the lower edge, with $m_0 = 20$~GeV and $\tan \beta = 23.7$. Here the 
${\tilde
\tau}_1$ NLSP has a mass of 322~GeV and a lifetime of $6.8 \times 10^4$~s.  
Both these points have rather larger values of $\chi^2$ than the best fit
in the CMSSM, as also seen in Table~\ref{tab:points}, but these points 
cannot be excluded on these grounds.

The changes in the spectra due to the shift in $m_t$ would be as follows:
$\tan \beta$ would be increased by $\sim 4$, $|\mu|$ would be lowered by
4~\% and $m_A$ would be lowered by 5-7~\%, with corresponding changes in
the heavy Higgses, neutralinos and charginos.  The light Higgs mass would
be lowered by 3-4 GeV, and changes in the sfermion masses would typically
be less than 1~\%, with the exception of the lighter stau, whose mass
would drop by about 7~\%.

\subsection{Discussion of Spectra}

As in our previous papers on CMSSM benchmarks~\cite{Bench,Bench2}, the 
parameters of the NUHM and GDM benchmarks were first specified using the 
code {\tt SSARD}~\cite{SSARD}. In order to facilitate the interfaces with 
standard 
simulation packages, the spectra calculated with {\tt SSARD} were then 
matched using parameters of the {\tt ISASUGRA~7.69} code~\cite{ISASUGRA}
to reproduce the main features of the {\tt SSARD} spectra.
The values of $m_0$ and $m_{1/2}$ were adjusted to give the same masses 
for the lightest neutralino $\chi$ and the lighter stau $\tilde{\tau}_1$.
Then the Higgs mass parameters $m_{H_u}$ and $m_{H_d}$ were varied
to reproduce the {\tt SSARD} values of $m_A$ and $\mu$.
As these choices altered slightly the values of $m_\chi$ and 
$m_{{\tilde \tau}_1}$, the procedure was then iterated. The final {\tt 
ISASUGRA~7.69} parameters are listed in Table~\ref{tab:isatable}.
This procedure was not followed for the GDM points, as our results
are less sensitive to the exact spectra, and here the {\tt SSARD} inputs 
were used. Note the difference in the sign convention for $A_0$ between the two codes.

\begin{table}[p!]
\centering
\renewcommand{\arraystretch}{0.90}
{\bf Supersymmetric spectra in NUHM and GDM benchmarks\\ calculated with
{\tt ISASUGRA~7.69}}\\
{~}\\
\begin{tabular}{|c||r|r|r||r|r|r|r|}
\hline
Model          & $\alpha$ & $\beta$ & $\gamma$ & $\delta$ & $\epsilon$ &  
$\zeta$ & $\eta$ \\ 
\hline
$m_{1/2}$      & 293 & 370 & 247 & 750 & 440 & 1000 & 1000 \\
$m_0$          & 206 & 225 & 328 & 500 &  20 &  100 &   20 \\
$\tan{\beta}$  & 10  & 10  & 20  & 10  & 15  & 21.5 & 23.7 \\
sign($\mu$)    & $+$ & $+$ & $+$ & $+$ & $+$ & $+$  & $+$ \\ 
$A_0$          & 0   & 0   &  0  &  0  &-25  &-127  &-25   \\
$m_t$          & 178 & 178 & 178 & 178 & 178 & 178  & 178 \\ \hline
Masses         &     &     &     &     &     &      &      \\ \hline
$|\mu |$       & 375 & 500 & 325 & 920 & 569 &1186  &1171  \\ \hline
h$^0$          & 115 & 117 & 115 & 122 & 119 & 124  & 124  \\
H$^0$          & 267 & 328 & 241 &1159 & 626 &1293  &1261  \\
A$^0$          & 265 & 325 & 240 &1152 & 622 &1285  &1253  \\
H$^{\pm}$      & 278 & 337 & 255 &1162 & 632 &1296  &1264  \\ \hline
$\chi^0_1$     & 113 & 146 &  95 & 310 & 175 & 417  & 417  \\
$\chi^0_2$     & 215 & 282 & 180 & 600 & 339 & 805  & 804  \\
$\chi^0_3$     & 380 & 503 & 332 & 925 & 574 &1192  &1176  \\
$\chi^0_4$     & 400 & 518 & 352 & 935 & 587 &1200  &1184  \\
$\chi^{\pm}_1$ & 215 & 283 & 180 & 601 & 340 & 807  & 806  \\
$\chi^{\pm}_2$ & 399 & 518 & 352 & 935 & 587 &1200  &1184  \\ \hline
$\tilde{g}$    & 711 & 880 & 619 &1691 &1026 &2191  &2191  \\ \hline
$e_L$, $\mu_L$ & 299 & 351 & 378 & 713 & 306 & 684  & 677  \\
$e_R$, $\mu_R$ & 216 & 241 & 328 & 572 & 171 & 387  & 374  \\
$\nu_e$, $\nu_{\mu}$
               & 287 & 340 & 368 & 703 & 290 & 669  & 662  \\
$\tau_1$       & 213 & 239 & 315 & 565 & 153 & 338  & 319  \\
$\tau_2$       & 300 & 352 & 378 & 712 & 309 & 677  & 670 \\
$\nu_{\tau}$   & 287 & 340 & 365 & 700 & 288 & 660  & 653  \\ \hline
$u_L$, $c_L$   & 674 & 826 & 636 &1604 & 935 &1991  &1998  \\
$u_R$, $c_R$   & 661 & 808 & 629 &1550 & 902 &1911  &1908  \\
$d_L$, $s_L$   & 679 & 831 & 642 &1606 & 938 &1993  &1990  \\
$d_R$, $s_R$   & 652 & 797 & 621 &1544 & 899 &1903  &1900  \\
$t_1$          & 492 & 622 & 453 &1219 & 710 &1545  &1553  \\
$t_2$          & 662 & 800 & 611 &1486 & 900 &1842  &1840  \\ 
$b_1$          & 609 & 752 & 558 &1456 & 852 &1807  &1804  \\
$b_2$          & 641 & 785 & 603 &1516 & 883 &1851  &1846  \\ 
\hline
\end{tabular}
\caption[]{\it 
Proposed NUHM and GDM benchmark points and mass spectra (in GeV), as
calculated using {\tt ISASUGRA 7.69}~\protect\cite{ISASUGRA} and adapting 
the input parameters to give the best match to the {\tt 
SSARD}~\protect\cite{SSARD}
spectra shown in Table~\ref{tab:points}, as described in the 
text.}
\label{tab:isatable}
\end{table}

As already mentioned, the first three of the new benchmarks, $\alpha,
\beta, \gamma$, are NUHM points chosen to yield rather low-mass spectra,
observable at an early stage of the LHC running, as might also point 
$\epsilon$. They also offer
interesting physics opportunities for the ILC. These points complement the
previous benchmark points B$^\prime$, C$^\prime$ and I$^\prime$
of~\cite{Bench,Bench2}, as they give rise to different search topologies.
On the other hand, points $\delta, \zeta$ and $\eta$ have heavier 
sparticles, and hence are much more challenging for both the LHC and the 
ILC.

\section{Sparticle Decay Branching Ratios}

One of the key particles appearing in sparticle decay chains is the second
neutralino $\chi_2$, whose branching ratios are quite model-dependent and
have significant impact on sparticle detectability at future 
colliders~\cite{oldBench,Bench,SPS,Bench2,AHiggs,CMSTP}.  
In particular, $\chi_2$ decays play crucial roles in reconstructing the
masses of heavier sparticles such as squarks and gluinos via cascade
decays. Moreover, $\chi_2$ decays may offer new ways to discover or
measure other new particles, such as the lightest MSSM Higgs boson $h$.
Therefore, we now use {\tt ISASUGRA~7.69} to discuss the principal
branching ratios of the $\chi_2$ in the various NUHM and GDM benchmark
scenarios introduced above, comparing them in particular with those in
the CMSSM at different points along the WMAP line for $\tan \beta = 10$
and $\mu > 0$, as discussed in~\cite{Bench2}.

We first recall the principal branching ratios of the $\chi_2$ in the
low-mass CMSSM benchmarks considered previously. In the case of point
B$^\prime$, the dominant decay mode was $\chi_2 \rightarrow \tilde{\ell}_R
\ell$ (11~\%) followed by $\tilde{\ell}_R \rightarrow \chi \ell$,
whereas in the case of point C$^\prime$ the dominant decay mode was
$\chi_2 \rightarrow \tilde{\ell}_L \ell$ (11~\%) followed by
$\tilde{\ell}_L \rightarrow \chi \ell$. On the other hand, in the case
of point I$^\prime$, which has a relatively large value of $\tan\beta =
35$, the dominant decay was $\chi_2 \rightarrow {\tilde \tau}_1 \tau$
(96~\%), followed by ${\tilde \tau}_1 \rightarrow \chi \tau$.

The new points $\alpha, \beta, \gamma$ provide qualitatively new
signatures, as shown in Fig.~\ref{fig:Luc}. At the point $\alpha$, the
$\chi_2$ mainly decays via $\chi_2 \rightarrow Z \chi$ (96~\%), which is
observable through the $Z$ leptonic decay mode. At the point $\beta$, the
main decay signature is $\chi_2 \rightarrow h \chi$ (64~\%), where the
Higgs boson can be reconstructed from its decay to $\bar{b} b$. In
addition, there are smaller branching ratios for $\chi_2 \rightarrow Z 
\chi$ (8~\%) and $\chi_2 \rightarrow {\tilde \tau}_1 \tau$ (23~\%). On 
the
other hand, point $\gamma$, which is just above the Higgs mass limit from
LEP, has direct three-body leptonic decays $\chi_2 \rightarrow \ell
\bar{\ell} \chi$ (4~\%) and $\chi_2 \rightarrow \tau \bar{\tau} \chi$
(3~\%), and the other decays are mainly $\chi_2 \rightarrow q {\bar q} 
\chi$ mediated by virtual $Z$ exchange.

At all three points, the chargino decays dominantly into $W \chi$. At
points $\alpha$ and $\beta$, the gluino is heavier than any of the squarks
and decays to $\tilde{q} q$. The $\tilde{q}_R$ decays directly to $q
\chi$, whereas the $\tilde{q}_L$ leads to cascade decays such as
$\tilde{q}_L \rightarrow q \chi_2$ (typically $\sim$ 30~\%) and
$\tilde{q}_L \rightarrow q' \chi^{\pm}$ (typically $\sim$ 60~\%). On the
other hand, at point $\gamma$ the gluino is lighter than the squarks of
the first two generations, and its dominant decay is $\tilde{g}
\rightarrow b \tilde{b}_1$ (81~\%), followed by $\tilde{b}_1 \rightarrow
b \chi_2$ (26~\%), $\tilde{b}_1 \rightarrow t \chi^{\pm}$ (36~\%) or
$\tilde{b}_1 \rightarrow t \chi^{\pm}_2$ (26~\%). The squarks of the
first two generations decay similarly as at points $\alpha$ and $\beta$
and with similar branching ratios.

As already remarked, at point $\delta$ the NLSP is the neutralino, which
looks stable from the point of view of a collider detector, and gives rise
to the usual missing-energy signature. As seen in Fig.~\ref{fig:Luc}, a
further signature is provided by $\chi_2 \rightarrow h \chi$ (91~\%),
with a smaller branching ratio for $\chi_2 \rightarrow Z \chi$ (5~\%). 
At
this point also, the gluino is heavier than any of the squarks, whose
decays are similar to those at points $\alpha$ and $\beta$.

At the last three points $\epsilon, \zeta, \eta$ with a ${\tilde \tau}_1$
NLSP, the gluino is heavier than any squark and decays to $\tilde{q} q$
with some preference for $\tilde{t}_1 t$ ($\sim$ 20~\%).  The most
important products of the subsequent squark cascade decays
are displayed in Table~\ref{tab:Luc}, which we
now explain. It is a general feature of these models that ${\tilde
q}_R \to q \chi$ with branching ratios $\simeq 100$~\%. Then $\chi \to
{\tilde \tau}_1 \tau$ with large branching ratios of 92/75/69~\% in models
$\epsilon, \zeta, \eta$, respectively, with essentially all the other
decays being $\chi \to {\tilde \ell}_R \ell$ followed by ${\tilde \ell}_R
\to \ell {\tilde \tau}_R \tau$. As a result, the dominant ${\tilde q}_R$
final states are $q {\tilde \tau}_1 \tau$, with somewhat smaller fractions
of $q \ell \ell {\tilde \tau}_1 \tau$. Analogously, in many cases ${\tilde
q}_L \to q \chi_2$, with branching ratios of 32/33/33~\% in models
$\epsilon, \zeta, \eta$, respectively, the other decays mainly being
${\tilde q}_L \to q^\prime \chi^\pm$. Many of the subsequent $\chi_2$
decays are also to ${\tilde \tau}_1 \tau$, or else ${\tilde \ell}_R \ell$
followed again by ${\tilde \ell}_R \to \ell {\tilde \tau}_R
\tau$~\footnote{There may also be some ${\tilde q}_L \to q {\tilde \tau}_2
\tau$ decays, followed by ${\tilde \tau}_2 \to {\tilde \tau}_1 Z/h$
decays.}. Thus, ${\tilde q}_L$ decays via $\chi_2$ populate the final
states $q {\tilde \tau}_1 \tau$, $q \ell \ell {\tilde \tau}_1 \tau$ and $q
\ell \ell \ell^\prime \ell^\prime {\tilde \tau}_1 \tau$. On the other
hand, ${\tilde q}_L \to q^\prime \chi^\pm$ decays mainly populate final
states containing neutrinos that would be more difficult to reconstruct,
with the possible exception of some $q^\prime \tau W {\tilde \tau}_1$ 
final states.

\begin{table}[htb]
\centering
\begin{tabular}{|l||c|c|c|}
\hline
Final state & $\epsilon$ & $\zeta $ &$\eta$   \\ 
\hline
via $\chi_2$ & \multicolumn{3}{|c|}{ }  \\ 
\hline
$\tilde{q}_L \rightarrow q l l \tilde{\tau}_1 \tau$ 
              & 6~\%      & 7~\%      & 6~\%  \\
$\tilde{q}_L \rightarrow q l l l' l' \tilde{\tau}_1 \tau$ 
              & 0.5~\%      & 2.3~\%      & 2.9~\%  \\
$\tilde{q}_L \rightarrow q (Z,h) \tilde{\tau}_1 \tau$ 
              & 1.3~\%      & 4~\%      & 4~\%  \\
$\tilde{q}_L \rightarrow q \tau \tau  \tilde{\tau}_1 \tau$ 
              & 1.2~\%      & 0.8~\%      & 0.6~\%  \\
$\tilde{q}_L \rightarrow q \tau \tau l l \tilde{\tau}_1 \tau$ 
              & 0.1~\%      & 0.3~\%      & 0.3~\%  \\
$\tilde{q}_L \rightarrow q \tilde{\tau}_1 \tau$ 
              & 4~\%      & 1.3~\%      & 1.5~\%  \\
decays with $\nu$'s
              & 18~\%      & 17~\%      & 17~\%  \\
\hline
via $\chi^{\pm}$ & \multicolumn{3}{|c|}{ }   \\ 
\hline
$\tilde{q}_L \rightarrow q' W \tilde{\tau}_1 \tau$ 
              & 6~\%      & 10~\%      & 10~\%  \\
decays with $\nu$'s
              & 57~\%      & 56~\%      & 54~\%  \\
\hline
via $\chi$ & \multicolumn{3}{|c|}{ }   \\ 
\hline
$\tilde{q}_R \rightarrow q \tilde{\tau}_1 \tau$ 
              & 92~\%      & 75~\%      & 69~\%  \\
$\tilde{q}_R \rightarrow q l l \tilde{\tau}_1 \tau$ 
              & 8~\%      & 25~\%      & 31~\%  \\
\hline
\end{tabular}
\caption{\it 
Final states for the benchmark points with Gravitino Dark Matter (GDM),
as calculated with {\tt ISASUGRA 7.69}~\protect\cite{ISASUGRA}.
}
\label{tab:Luc}
\end{table}

The general conclusion is that LHC final states in these GDM models
with ${\tilde \tau}_1$ NLSPs contain a pair of $\tau$ leptons and quite
possibly additional lepton pairs, as seen in Table~\ref{tab:Luc}. In the
case of benchmark scenario $\epsilon$, $m_{\chi_2} - m_{{\tilde \ell}_L}
\simeq 35$~GeV and $m_{\chi} - m_{{\tilde \tau}_1} \simeq 23$~GeV, so the
efficiency for picking up the additional cascade decay leptons may be
reduced at the LHC~\footnote{We note, however, that it will not be
necessary to trigger on these leptons, since the sparticle production
events will generally contain many energetic hadronic jets.}, but in
benchmark scenarios $\zeta, \eta$ these mass differences exceed 100~GeV,
and these cascade decay leptons should be readily detectable.

\section{Observability at Different Accelerators}

\subsection{Detectability at the LHC}

We now provide rough estimates of the numbers of different species of
supersymmetric particles that may be detectable at the LHC in the various
NUHM and GDM benchmark scenarios introduced above. We assume the 
production cross sections for squarks and gluinos at the
LHC that are listed in Table~\ref{tab:LHCxsec}.  The physics objects shown
in the figures below are obtained in the following way.

$\bullet$ Jets are reconstructed from particles generated by the {\tt
PYTHIA}~\cite{PYTHIA} Monte Carlo, using an iterative cone algorithm with
a cone size of 0.5 radians.  In order to model a typical LHC detector
acceptance, we require each jet to have a pseudorapidity $|\eta| < 3.0$
and a transverse energy $E_T > 20$~GeV. These jets include hadronic tau
decays.

$\bullet$ The missing transverse energy is calculated from the
transverse energies $E_T$ of the visible particles.

$\bullet$ Charged leptons $e, \mu$ are accepted if their transverse
momenta $p_T > 10$~GeV and their pseudorapidities $|\eta| < 2.4$. Their
momenta are smeared with a Gaussian error between 1~\% and
10~\%, depending on the momentum.

$\bullet$ We assume a 50~\% efficiency for identifying $b$ jets, with
mis-tagging rates of 15~\% for charm jets and 5~\% for light quarks and
gluons.

$\bullet$ We assume a 50~\% efficiency for identifying hadronic $\tau$
decays, with a 6~\% mis-tagging rate for jets with $E_T < 30$~GeV and a
1~\% mis-tagging rate for jets with $E_T > 30$~GeV~\cite{AHiggs}.

\noindent
More complete and solid results should be obtained from detailed
experimental simulations.

\begin{table}[htb]
\centering
\begin{tabular}{|c||r|r|r|r|r|r|r|}
\hline
Model          & $\alpha$ & $\beta$ & $\gamma$ & $\delta$ & $\epsilon$ &  
$\zeta$ & $\eta$ \\ 
\hline
$\sigma({\tilde g} {\tilde g})$ 
& 5.8 & 1.4 & 16 & 0.008 & 0.45 & 0.001 & 0.001 \\
$\sigma({\tilde q} {\tilde g})$   
& 16 & 4.9 & 29 & 0.062 & 2.0 & 0.008 & 0.008 \\
$\sigma({\tilde q} {\bar {\tilde q}})$  
& 4.3 & 1.4 & 5.6 & 0.017 & 0.65 & 0.003 & 0.003 \\
$\sigma({\tilde q} {\tilde q})$   
& 3.9 & 1.6 & 5.2 & 0.050 & 0.85 & 0.012 & 0.012 \\
$\sigma_{tot}({\tilde g})$       
& 27 & 7.7 & 62 & 0.078 & 2.9 & 0.010 & 0.010 \\
$\sigma_{tot}({\tilde q})$       
& 32 & 11 & 51 & 0.20 & 5.0 & 0.038 & 0.038 \\
$\sigma({\tilde t}_1)$   
& 1.1 & 0.29 & 1.7 & 0.004 & 0.13 & 0.001 & 0.001 \\
$\sigma({\tilde t}_2)$   
& 0.17 & 0.055 & 0.28 & 0.001 & 0.026 & 0.000 & 0.000 \\
\hline
\end{tabular}
\caption{
\it Cross sections in pb for models with non-universal 
Higgs masses (NUHM)
or Gravitino Dark Matter (GDM)
calculated with {\tt PROSPINO}~\cite{PROSPINO} at NLO
and with masses from {\tt SSARD}.
The squark cross section is computed for the first 5 flavours.
The quantity $\sigma_{tot}$ is the inclusive sum over all production 
mechanisms of 
the gluino or squark, e.g., $\sigma_{tot}({\tilde g}) = 2 \; 
\sigma({\tilde g} {\tilde g}) \sigma({\tilde q} {\tilde g})$.
Production of $\tilde{t}_1$ and $\tilde{t}_2$ are not included in the 
sums, but listed separately.
}
\label{tab:LHCxsec}
\end{table}

We start with the benchmark points $\alpha$ to $\delta$, adopting criteria
similar to those used previously in discussions of CMSSM benchmark
scenarios with a neutralino LSP~\cite{Bench,Bench2}.

$\bullet$ {\it Higgs bosons}: We generally follow the ATLAS and CMS
studies of the number of observable Higgs bosons as a function of $m_A$
and $\tan \beta$~\cite{AHiggs,CMSTP}, bearing in mind that they have no
significant `exotic' decay modes into non-Standard Model particles. The
lightest neutral Higgs boson $h$ is detectable at all four points, and the
heavier neutral Higgs bosons $H, A$ would be observable in scenarios
$\alpha$, $\beta$ and $\gamma$. In contrast, the charged Higgs bosons
$H^{\pm}$ would be observable only at point $\gamma$ where $\tan \beta =
20$ since, according to previous studies~\cite{AHiggs,CMSTP}, 
$H^{\pm}$
cannot be seen at the LHC when $\tan \beta = 10$, for any studied values
of the other MSSM parameters.

$\bullet$ {\it Gauginos}: The lightest neutralino $\chi$ is considered
always to be observable via the cascade decays of observed supersymmetric
particles~\footnote{We recall that, from the detector point of view, the 
$\chi$ is effectively stable in benchmark $\delta$, and hence has a 
missing transverse energy signature, as at points $\alpha, \beta, 
\gamma$.}. We consider the $\chi_2$ to be observable at the LHC if the
product of its production cross section and the relevant decay branching
ratio ($\chi h, Z$ or $\ell^+ \ell^-$) is at least 0.01~pb, corresponding
to 1000 events produced with 100~fb$^{-1}$ of integrated luminosity.
Thanks to the rather large production cross sections in
Table~\ref{tab:LHCxsec}, the $\chi_2$ is observable at all four points
in the cascade decays of squarks. On the other hand, as the lighter
chargino $\chi^{\pm}$ decays with a branching ratio $>90$~\% into $W 
\chi$,
it will be difficult to detect in cascade decays, so the possibility we
have considered is via direct production of $\chi_2 \chi^{\pm}$, leading
to tri-lepton final states. Previous studies have indicated that the
$\chi^{\pm}$ would not be observable in this mode for $m_{1/2} > 170$~GeV,
as in all the benchmark scenarios $\alpha, \beta, \gamma, \delta$. On the
other hand, we note that the associated $\chi_2 \chi^{\pm}$ production 
cross sections at points $\alpha,
\gamma$ are $\sim 2.5, 5.1$~fb, respectively, so these points would be 
worth further study.

\begin{figure}
\begin{center}
\begin{tabular}{c c}
\mbox{\epsfig{file=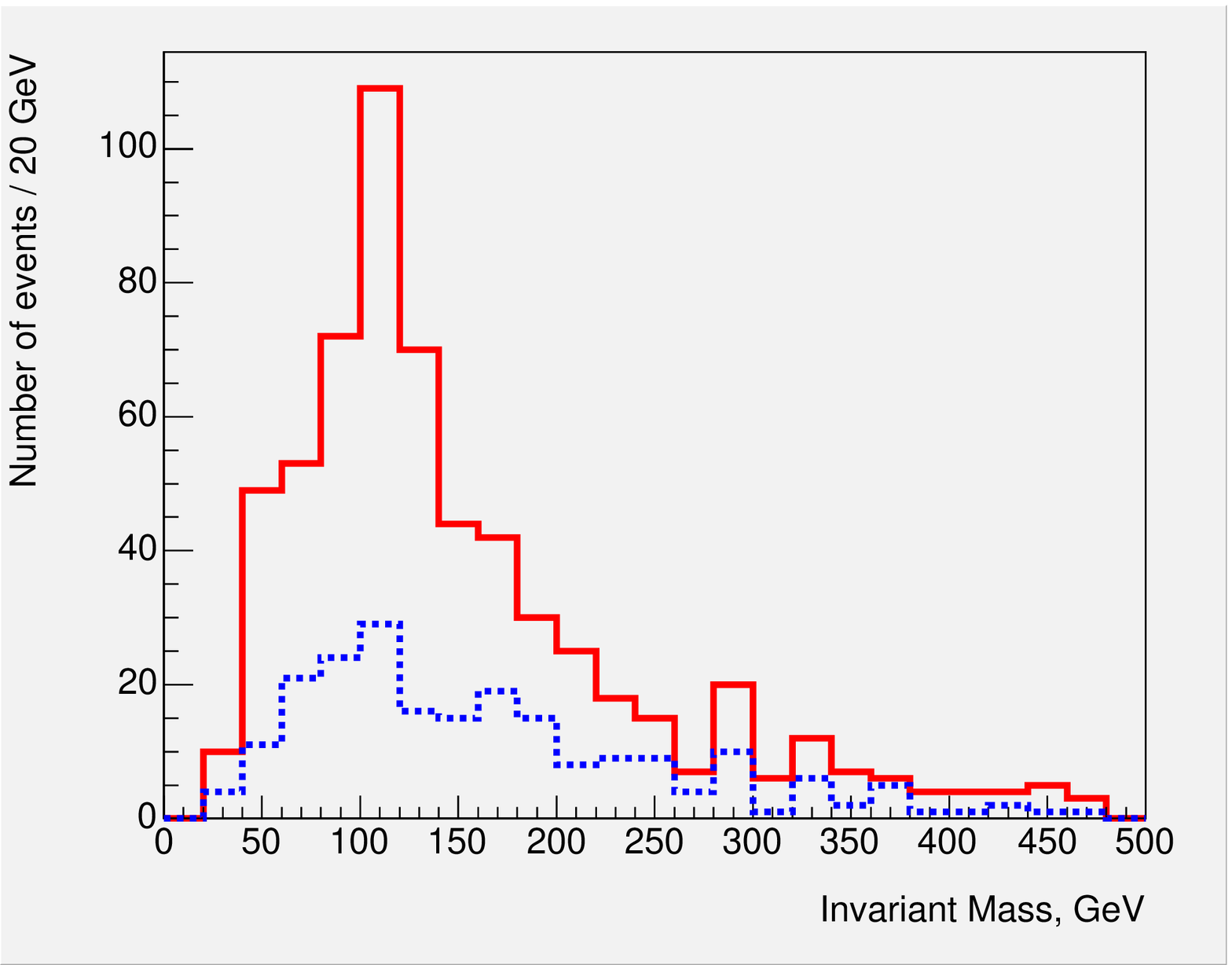,height=5.4cm}} &
\mbox{\epsfig{file=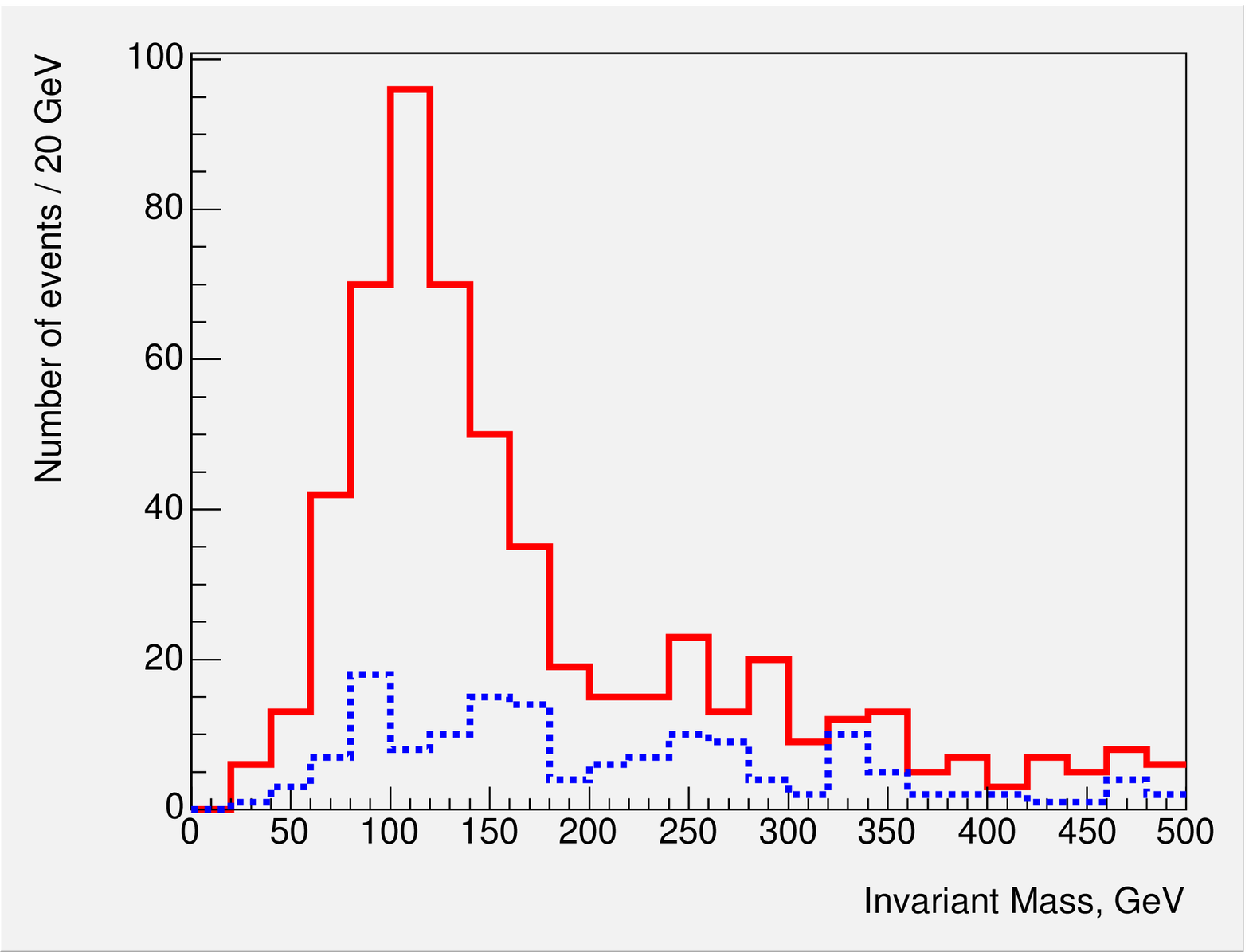,height=5.4cm}} \\
\end{tabular}
\begin{tabular}{c c}
\mbox{\epsfig{file=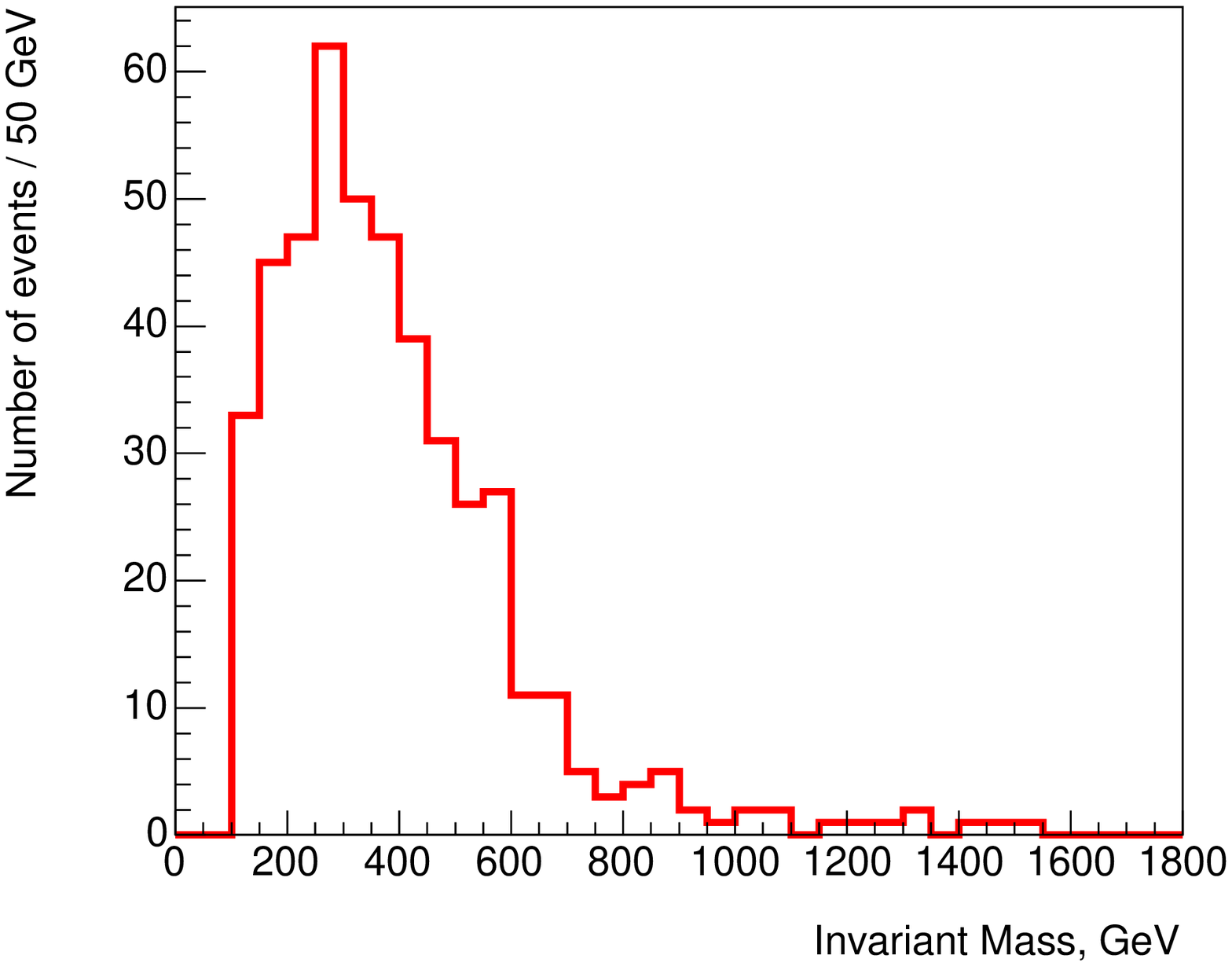,height=6cm}} &
\mbox{\epsfig{file=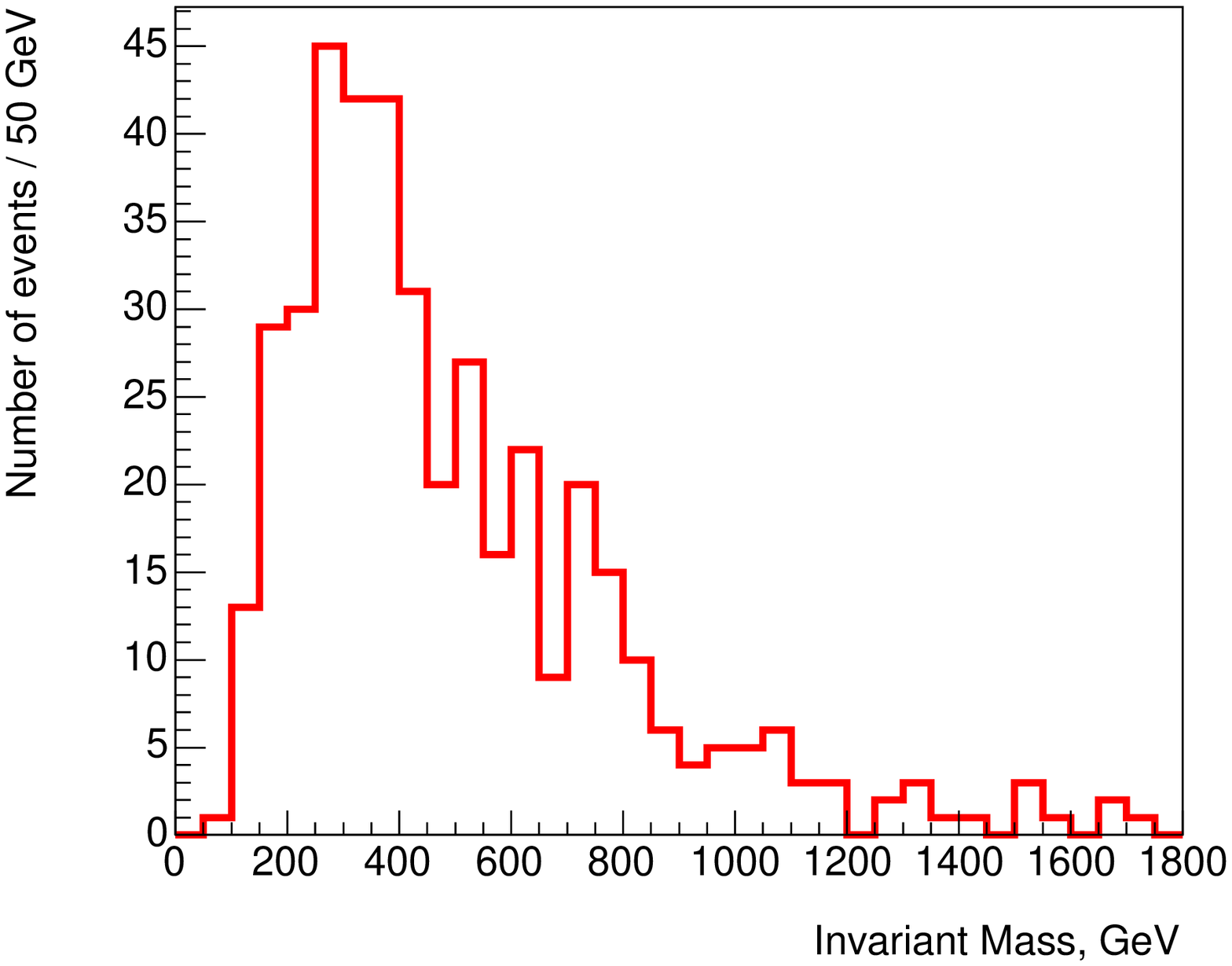,height=6cm}} \\
\end{tabular}
\begin{tabular}{c c}
\mbox{\epsfig{file=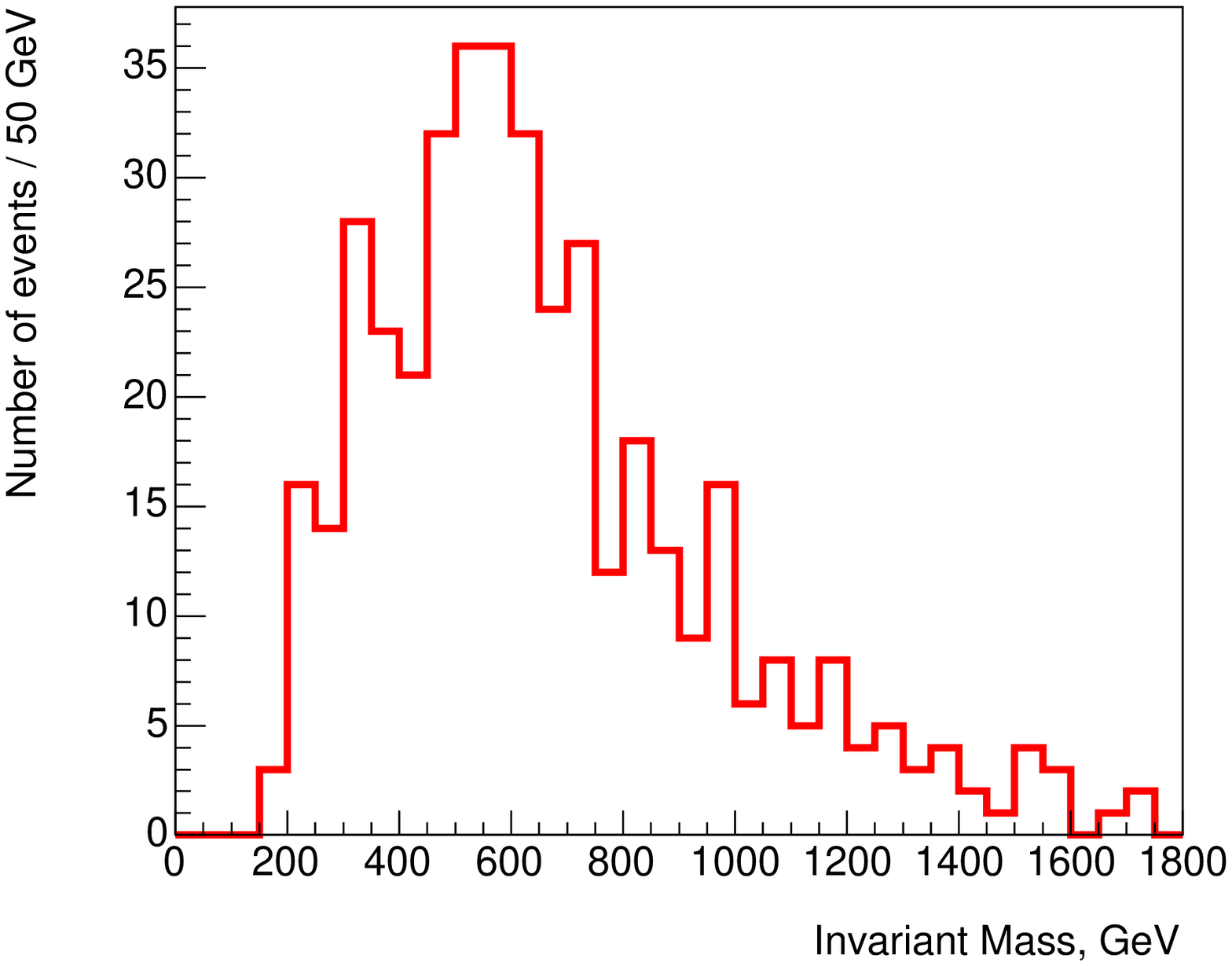,height=6cm}} &
\mbox{\epsfig{file=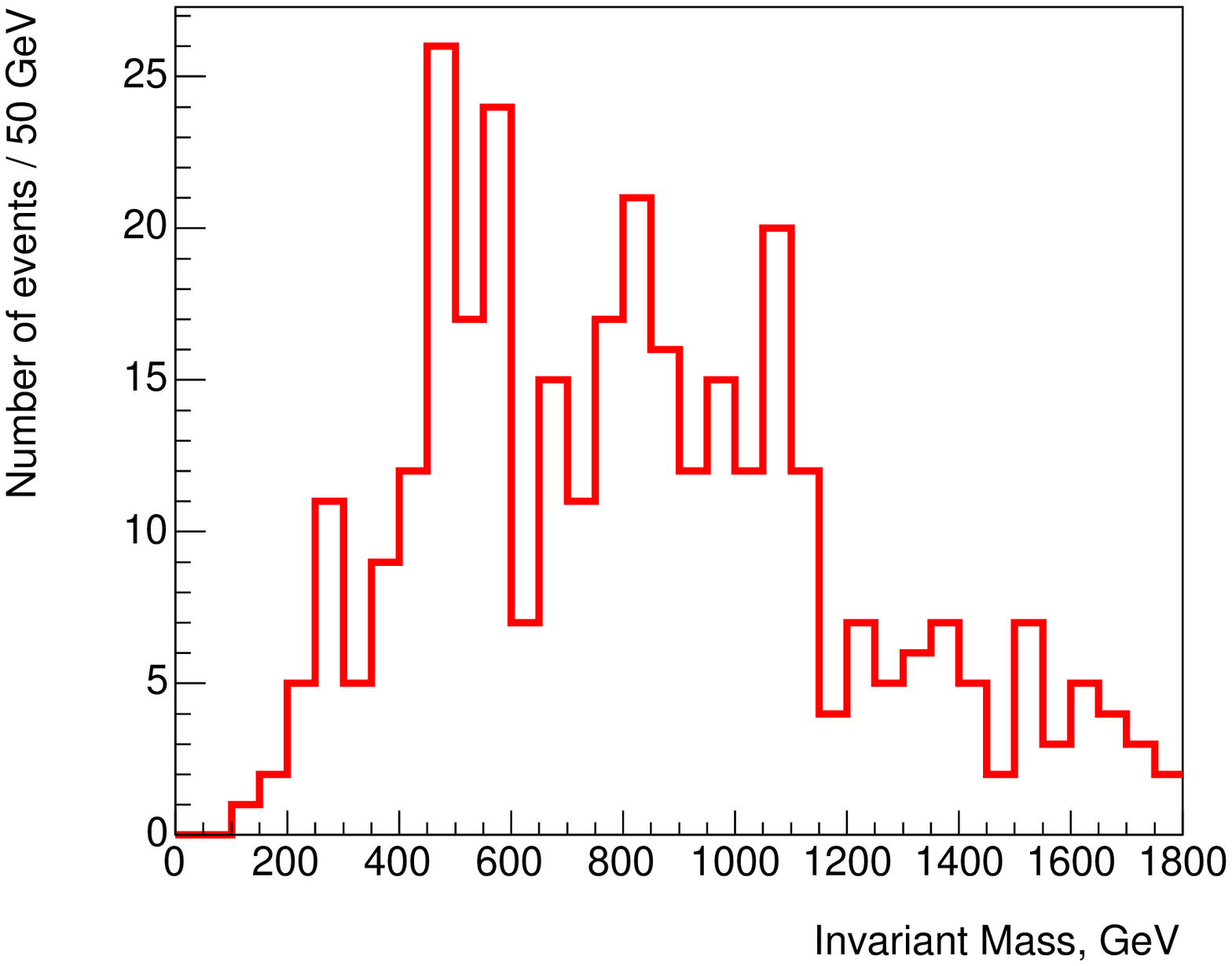,height=6cm}} \\
\end{tabular}
\end{center}   
\vspace*{-0.25in}
\caption{\label{fig:chi2h}\it
The signals from the lightest Higgs boson $h$ in $\chi_2 \to \chi h$ 
cascade decays in scenarios (a) $\beta$ and (b) $\delta$, respectively, 
as found using {\tt PYTHIA}~\cite{PYTHIA} interfaced with {\tt 
ISASUGRA}~\cite{ISASUGRA}. In 
each case, the solid line shows the sum of the supersymmetric signal 
and 
background, and the dashed line is the supersymmetric background alone. 
The Standard Model background is much smaller.
The $q h$ invariant mass distributions in scenarios (c) $\beta$ and (d) 
$\delta$, respectively, each exhibit an end-point
corresponding to ${\tilde q}_R \to q \chi_2$ cascade decays followed by 
$\chi_2 \to \chi h$. The $q q h$ invariant mass distributions in 
scenarios (e) $\beta$ and (f) $\delta$, respectively, each exhibit a 
feature corresponding to ${\tilde g} \to {\tilde q}_R q$ and ${\tilde 
q}_R \to q \chi_2$ cascade decays followed by $\chi_2 \to \chi h$.
}
\end{figure}

One of the motivations for specifying the parameters of scenarios $\beta$
and $\delta$ was to consider models with large branching ratios for the
cascade decays $\chi_2 \to \chi h$, which have been studied previously by
both ATLAS and CMS. Fig.~\ref{fig:chi2h} shows the cascade $h \to {\bar b}
b$ signals expected in scenarios $\beta$ and $\delta$.
We select events with missing 
$E_T > 150$~GeV and require the candidate $b$ jets to be separated by 
an `angle' $\Delta R \equiv \sqrt{ \Delta \eta^2 + \Delta \phi^2} < 2$. 
We have generated 5000 sample supersymmetric events in each scenario: 
these correspond to integrated luminosities of less than one fb$^{-1}$ for
scenario $\beta$ and about 50~fb$^{-1}$ for scenario $\delta$.
The solid lines are the signals and
the dashed lines are the supersymmetric backgrounds in the two scenarios
from events not containing Higgs bosons~\footnote{These are mainly due to
events containing ${\tilde b}$ squarks.}: the Standard Model backgrounds
are much smaller.  The $h$ is visible in $\chi_2$ decays in scenario
$\beta$ (we recall that the event sample corresponds to a very small 
integrated luminosity in this case) and the signal is even clearer at 
point $\delta$.  Previous, more detailed CMS and ATLAS
studies of low-mass points similar to $\beta$ have also concluded that the
$h$ should be observable in sparticle cascade decays, and a point very
similar to $\beta$ is being studied thoroughly for the CMS physics 
TDR~\cite{CMSh}. 
Our first examination of point $\delta$ is promising, but more detailed
studies of such a high-mass point would also be useful.

$\bullet$ {\it Squarks}: The spartners of the lighter quark flavours $u,
d, s, c$ are considered to be observable if $m_{\tilde{q}}
< 2.5$~TeV~\cite{squarks}, so they could be observed in all four benchmark
scenarios. However, their flavours could not be distinguished at the LHC.
We further assume that the stops and sbottoms ${\tilde t}, {\tilde b}$ are
identifiable only if they weigh below 1~TeV, unless the gluino weighs $<
2.5$~TeV and the stop or sbottom can be produced in its two-body decays.
As in scenarios $\alpha, \beta, \gamma$ the stops and sbottoms are 
relatively
light, we consider them to be observable. At point $\delta$, the branching
ratio for ${\tilde g} \to {\tilde t}_1 t \sim 40$~\%, whereas the decays
into ${\tilde t}_2 t, {\tilde b}_1 b$ and ${\tilde b}_2 b$ are each only
${\cal O}(10)$~\%. Accordingly, we consider only the $\tilde{t}_1$ to be
detectable at point $\delta$. A note of caution is in order: since the
detection of ${\tilde t}$ or ${\tilde b}$ production is difficult to
assess without simulation studies, these conclusions should be taken with
care.

We have considered the detectability of ${\tilde q}_R$ via their decays
${\tilde q}_R \to \chi_2 q$ followed by $\chi_2 \to \chi h$. 
Selecting the
events in the $h$ peaks in $\chi_2$ decays shown in 
Fig.~\ref{fig:chi2h}(a,b),
and then combining the reconstructed $h$ boson with the hadronic jet
${J_1}$ that maximizes the product
${\mathbf p_h}\cdot{\mathbf p_{J_1}}$, we obtain the candidate $q
h$ invariant mass distributions shown in panels (c, d) of
Fig.~\ref{fig:chi2h} for points $\beta, \delta$, respectively. 
In principle, ${\tilde
q}_R$ decays give end-points in these distributions, because of
the two-body decay phase space, which are less distinctive than 
the
corresponding dilepton edges in ${\tilde q}_R \to q \ell^+ \ell^- \chi$
decays. Edge features are visible in both scenarios
$\beta, \delta$, close to the expected positions at $\sim
600, 1200$~GeV, respectively~\footnote{Similar distributions are being 
studied in more
detail for the CMS Physics TDR~\cite{CMSh}.}.
 
$\bullet$ {\it Gluinos}: These are generally considered to be observable
for masses below 2.5~TeV~\cite{squarks}, and hence can be discovered at
all four points.

As seen in panels (e, f) of Fig.~\ref{fig:chi2h}, we have also considered
specific features of the searches for gluinos at points $\beta, \delta$,
respectively. In each case, we have selected the hadronic jet ${J_2}$ that
maximized the product ${\mathbf p_{J_1}}\cdot{\mathbf p_{J_2}}$ and 
plotted the $h J_1
J_2$ mass distribution. Gluino decays should give distinctive `edge'
features at $\sim 640, 1280$~GeV, respectively, corresponding to the
multi-body phase space. This feature is not very apparent for point 
$\beta$ with the small sample generated here, but more
apparent for point $\delta$. A point similar to the former is also under 
study for the CMS Physics TDR~\cite{CMSh}. A more detailed study of 
benchmark $\delta$ with optimized cuts would also be desirable.

We have also considered the decays ${\tilde g} \to {\tilde b} b$ at points
$\alpha, \beta, \gamma$ and $\delta$. In each case, the product of the 
${\tilde g}$ production cross section and branching ratio appears high
enough to enable $m_{\tilde g} - m_{\tilde b}$ to be measured with
sufficient accuracy to verify that this point has a value of $m_0$
significantly different from that on the CMSSM WMAP line for the same
value of $m_{1/2}$. However, more refined studies of gluino search
strategies would clearly be useful.

$\bullet$ {\it Charged Sleptons}: Since the mass differences
$m_{\tilde{\ell}}-m_{\chi} \sim 100$~GeV at all these benchmarks, we
consider ${\tilde{\ell}}$ decays into leptons always to be observable,
provided that the slepton mass is light enough and hence the production
cross section is large enough.  We note that all four points have
negligible branching ratios for the decays of $\chi_2$ to sleptons, which
implies that cascades will not contribute to slepton observability and
that sleptons can only be detected via their direct production.
Following~\cite{Bench}, we consider the direct production rates to be
large enough if $m_{\tilde{\ell}} < 350$~GeV. According to this criterion,
all the charged sleptons would be observable in scenarios $\alpha$ and
$\beta$ (though ${\tilde \ell}_L$ and ${\tilde \tau}_2$ signals would be 
very
marginal at the latter point), and ${\tilde e}_R, {\tilde \mu}_R, {\tilde
\tau}_1$ would be observable at point $\gamma$. However, we consider the
observability of the $\tilde{\tau}_{1,2}$ to be difficult to assess
without a detailed study and, conservatively, we do not count them as
observable in any scenario. The sleptons are all too heavy to be observed 
at point $\delta$.

$\bullet$ {\it Sneutrinos}: We do not consider sneutrinos to be observable
at the LHC. Although many of the sneutrino decays are into visible
particles at points $\alpha, \beta, \gamma$, the modes $(\mu^\pm \chi^\mp,
\nu \chi_2)$ having branching ratios (46, 20), (37, 17), (55, 33)~\%
respectively. Moreover, the associated ${\tilde \ell} {\tilde \nu}$
production cross sections are quite large in scenarios $\alpha, \beta,
\gamma$: 220/110/80~fb, respectively. Nevertheless, no viable discovery
strategy has yet been developed.

\begin{figure}
\begin{center}
\begin{tabular}{c c}
\mbox{\epsfig{file=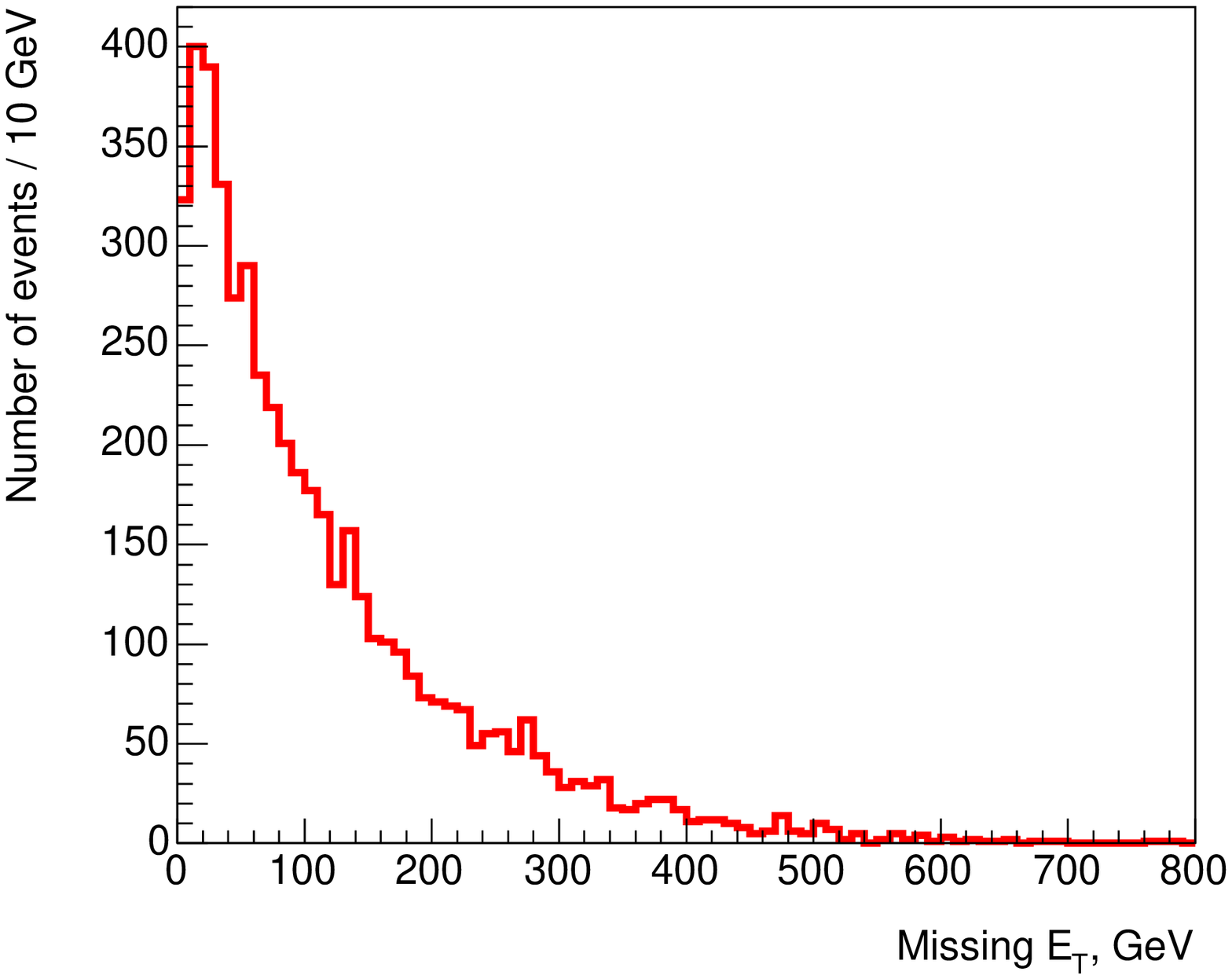,height=6cm}} &
\mbox{\epsfig{file=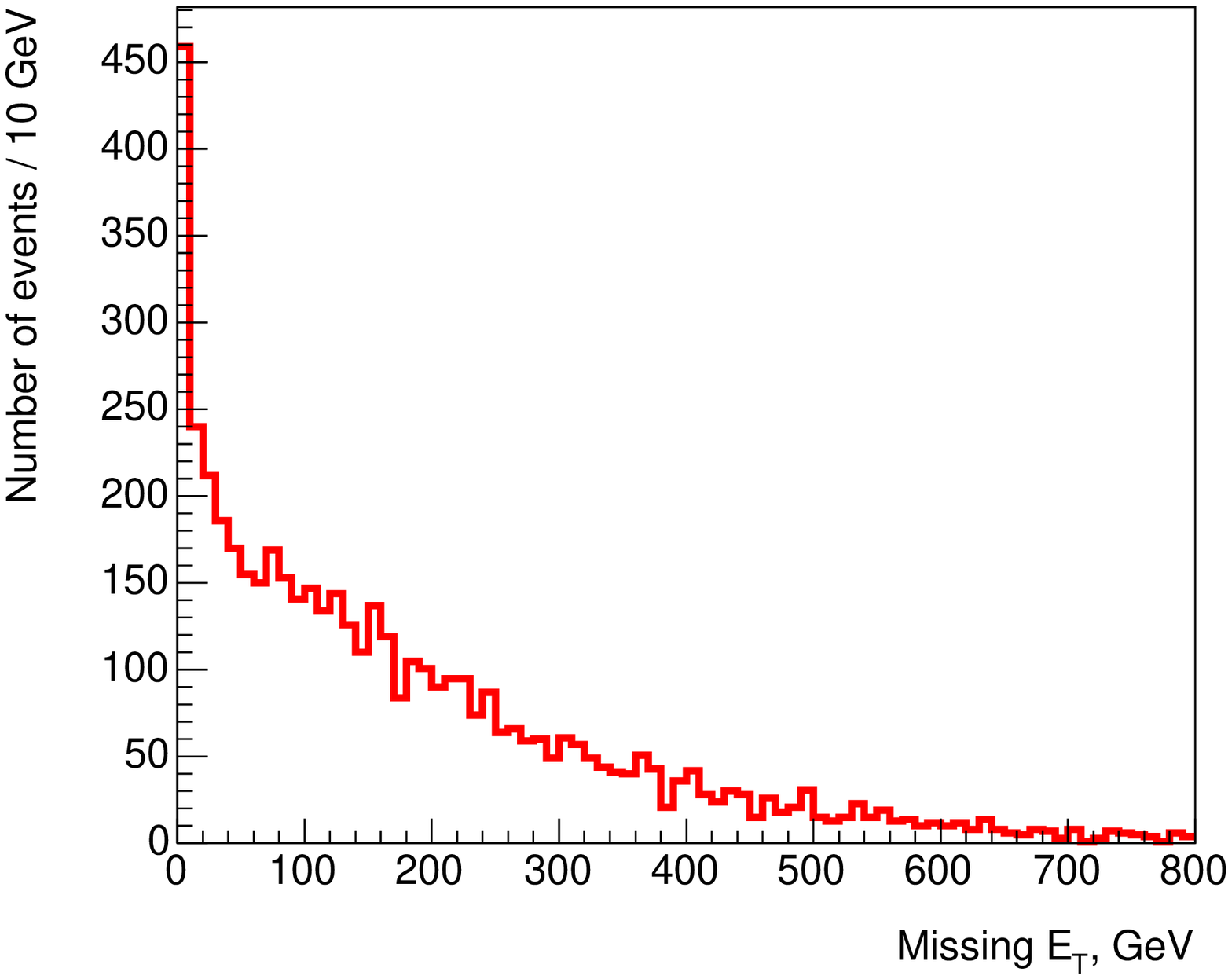,height=6cm}} \\
\end{tabular} 
\end{center}  
\caption{\label{fig:met}\it
The missing $E_T$ spectra in scenarios $\epsilon$ and $\zeta$: that for 
point $\eta$ is similar to the latter.
These plots were obtained with {\tt PYTHIA}~\cite{PYTHIA} interfaced with 
{\tt ISASUGRA}~\cite{ISASUGRA}.}
\end{figure}

We next discuss particle observabilities in the GDM scenarios $\epsilon,
\zeta, \eta$, where $\tilde{\tau}_1$ is the NLSP. The branching ratios for
final states resulting from the interesting cascade decays of squarks at
the LHC are shown in Table~\ref{tab:Luc}. We first note that sparticle
pair-production in these scenarios gives rise to substantial missing 
$E_T$, as seen
in Fig.~\ref{fig:met} for scenarios $\epsilon$
and $\zeta$ (point $\eta$ is very similar to the latter). We assume
that the metastable ${\tilde \tau}_1$'s are measured in the detector, as 
discussed below. The
missing $E_T$ is traceable to the many neutrinos in the final states,
e.g., from $\tau$ decays and/or the many ${\tilde q}_L$ decays with other
neutrinos.

\begin{figure}
\begin{center}
\begin{tabular}{c c}
\mbox{\epsfig{file=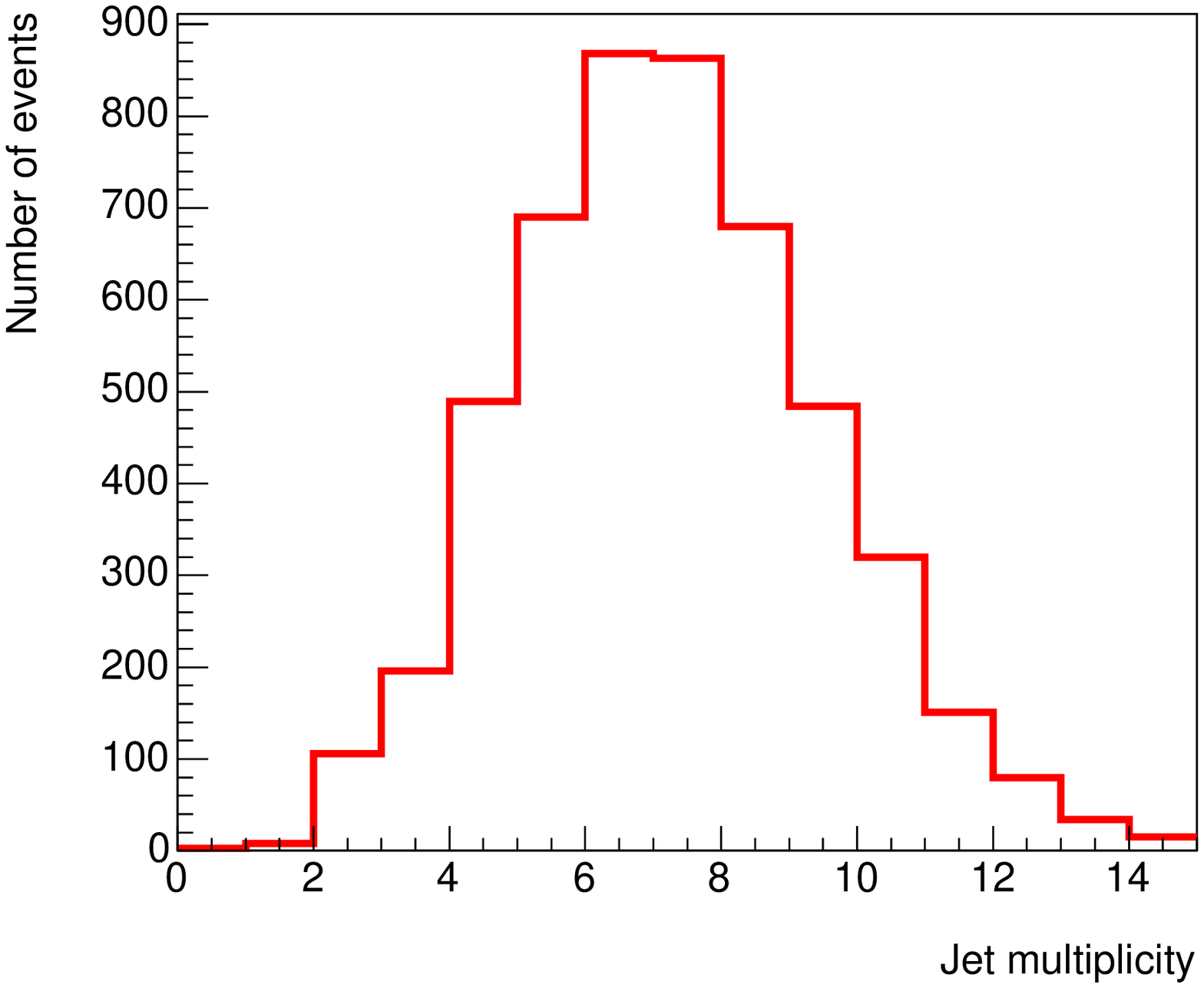,height=6cm}} &
\mbox{\epsfig{file=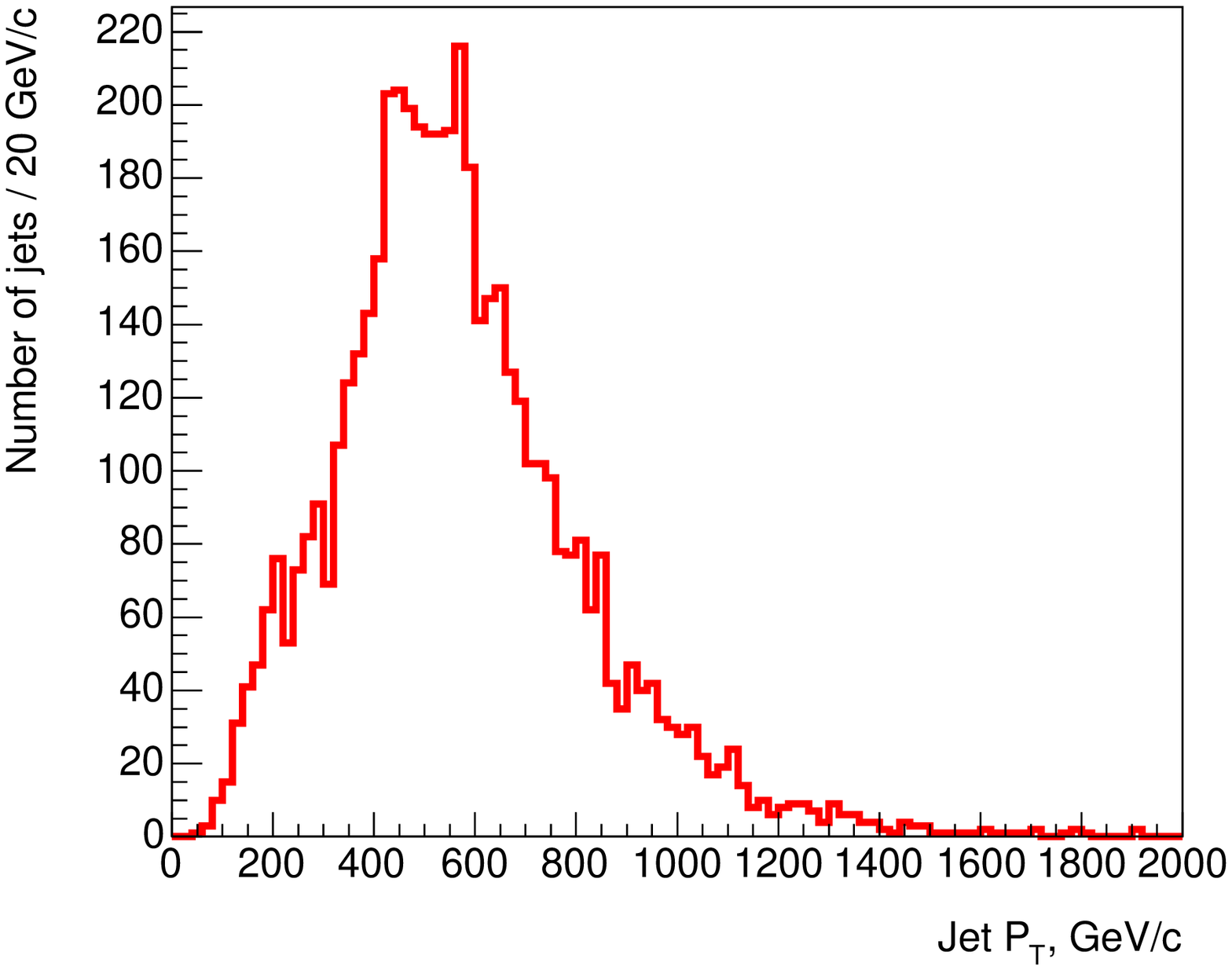,height=6cm}} \\
\end{tabular}
\end{center}   
\begin{center}
\begin{tabular}{c c}
\mbox{\epsfig{file=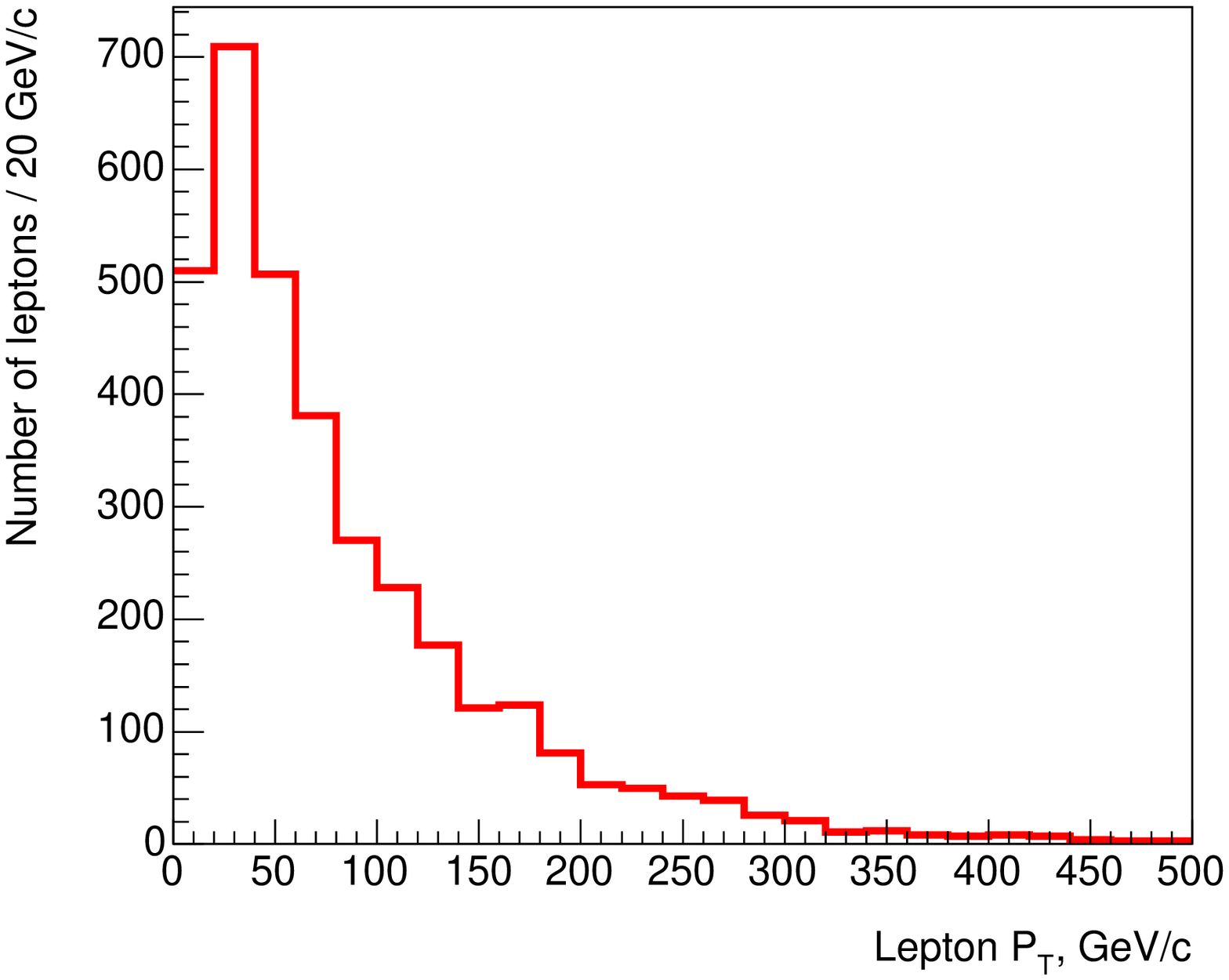,height=6cm}} &
\mbox{\epsfig{file=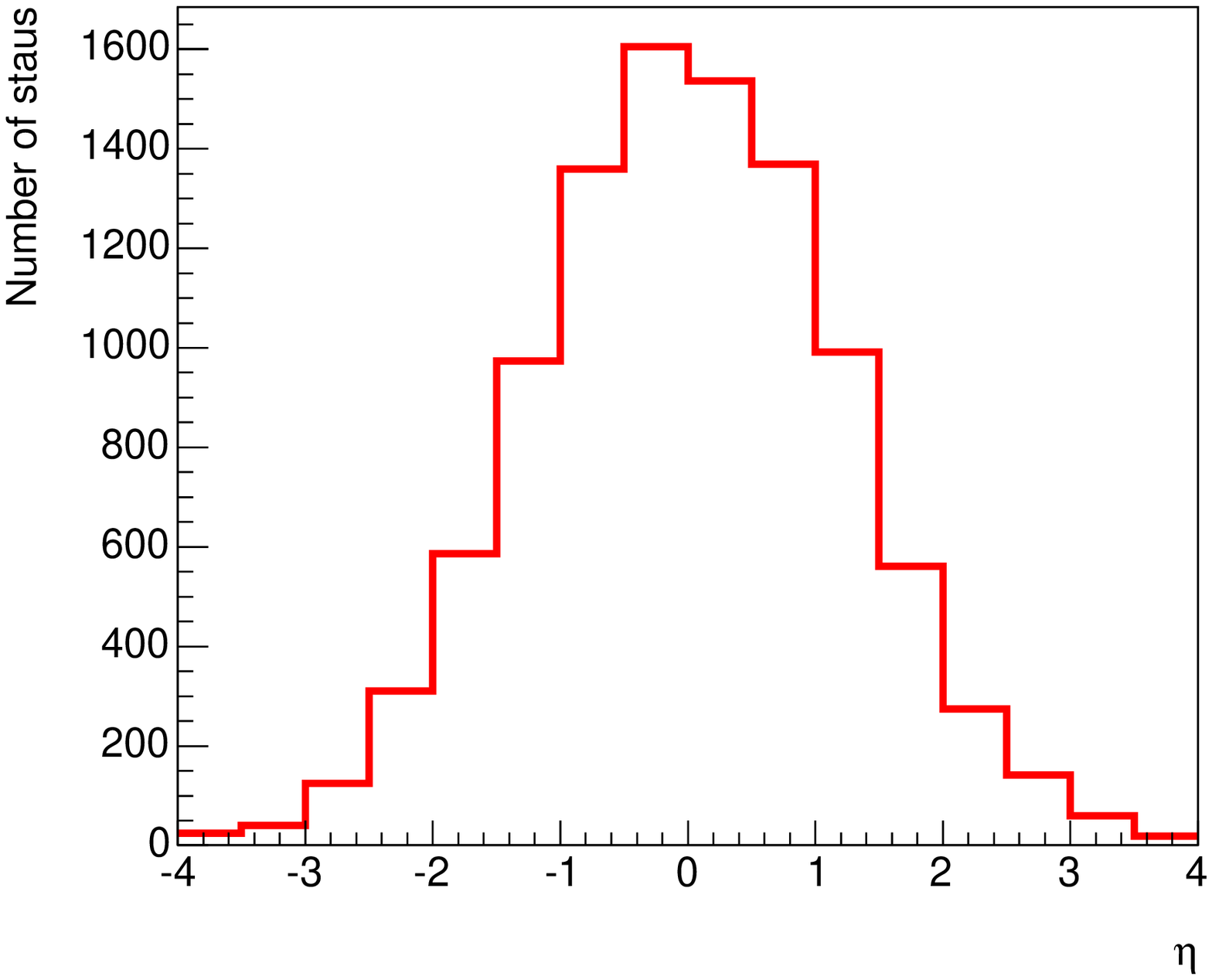,height=6cm}} \\
\end{tabular}
\end{center}   
\caption{\label{fig:trigger5}\it
Kinematic distributions in events produced in the GDM 
benchmark scenario $\epsilon$: (a) hadron jet multiplicity, (b) the 
transverse energy $E_T$ 
of the most energetic jet, (c) the transverse energy $p_T$ of the most 
energetic lepton, and 
(d) the pseudorapidity distribution for ${\tilde \tau}_1$ production. 
These plots were obtained with {\tt PYTHIA}~\cite{PYTHIA} interfaced with 
{\tt ISASUGRA}~\cite{ISASUGRA}.} 
\end{figure}

\begin{figure}
\begin{center}
\begin{tabular}{c c}
\mbox{\epsfig{file=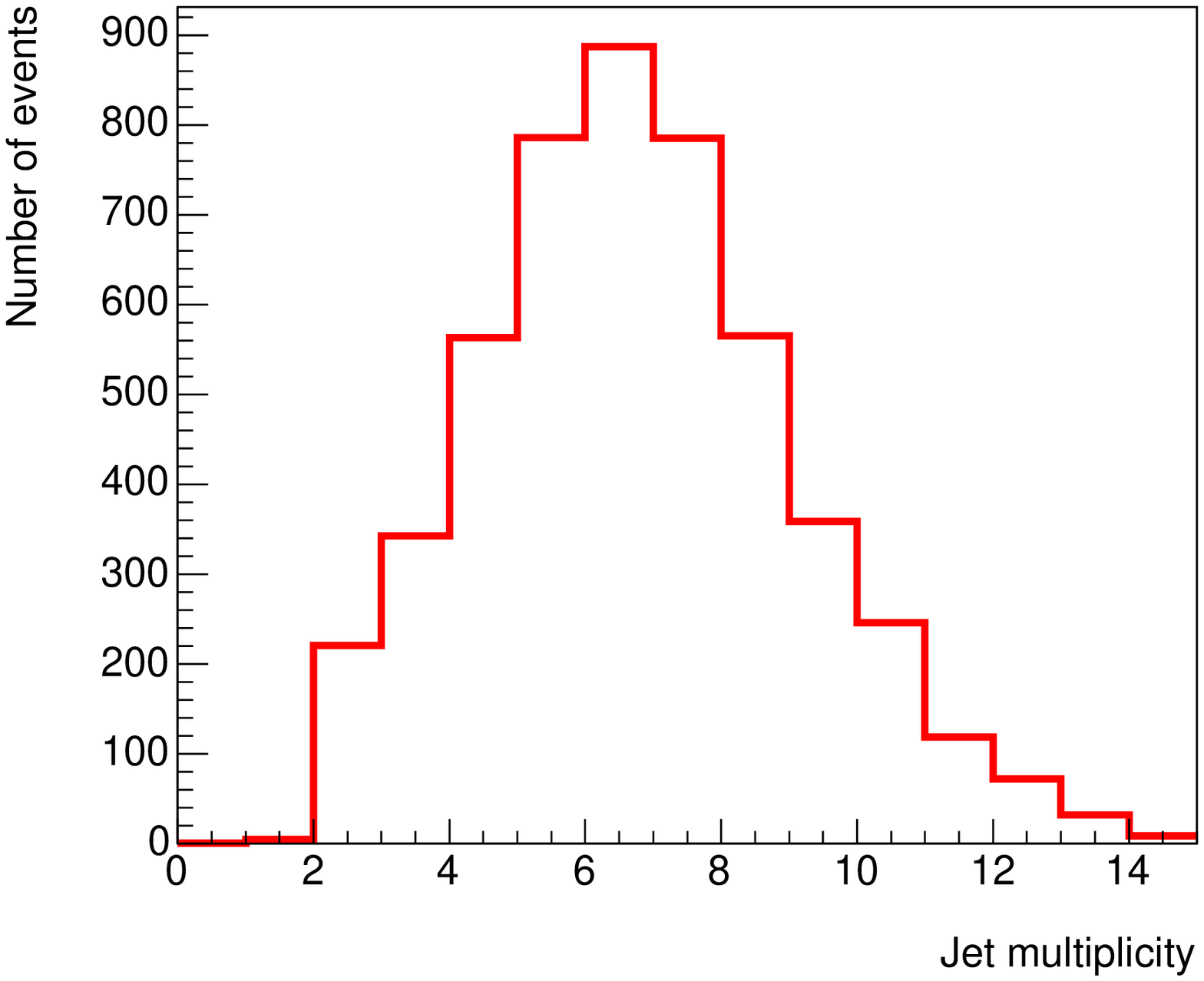,height=6cm}} &
\mbox{\epsfig{file=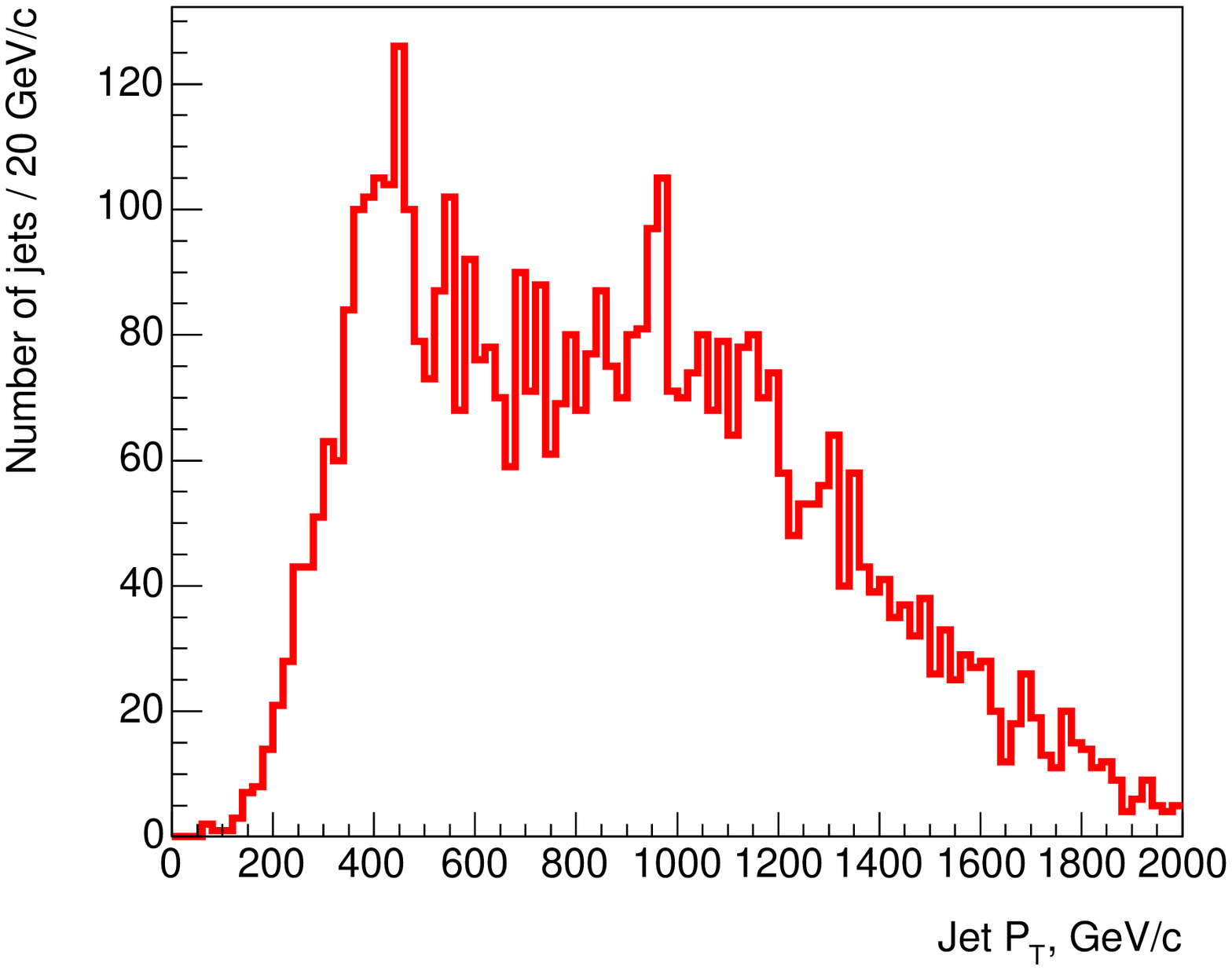,height=6cm}} \\
\end{tabular}
\end{center}   
\begin{center}
\begin{tabular}{c c}
\mbox{\epsfig{file=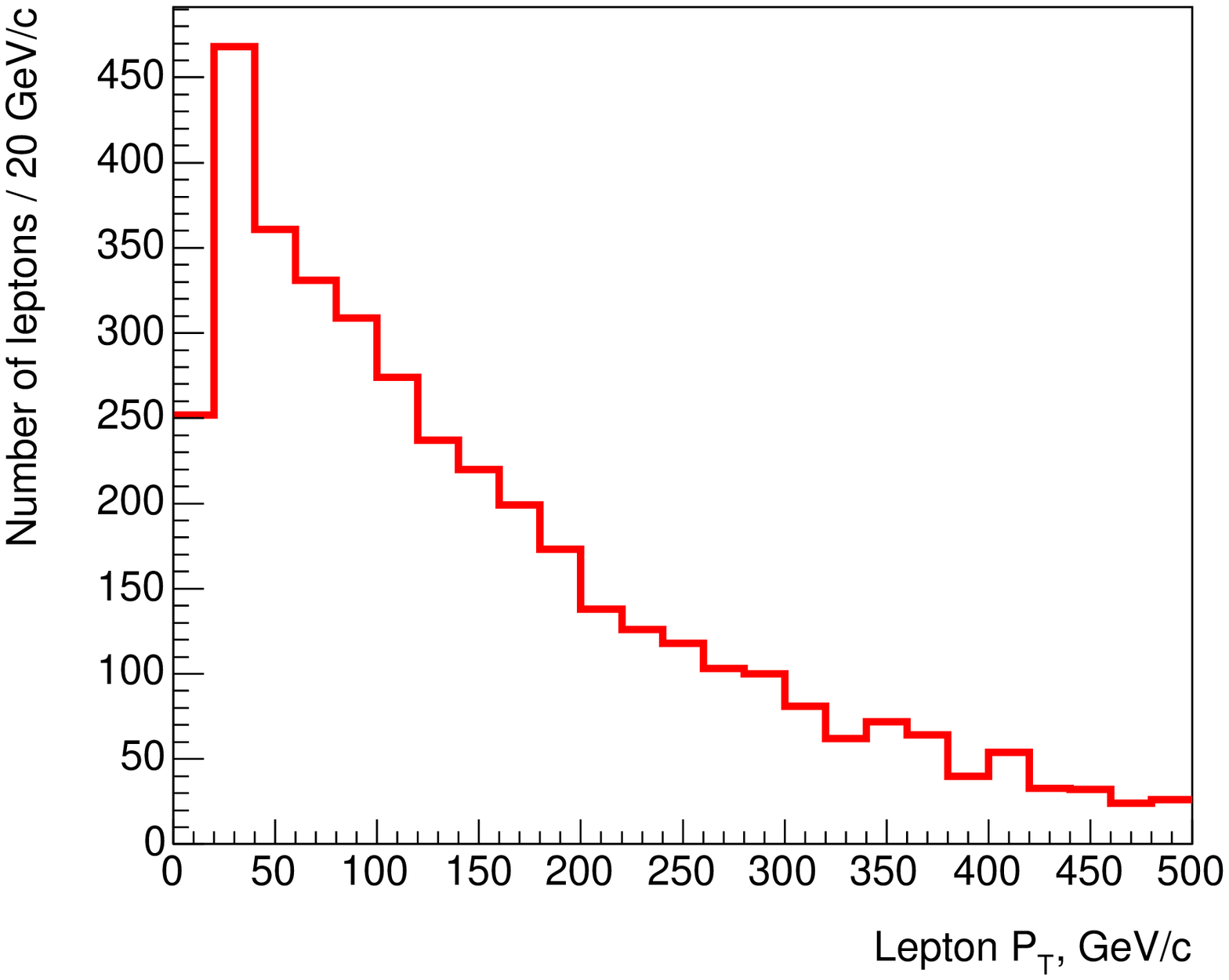,height=6cm}} &
\mbox{\epsfig{file=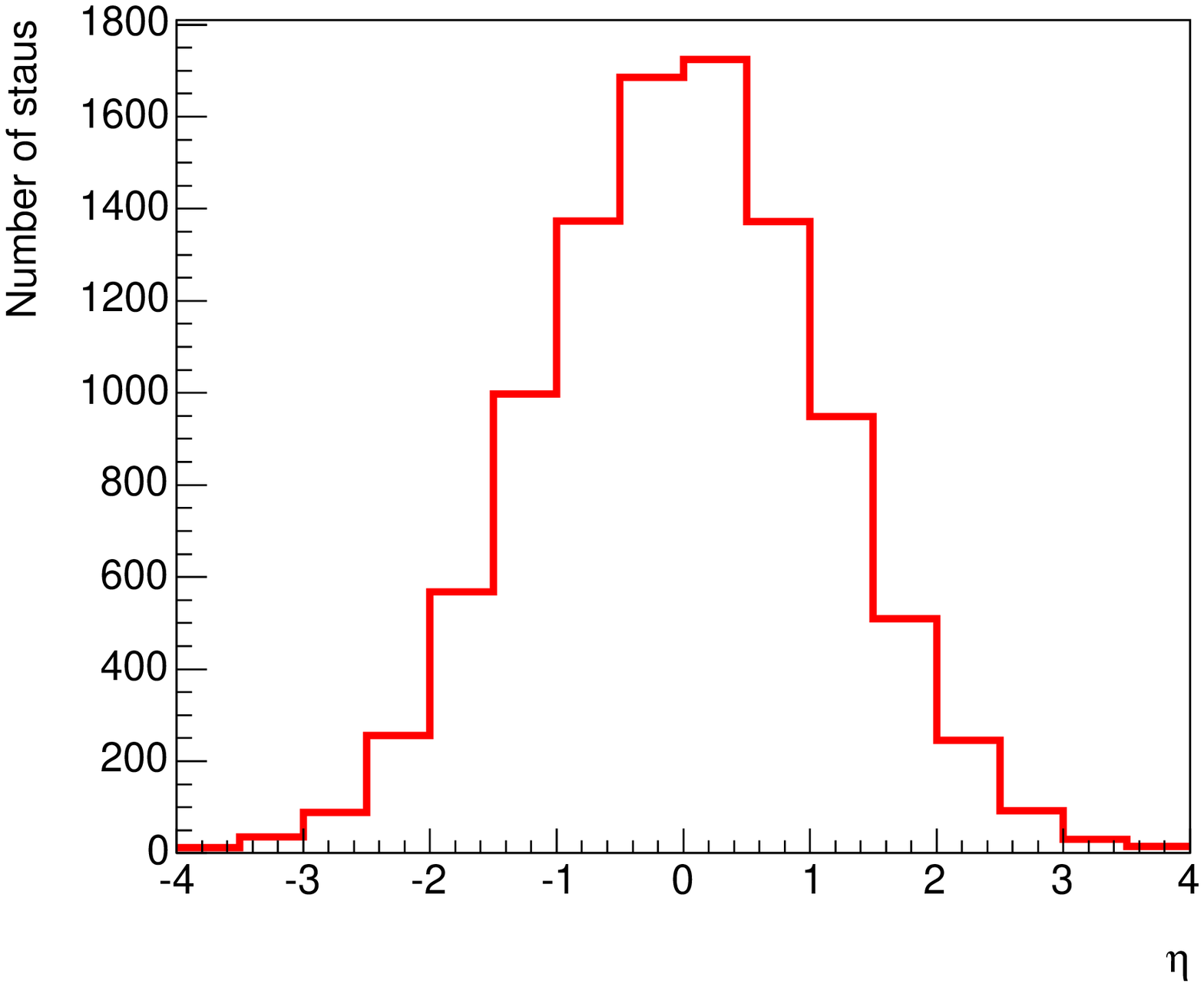,height=6cm}} \\
\end{tabular}
\end{center}   
\caption{\label{fig:trigger6}\it
Kinematic distributions in events produced in the GDM
benchmark scenario $\zeta$, as in Fig.~\protect\ref{fig:trigger5}.}
\end{figure}

\begin{figure}
\begin{center}
\begin{tabular}{c c}
\mbox{\epsfig{file=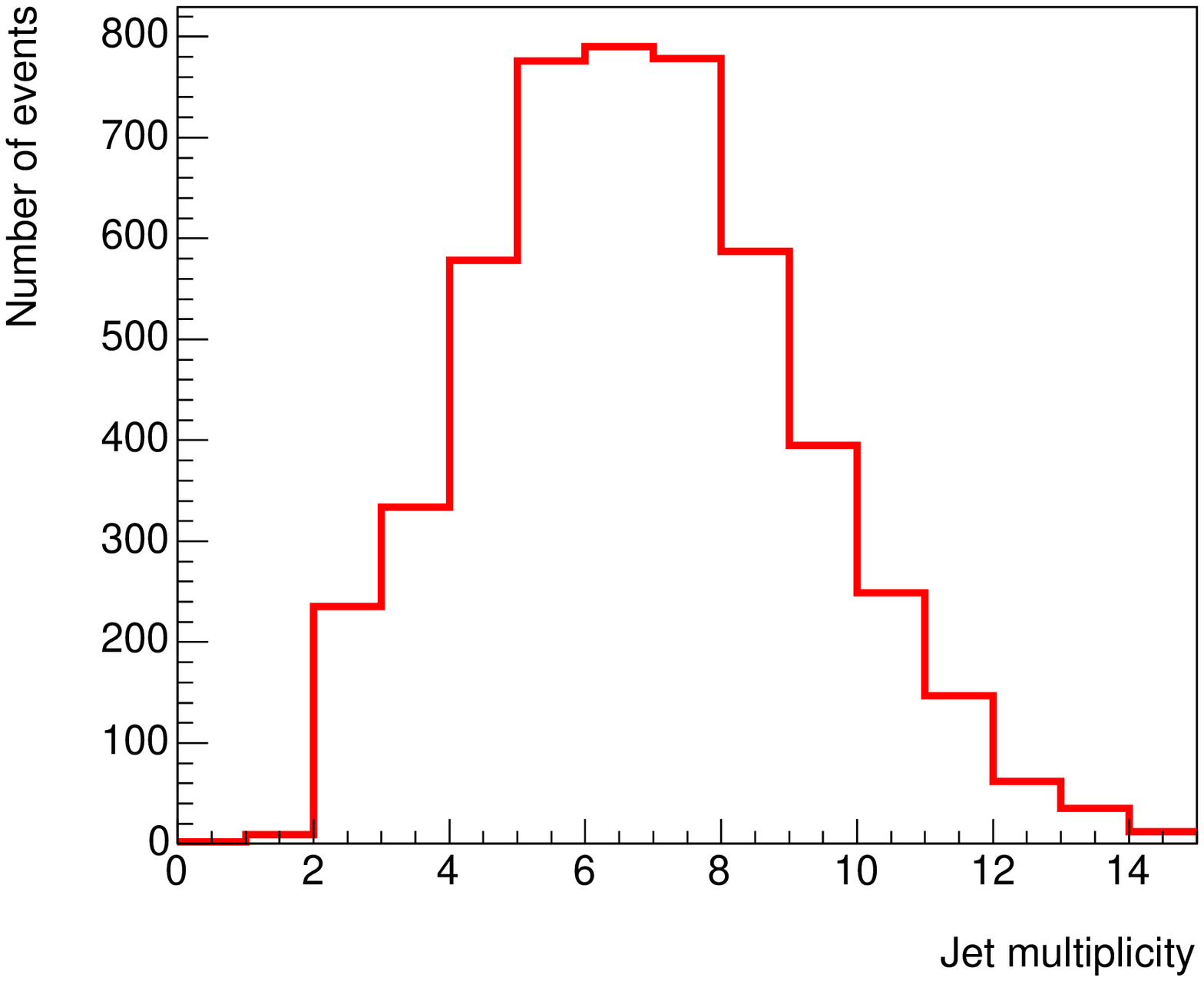,height=6cm}} &
\mbox{\epsfig{file=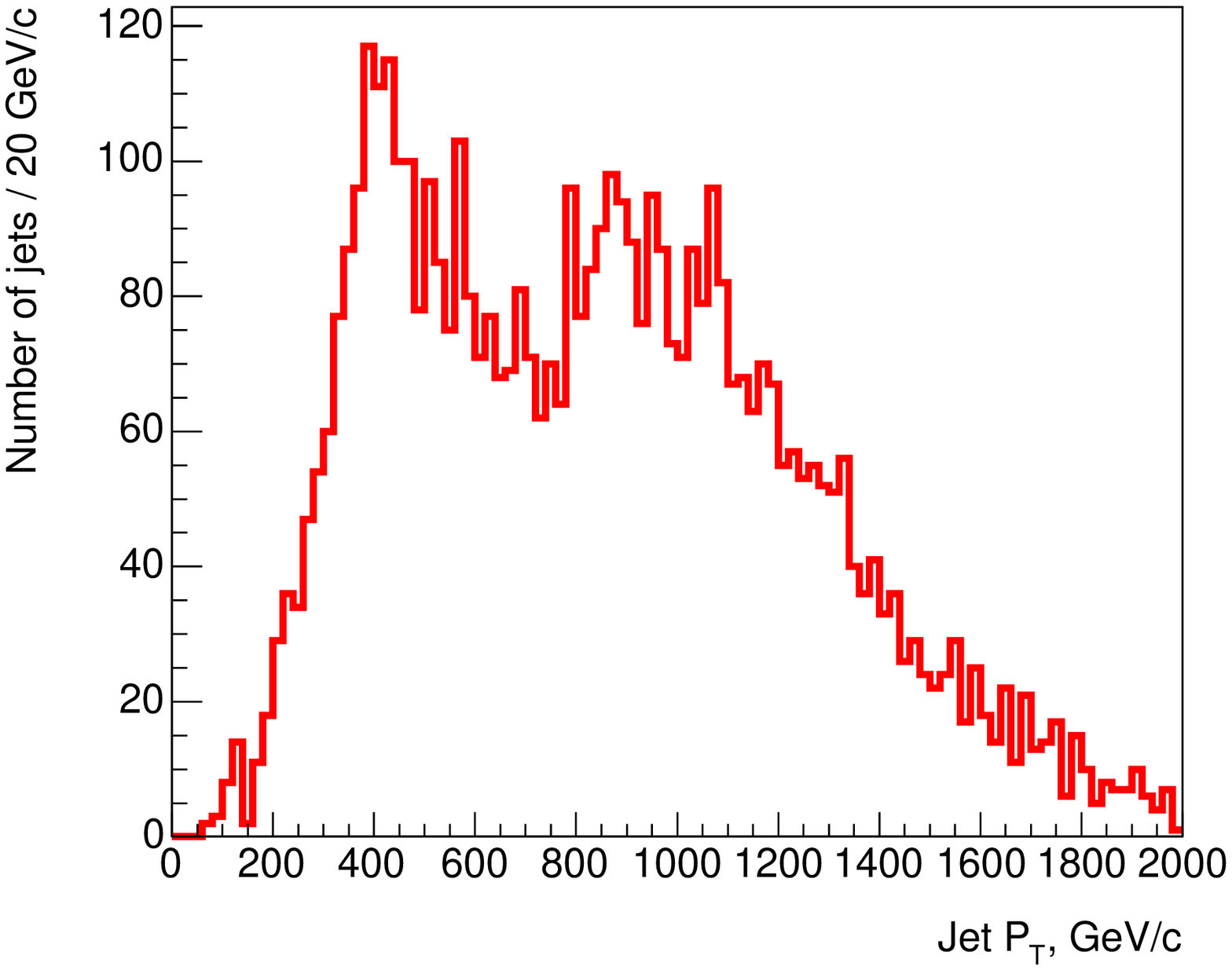,height=6cm}} \\
\end{tabular}
\end{center}   
\begin{center}
\begin{tabular}{c c}
\mbox{\epsfig{file=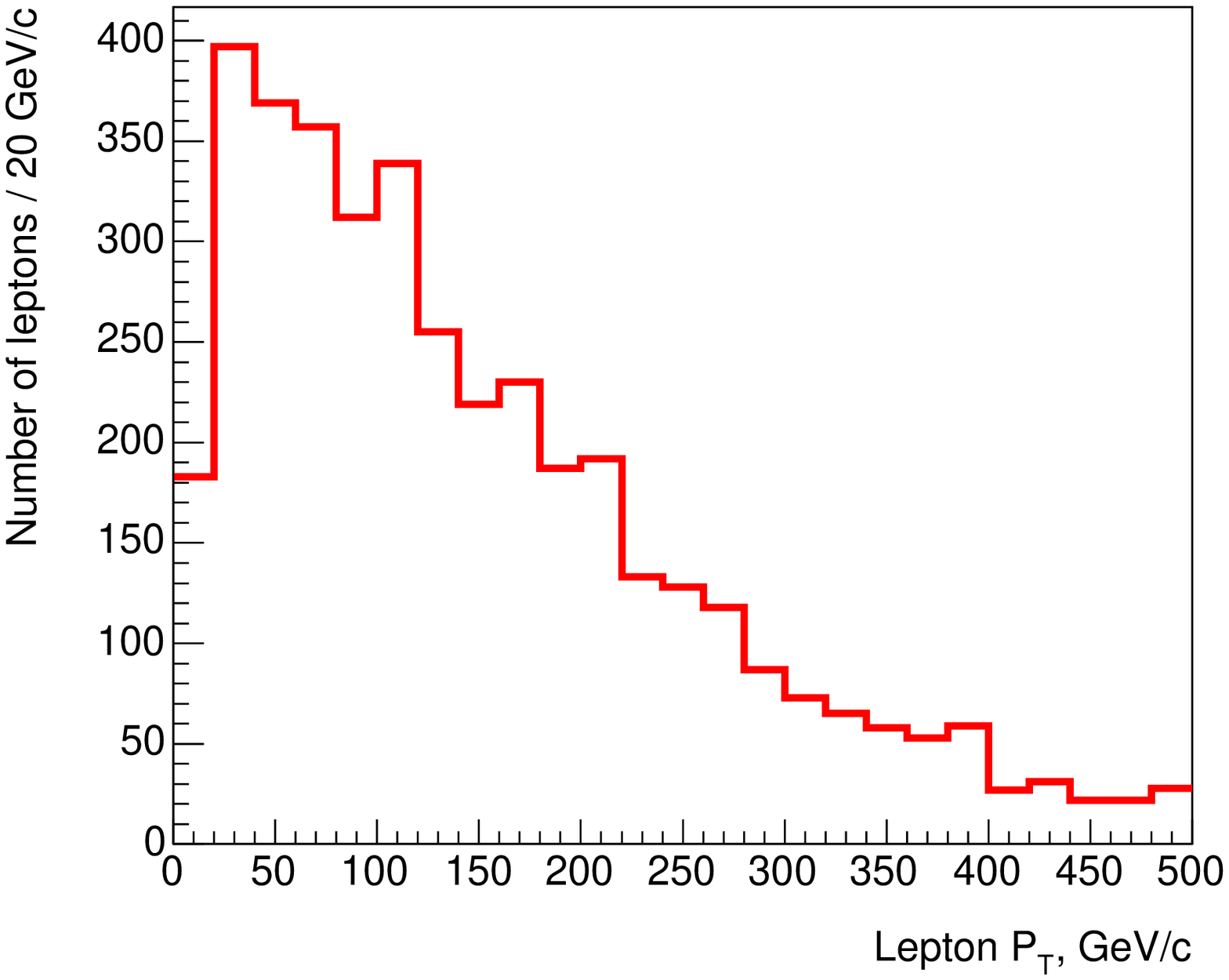,height=6cm}} &
\mbox{\epsfig{file=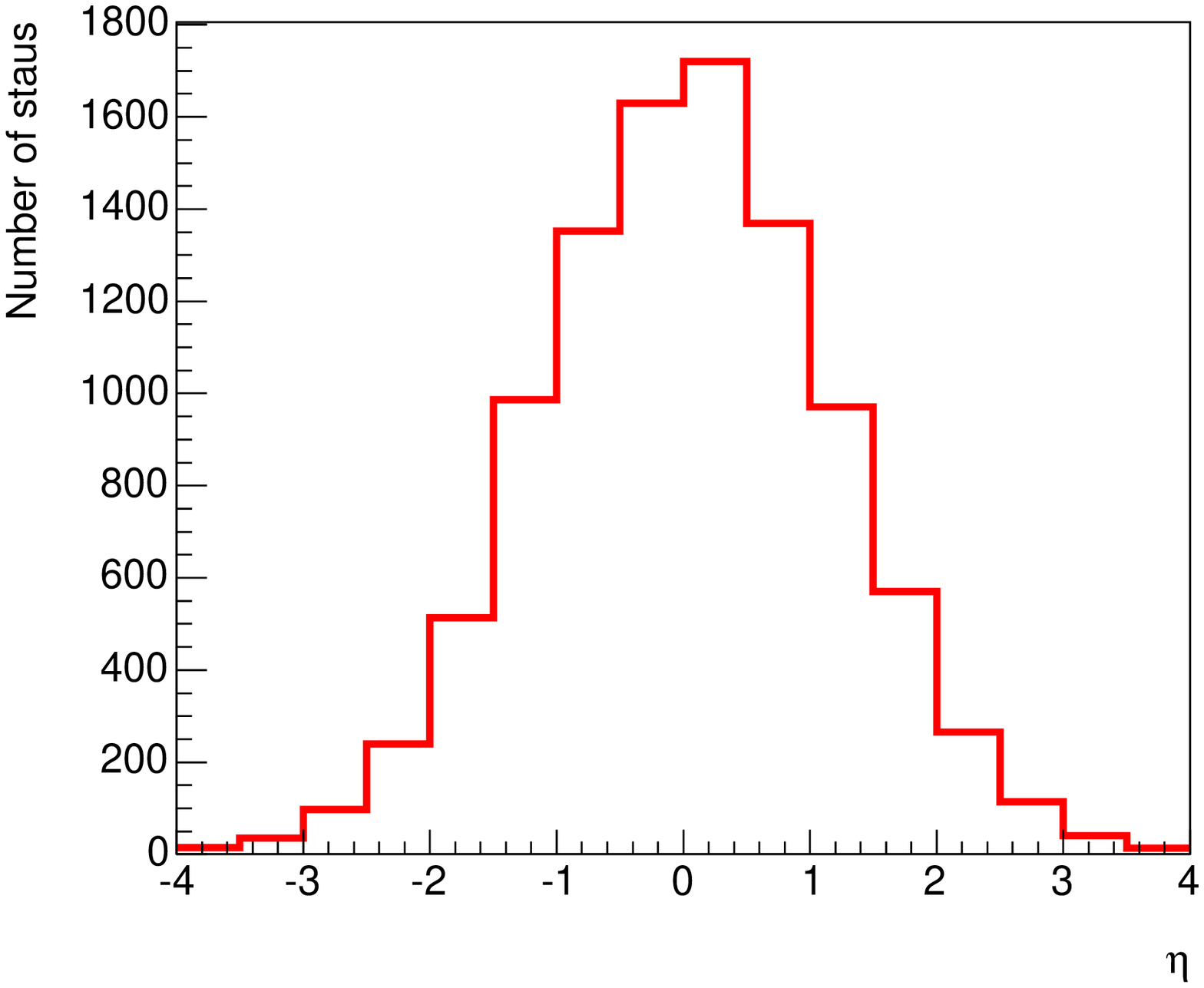,height=6cm}} \\
\end{tabular}
\end{center}   
\caption{\label{fig:trigger7}\it
Kinematic distributions in events produced in the GDM
benchmark scenario $\eta$, as in Fig.~\protect\ref{fig:trigger5}.} 
\end{figure}

We display in Figs.~\ref{fig:trigger5}, \ref{fig:trigger6} and
\ref{fig:trigger7} some other characteristics of events in these benchmark
mSUGRA GDM scenarios. Panels
(a) of Figs.~\ref{fig:trigger6} and \ref{fig:trigger7} show the jet
multiplicity distributions for points $\zeta, \eta$. Two-body decays of
the ${\tilde q}_R$ are responsible for the bimodal distibutions of the
leading jet transverse energies in panels (b) of both
Figs.~\ref{fig:trigger6} and \ref{fig:trigger7}. The peak at $E_T \sim
1$~TeV is due to the two-body ${\tilde q}_R$ decays, and the lower-$E_T$
peak is due to other sparticle decays. Panels (c) of
Figs.~\ref{fig:trigger6} and \ref{fig:trigger7} show barely visible
features in the leading lepton $E_T$ distributions due to slepton cascade
decays. There is no such feature in Fig.~\ref{fig:trigger5}, where the
cascade lepton energy is smaller. Nevertheless, we note that large
fractions of the cascade-decay leptons have transverse momenta large
enough to be detected with high efficiency, and could potentially be used
as event triggers in addition to the high-$E_T$ jets.

$\bullet$ {\it Gravitino}: We consider the gravitino to be observable in
GDM scenarios where the NLSP can be stopped and its decays observed.  
This was obviously not the case in scenario $\delta$, considered above,
and perhaps not in scenarios $\zeta, \eta$ either, as we discuss later.  
On the other hand (see below), it seems possible to obtain a
substantial sample of $\tilde{\tau}_1$ decays in scenario $\epsilon$, so
we consider the gravitino to be indirectly observable in this case.

$\bullet$ {\it Sleptons}: The ${\tilde \tau}_1$ has a distinctive
time-of-flight signature, and we consider that it could be detected with
some efficiency in both ATLAS and CMS in all the scenarios $\epsilon,
\zeta, \eta$. As seen in panels (d) of Figs.~\ref{fig:trigger5}, 
\ref{fig:trigger6} and \ref{fig:trigger7}, it is generally produced quite 
centrally
and in association with a considerable number of high-$E_T$ jets and/or
leptons. As discussed below, its mass can probably be measured with an
accuracy $\sim 1$~\% in all three scenarios, and a significant sample of
decays of stopped ${\tilde \tau}_1$'s should enable its lifetime to be
measured at point $\epsilon$.  Even in scenarios $\zeta, \eta$, one
expects a sample of several hundred events with ${\bar {\tilde q}_R}
{\tilde q}_R$ production followed by ${\tilde q}_R \to q (\chi \to {\tilde
\tau}_1 \tau)$ decay on one side, and ${\tilde q}_R \to q (\chi \to
{\tilde \ell}_R \ell)$ decay on the other side, followed by ${\tilde
\ell}_R \to \ell {\tilde \tau}_1 \tau$ decay.  We therefore expect the
${\tilde \ell}_R$ to also be observable at all three points. In scenario
$\epsilon$, one should also be able to reconstruct the cascade ${\tilde
q}_L \to q \chi_2$ followed by $\chi_2 \to {\tilde \ell}_L \ell$ and then
${\tilde \ell}_L \to \chi \ell$. Knowing already the $\chi$ mass from the
analysis of ${\tilde q}_R$ decays, one should be able to reconstruct the
${\tilde \ell}_L$ and $\chi_2$ masses at point $\epsilon$. At points
$\zeta, \eta$, we expect a sample of a few dozen events, which might be
sufficient to argue that $m_{{\tilde \ell}_L} > m_\chi > m_{{\tilde
\ell}_R} > m_{{\tilde \tau}_1}$, and thereby provide some discrimination
against GMSB models~\cite{GMSB}~\footnote{As we discuss later, the 
potential
discriminators between gravity-mediated GDM models and GMSB models include
the sequence of sparticle masses and the pattern of ino mixing.}, but
would be insufficient to reconstruct the heavier slepton masses. In
summary, therefore, we consider all the charged sleptons to be observable
at point $\epsilon$, but only ${\tilde \tau}_1$ and ${\tilde \ell}_R$ at
points $\zeta, \eta$.

$\bullet$
{\it Sneutrinos}: We do not consider these to be observable in any of the 
the three scenarios.

$\bullet$ 
{\it Gauginos}: The lightest neutralino $\chi$ should be observable in all
three scenarios $\epsilon, \zeta, \eta$, as a resonance in ${\tilde
\tau}_1 \tau$ combinations, for example in ${\tilde q}_R \to q (\chi \to
{\tilde \tau}_1 \tau)$ cascade decays which have branching ratios of
92/75/69~\% in the three scenarios. On the other hand, the second
neutralino $\chi_2$ is probably observable only in ${\tilde q}_L$
decays, and only in scenario $\epsilon$. We do not consider the
charginos and heavier neutralinos to be observable at any of these points.

$\bullet$ {\it Higgs bosons}: The $h$ should be observable in the
three scenarios $\epsilon$ to $\eta$, but the heavier Higgs bosons are 
not expected to be observable in any of them.

$\bullet$
{\it Squarks}: At point $\epsilon$, all squark flavours, excluding the
${\tilde t}_2$ but including the ${\tilde t}_1$ and ${\tilde b}_{1,2}$
(which appear in 10/15/11~\% of ${\tilde g}$ decays), should be
observable. The spartners of the $u, d, s, c$ quarks could also be
observed at points $\zeta, \eta$, but not the ${\tilde t}_{1,2}$ and
${\tilde b}_{1,2}$.

$\bullet$ {\it Gluinos}: According to our standard criteria, these should
be observed in all three scenarios $\epsilon, \zeta, \eta$.

\begin{figure}
\begin{center}
\begin{tabular}{c c}
\mbox{\epsfig{file=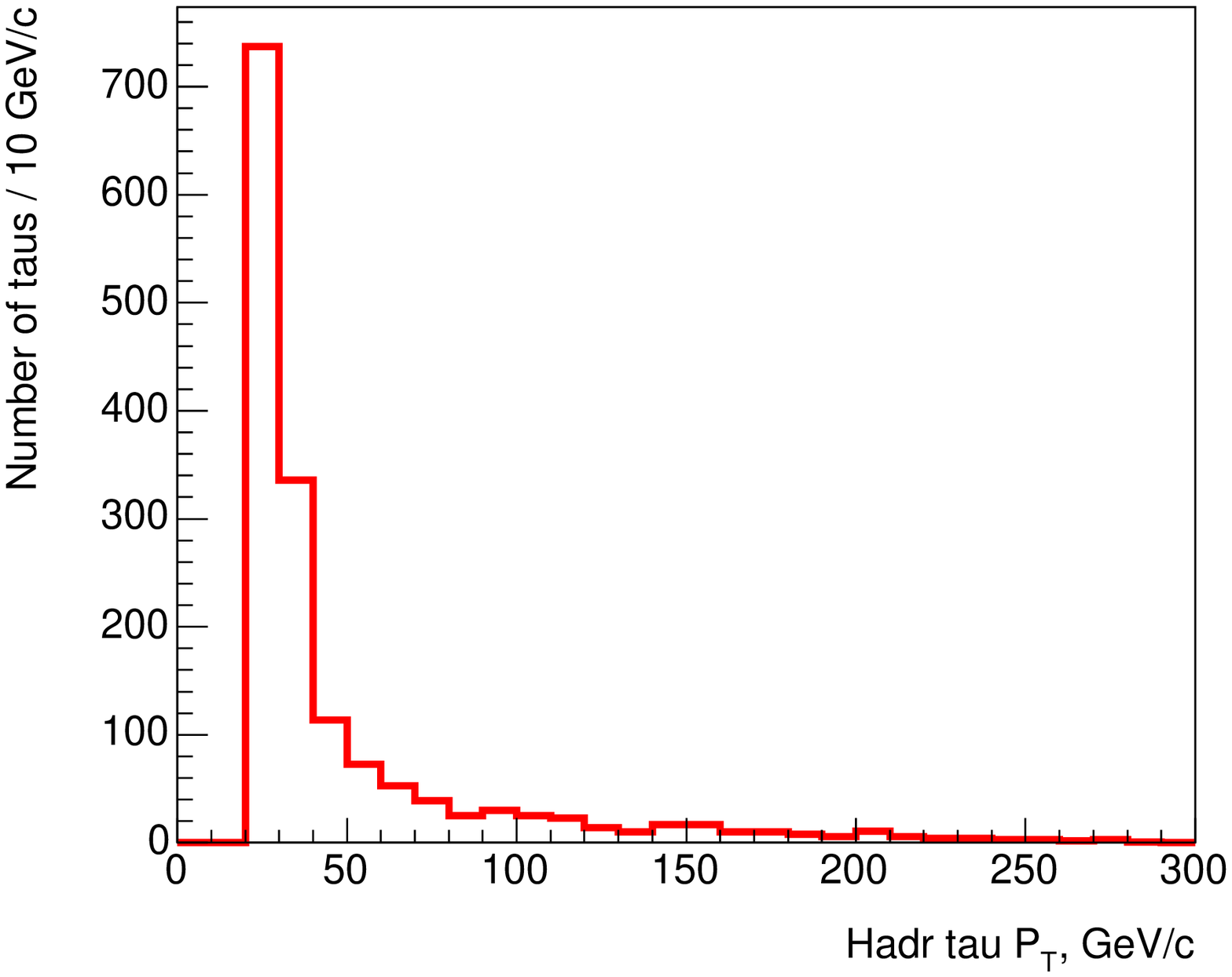,height=6cm}} &
\mbox{\epsfig{file=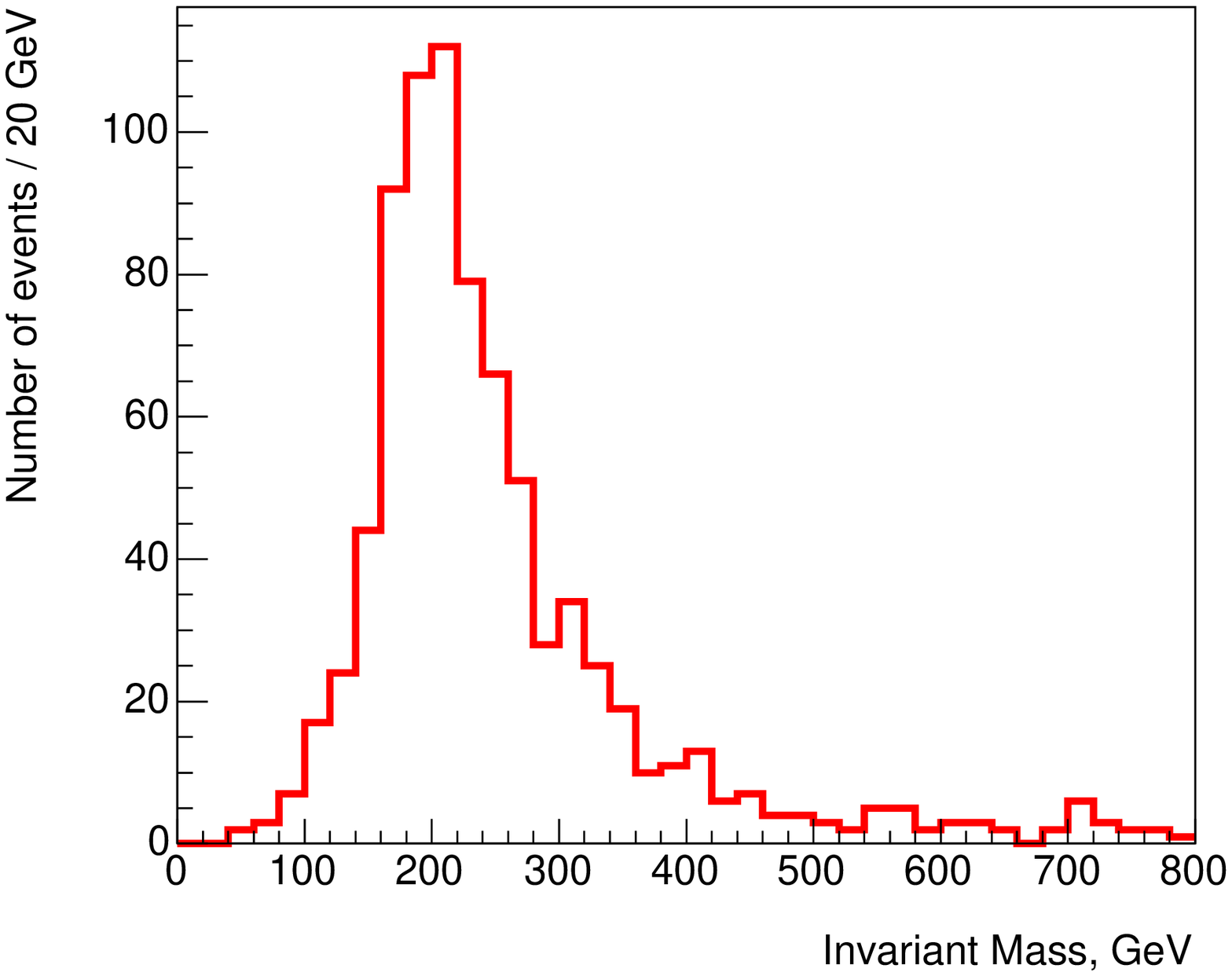,height=6cm}} \\
\end{tabular}
\begin{tabular}{c c}
\mbox{\epsfig{file=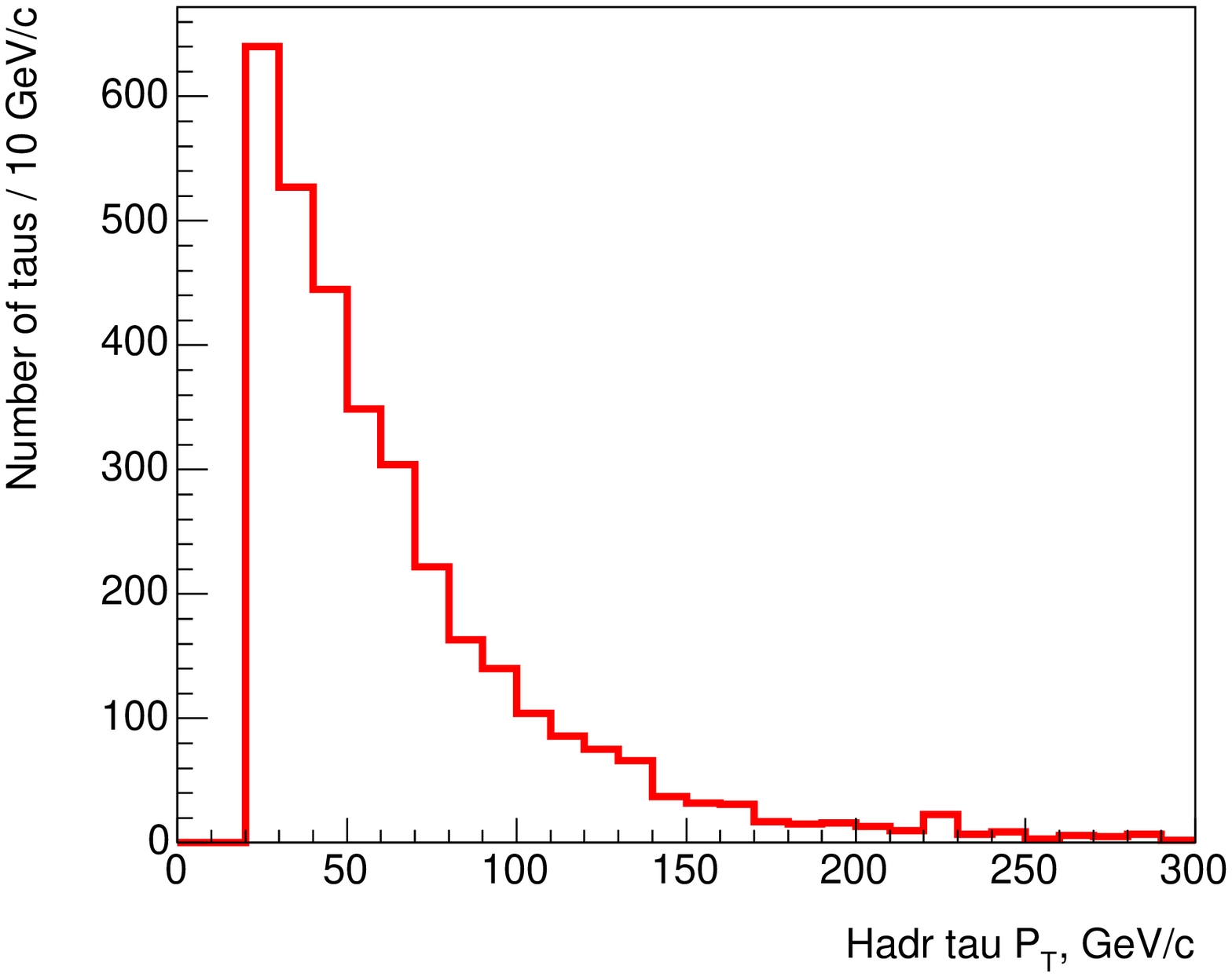,height=6cm}} &
\mbox{\epsfig{file=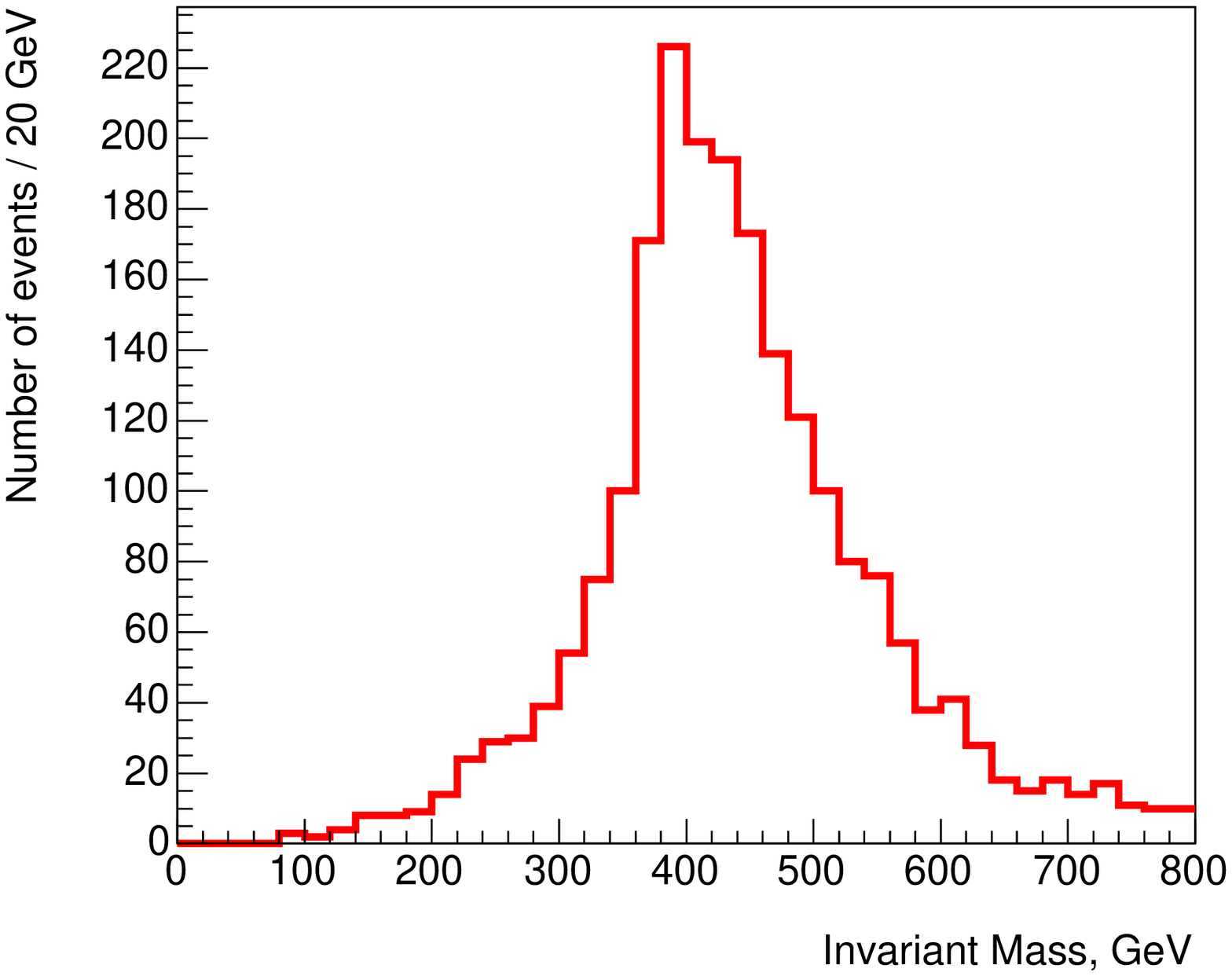,height=6cm}} \\
\end{tabular}
\begin{tabular}{c c}
\mbox{\epsfig{file=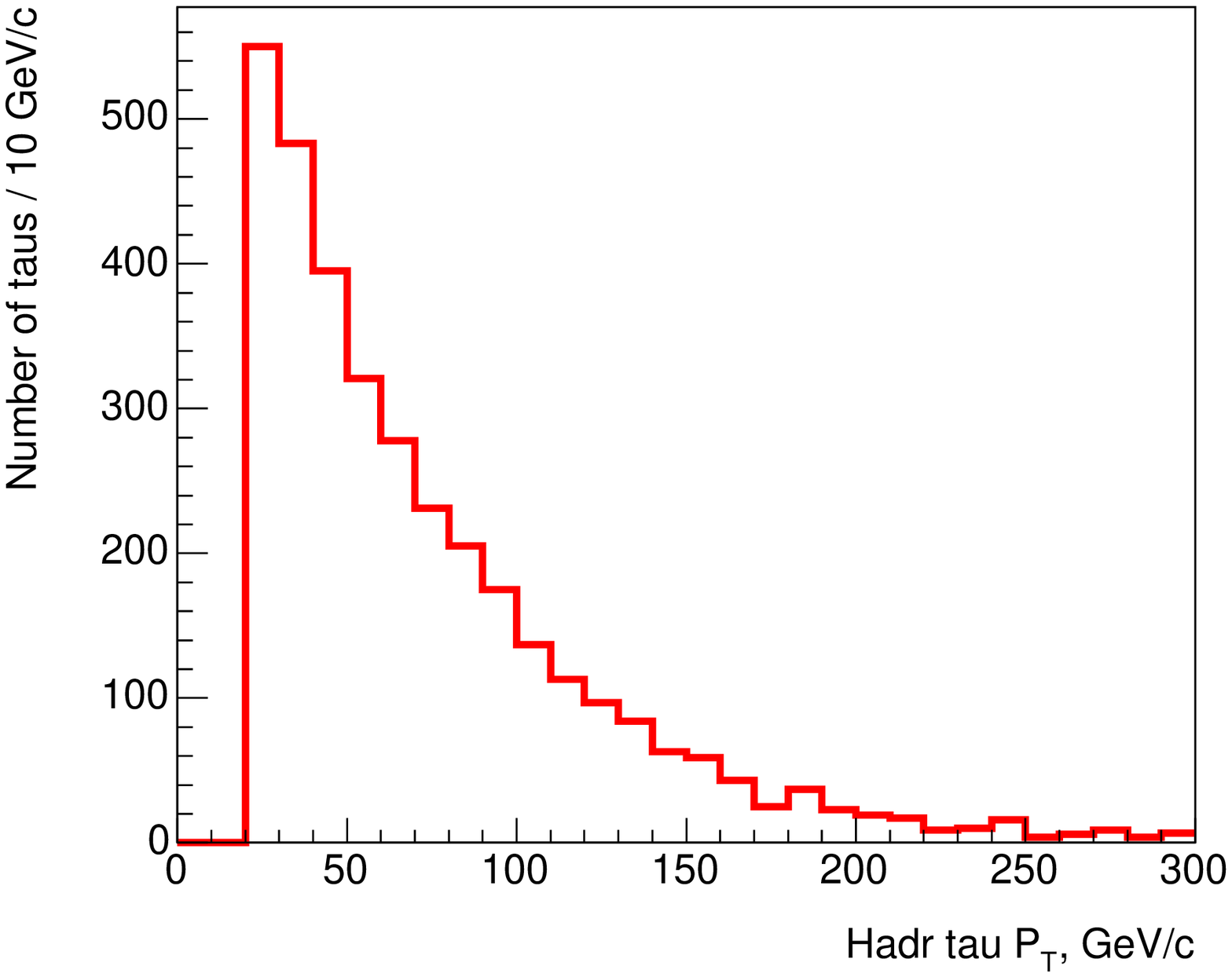,height=6cm}} &
\mbox{\epsfig{file=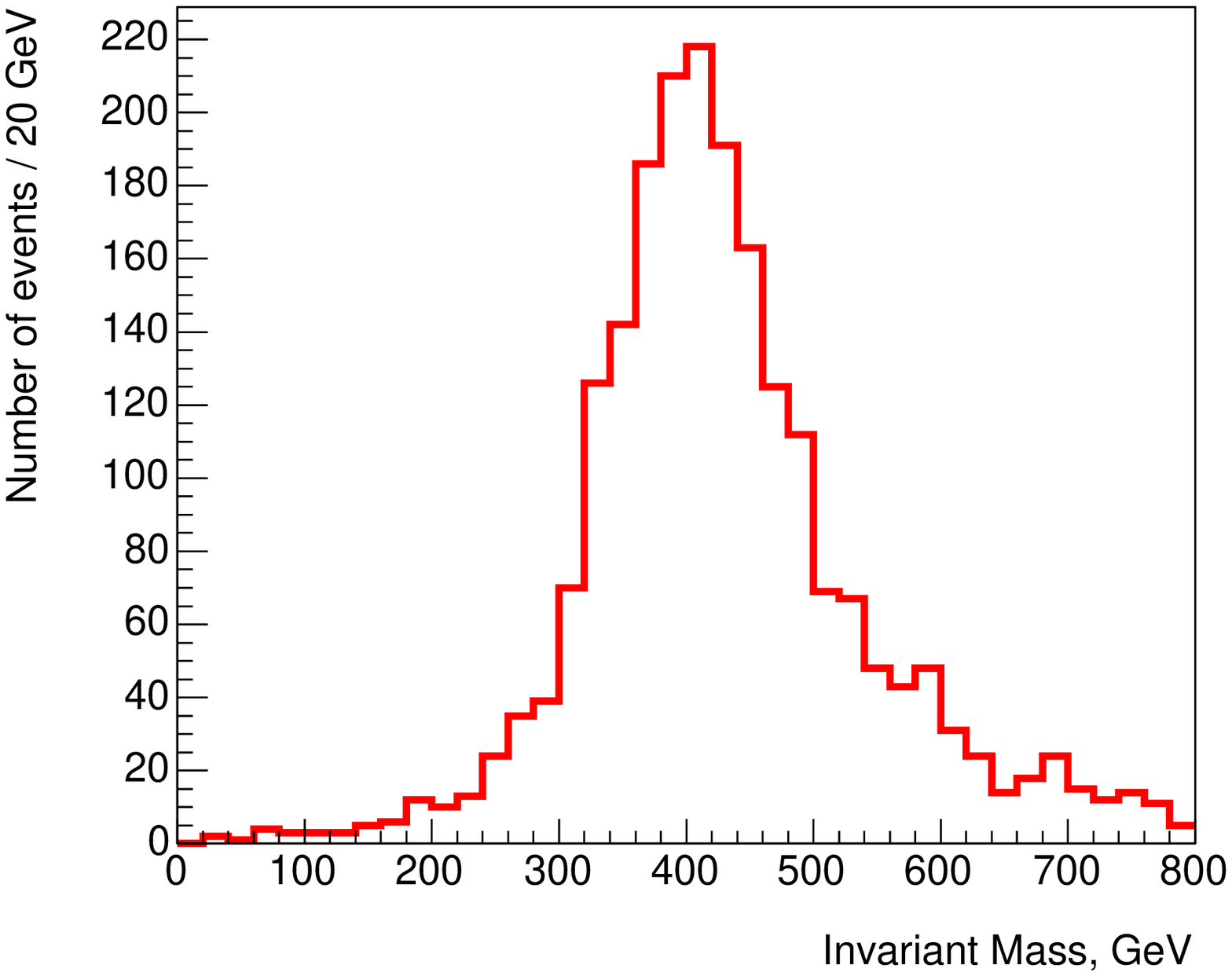,height=6cm}} \\
\end{tabular}
\end{center}   
\caption{\label{fig:stautau}\it
The $\tau$ $p_T$ distributions and the $\tau - {\tilde \tau}_1$ 
invariant-mass 
distributions for (a, b) benchmark scenario $\epsilon$, (c, d) benchmark 
scenario $\zeta$ and (e, f) benchmark scenario $\eta$.
}
\end{figure}

We have made a first examination of issues in the reconstruction of
sparticle cascade decays in GDM benchmark scenarios $\epsilon, \zeta,
\eta$, using as initial building-blocks the final-state ${\tilde \tau}_1$
and $\tau$. The $p_T$ distributions for hadronic $\tau$-decay jets at the
three points are shown in Fig.~\ref{fig:stautau}(a, c, e), where we see
that those at points $\zeta, \eta$ are significantly harder than at point
$\epsilon$. 
The mis-tag probabilities we take from~\cite{AHiggs} are adequate
for identifying the large-$p_T$ $\tau$ hadronic jets in GDM
sparticle-pair-production events, which would have been collected by the 
normal large-$p_T$ trigger at the LHC. Using
these tagging estimates, and combining the candidate $\tau$ hadronic jets
with the ${\tilde \tau}_1$ tracks, which are assumed to be measured with
an accuracy $\delta p/p = 5$~\%, we obtain the $\tau - {\tilde \tau}_1$
invariant-mass distributions shown in panels (b, d, f) of
Fig.~\ref{fig:stautau}, respectively. In each case, we see a clear signal
due to $\chi \to \tau {\tilde \tau}_1$ decays.

We have then considered the reconstruction of higher cascade decays in
scenarios $\epsilon$ and $\zeta$, the latter being very similar to point
$\eta$.  We find peaks in ${\tilde \tau}_1 - \tau - \ell$ combinations
corresponding to the ${\tilde \ell}_{L,R}$, and in ${\tilde \tau}_1 - \tau
- \ell - \ell$ combinations corresponding to the $\chi_2$. However, the
combinatorial backgrouds have shapes quite similar to the signals. Full
studies of these scenarios lie beyond the scope of this survey, but it
does seem that sparticle cascades can be reconstructed in these scenarios,
analogously to what was shown previously for scenarios with a $\chi$ LSP.

\subsection{Detectability at $e^+ e^-$ Linear Colliders}

As in our previous studies~\cite{Bench,Bench2}, our criteria for the
observability of supersymmetric particles at linear colliders are based on
their pair-production cross sections.

$\bullet$ Particles with cross sections in excess of 0.1~fb are considered
as observable, because they would give rise to more than 100 events with 
an
integrated luminosity of 1~ab$^{-1}$.

$\bullet$ The lightest neutralino $\chi$ is considered to be observable
only through the decays of heavier supersymmetric
particles.

$\bullet$ Sneutrinos are considered to be detectable when
the sum of the branching fractions for decays which lead to clean
experimental signatures, such as $\tilde{\nu}_{\ell} \to \chi^{\pm}
\ell^{\mp}$ ($\ell$ = $e$, $\mu$, $\tau$) and $\tilde{\nu}_{\tau} \to W^+
\tilde{\tau_1}^-$, exceeds 15~\%.

$\bullet$ 
The $\gamma \gamma$ collider option at a linear collider would
allow single production of  heavy neutral Higgs bosons via the $s$-channel
processes $\gamma \gamma \to A$ and $\gamma \gamma \to H$, extending the
reach up to 375~GeV for 0.25~TeV $e^\pm$ beams, 750~GeV for 0.5~TeV
$e^\pm$ beams, 2.0~TeV for 1.5~TeV $e^\pm$ beams and 3.75~TeV for 2.5~TeV 
$e^\pm$ beams. A $\gamma
\gamma$ collider may also be used to look for gluinos~\cite{Klasen}, but 
we do not
include this possibility in our analysis.

$\bullet$ Finally, we assume that a metastable ${\tilde \tau_1}$ could be
detected at any linear $e^+ e^-$ collider with more than 100 events, and
note that the mass could be measured more accurately than at the LHC, by
measuring the production threshold as well as $1/\beta$ (see the next 
Section for further discussion).

As previously~\cite{Bench,Bench2}, we consider $e^+ e^-$ collision
energies $\sqrt{s}$ = 0.5~TeV, 1.0~TeV, 3~TeV and 5~TeV, and also the
combined capabilities of the LHC and a 1-TeV linear collider.

$\bullet$ For $\sqrt{s}$ = 0.5~TeV, at NUHM
benchmark point $\alpha$ the $\chi, \chi_2, \chi_1^\pm$ and the lighter 
sleptons ${\tilde \mu}_R,
{\tilde e}_R, {\tilde \tau}_1$ would be observable. The prospects at
benchmark point $\beta$  are similar, except that the
$\chi^\pm$ would not be observable. At benchmark point $\gamma$, all the 
inos except the $\chi_2^\pm$ would be observable, and the 
sleptons ${\tilde \mu}_R,
{\tilde e}_R, {\tilde \tau}_1$ would be observable in $\chi_4$ decays. 
The lightest $h$ is always observable and the  $H, A$ can be 
produced in $\gamma \gamma$
collisions for these benchmark points.

The prospects for the GDM benchmark points are not so good: apart from
$h$, only $ {\tilde \mu}_R, {\tilde e}_{L,R}, {\tilde \tau}_{1,2}$ are
observable~\footnote{The heavier sleptons are visible via associated
${\tilde e}_R {\tilde e}_L$ and ${\tilde \tau}_1 {\tilde \tau}_2$
production.}, and then only in the low-mass scenario $\epsilon$ that was
chosen at the tip of the low-$m_0$ GDM wedge in mSUGRA parameter space.  
No squarks or gluinos are observable in any NUHM or GDM scenario at
$\sqrt{s}$ = 0.5~TeV.

$\bullet$ For $\sqrt{s}$ = 1~TeV, all the neutralinos, charginos and
sleptons (both charged and neutral) become observable in scenarios
$\alpha, \beta$ and $\gamma$, and the heavy Higgs bosons can now be pair
produced in $e^+e^-$ collisions directly. In the GDM scenario $\delta$,
associated $\chi \chi_2$ production becomes observable, albeit with a low
event rate, and ${\tilde \mu}_R, {\tilde e}_R, {\tilde \tau}_1$
production can be detected in $\chi_2$ decays. At point $\epsilon$, 
associated $\chi
\chi_2$ production should again be observable, and probably also
associated $\chi \chi_{3,4}$ production, as well as $\chi^\pm$ pair
production, associated $\chi^\pm \chi_2^\mp$ production and
pair-production of all the charged sleptons. Finally, at points $\zeta$
and $\eta$, only $\chi$, ${\tilde \mu}_R, {\tilde e}_R$ and ${\tilde
\tau}_1$ are expected to be observable.

$\bullet$ For $\sqrt{s}$ = 3~TeV, all the Higgs bosons, neutralinos,
charginos, sleptons and squarks would be observable in each of the
scenarios $\alpha, \beta, \gamma, \epsilon$. The same is true at benchmark
point $\delta$, with the exception of the left-handed first- and
second-generation squarks, and the right-handed squarks ${\tilde t}_2$ and
${\tilde b}_2$, because of their low rates. At benchmark points $\zeta,
\eta$, one should observe all the weakly-interacting sparticles, but (with
the exception of the ${\tilde t}_1$) the squarks would still be out of
kinematic reach. At benchmark point $\gamma$, also the gluino will be
observable in squark decays.

$\bullet$ For $\sqrt{s}$ = 5~TeV, the story is simple: all the sparticles 
except ${\tilde g}$ would be observable in all the proposed NUHM and GDM 
benchmarks, and also the ${\tilde g}$ at point $\gamma$.

\subsection{Summary}

\begin{figure} 
\epsfig{file=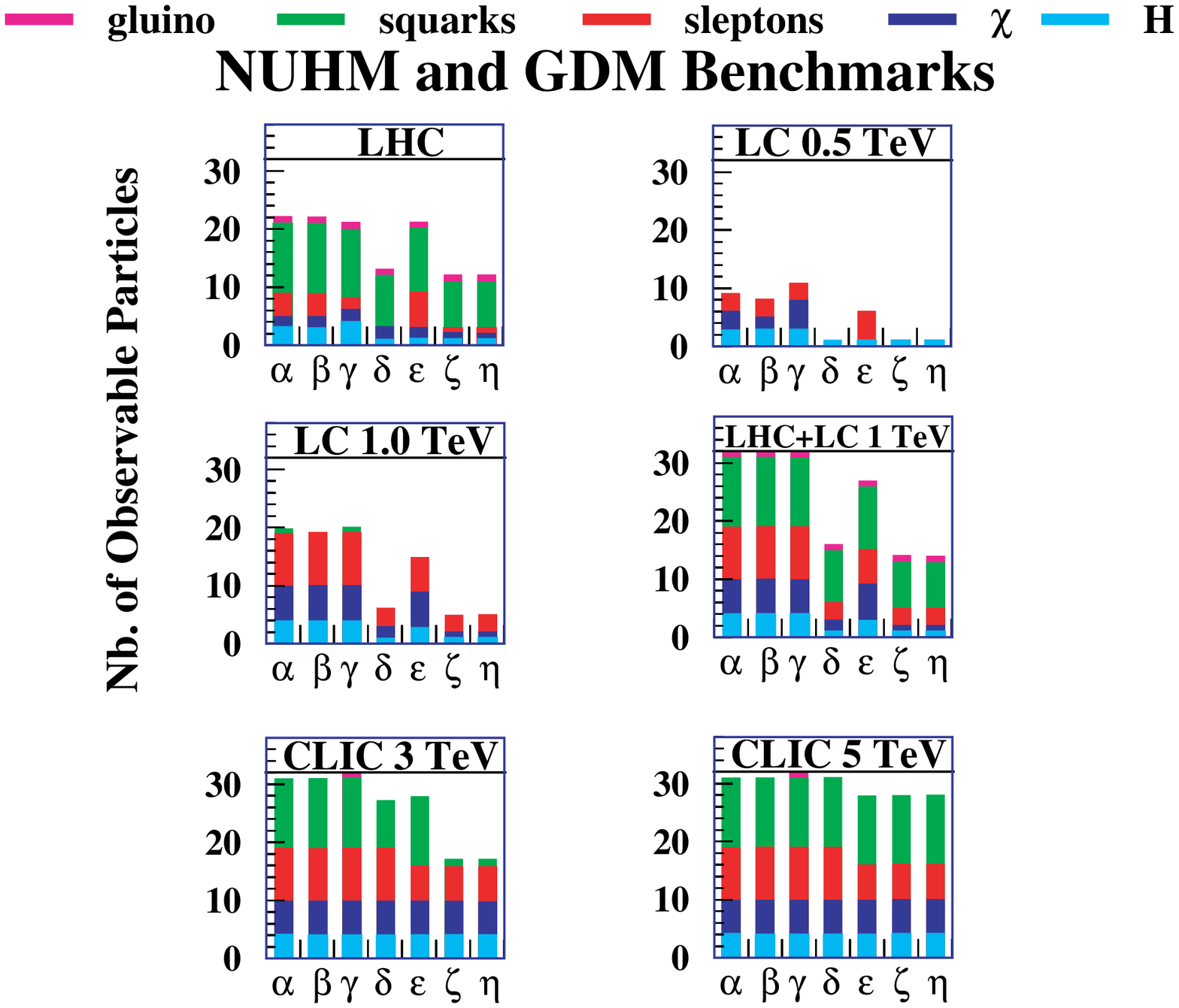,height=5.5in}
\caption{\label{fig:newM}
{\it 
Summary of the numbers and types of MSSM particles that may be detected
at various accelerators in the NUHM and GDM benchmark scenarios 
considered here. The gravitino may also be indirectly `observable' at 
point $\epsilon$. We emphasize that these numbers are estimates that 
need to be validated by experimental simulations. As 
in~\protect\cite{Bench,Bench2}, 
we see that the capabilities of the LHC and of linear $e^+ e^-$ colliders 
are largely complementary. We re-emphasize that mass and coupling measurements
at $e^+ e^-$ colliders are usually much cleaner and more precise than at
the LHC, where, for example, it is not known how to distinguish the light 
squark flavours. 
}}
\end{figure}

Fig.~\ref{fig:newM} summarizes the numbers of different species of MSSM
particles visible at different accelerators. We see that the LHC provides
good coverage for strongly-interacting sparticles (uppermost green and
pink bars) in all the NUHM and GDM benchmarks scenarios considered,
whereas its coverage for weakly-interacting sparticles (middle red and
blue bars) is rather uneven. The lightest Higgs boson is always detectable
(lowest light blue bars) and in some cases also heavier Higgs bosons. In
scenarios $\alpha, \beta, \gamma, \epsilon$, a linear $e^+e^-$ collider
with $\sqrt{s} = 0.5$~TeV would provide useful extra information about 
some
weakly-interacting sparticles.  In all cases, it would provide detailed
measurements of one or more Higgs bosons. A linear $e^+e^-$ collider with
$\sqrt{s} = 1.0$~TeV would provide better information on both
weakly-interacting sparticles and Higgs bosons, but still no information
on squarks or gluinos. The combination of the LHC and a 1.0-TeV linear
collider would provide good coverage overall, but this would still be
incomplete in scenarios $\delta, \zeta, \eta$, in particular. A linear
$e^+e^-$ collider such as CLIC with $\sqrt{s} = 3.0$~TeV would provide
detailed studies of all the weakly-interacting sparticles and Higgs bosons
in all the scenarios studied, and also provide new opportunities to study
squarks in scenarios $\alpha, \beta, \gamma, \delta, \epsilon$. Finally,
CLIC at 5.0~TeV would provide detailed measurements of all the MSSM
particles except possibly the gluino, for which one would still rely on 
the
LHC~\footnote{Unless one could also observe $\gamma \gamma \to {\tilde g}
{\tilde g}$~\cite{Klasen}, a possibility not considered here.}.

\section{The Stau NLSP in GDM Scenarios}

\subsection{Production and Detectability at the LHC}

In the mSUGRA GDM models studied here, all supersymmetric events yield a
pair of ${\tilde \tau}_1$ NLSPs. The astrophysical BBN/CMB constraint
prevents the ${\tilde \tau}_1$ lifetime from exceeding $\sim 3 \times
10^{6}$~s~\cite{CEFO}, and we do not discuss here ${\tilde \tau}_1$ 
lifetimes smaller than
$10^4$~s. Charged NLSPs with lifetimes in this range would appear to a
generic collider detector like massive stable charged particles, and the
three benchmark scenarios $\epsilon, \zeta, \eta$ studied here span this
range of lifetimes.

\begin{figure}
\begin{center}
\begin{tabular}{c c c}
\mbox{\epsfig{file=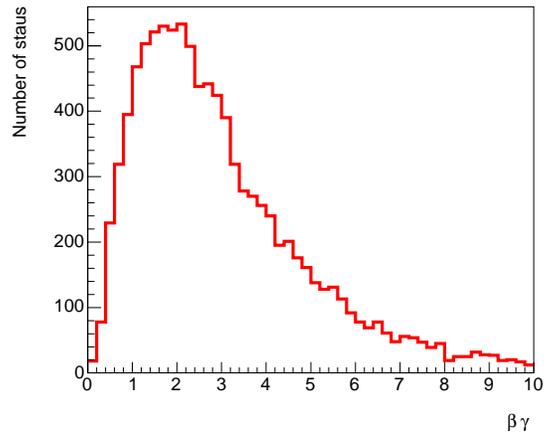,height=6.5cm}} \\
\mbox{\epsfig{file=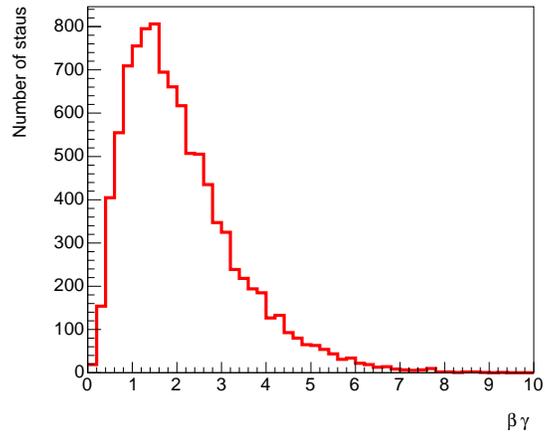,height=6.5cm}} \\
\mbox{\epsfig{file=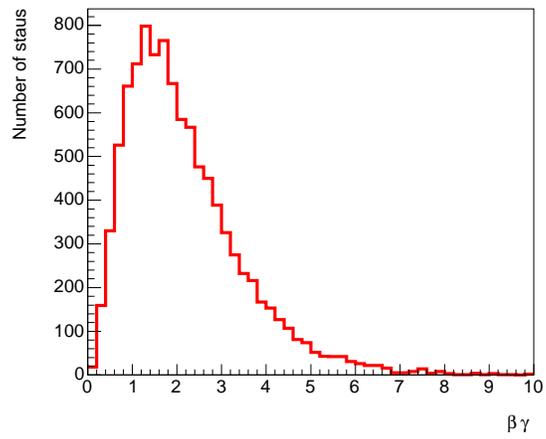,height=6.5cm}} \\
\end{tabular}
\end{center}   
\caption{\label{fig:betagamma}\it
The spectra of the non-relativistic factor $\beta \gamma$ for the 
${\tilde \tau}_1$'s produced at the LHC, in the benchmark scenarios 
$\alpha, \beta, \gamma$, respectively.
}
\end{figure}

Fig.~\ref{fig:betagamma} shows the distributions of the non-relativistic
factor $\beta \gamma$ expected for the ${\tilde \tau}_1$'s in these mSUGRA 
GDM scenarios from cascade
decays of squarks and gluinos at the LHC. The great majority
of the ${\tilde \tau}_1$'s produced at the LHC are far from being
ultra-relativistic, and so should yield exotic time-of-flight (TOF) and/or
$dE/dx$ signals~\footnote{We are not optimistic about the
prospects of detecting these signals in hadron-collider events without
other distinguishing features, such as the Drell-Yan production of
${\tilde \tau}_1$ pairs.}. The same would be true of ${\tilde \tau}_1$'s
produced at the ILC, if its centre-of-mass energy reaches above the
pair-production threshold in the corresponding scenarios, namely 310, 680
and 650~GeV in benchmarks $\epsilon, \zeta$ and $\eta$, respectively. CLIC
with a centre-of-mass energy of 3~TeV would be able to produce ${\tilde
\tau}_1$ pairs with masses $\le 1.5$~TeV, and hence probe the GDM wedge
out to $m_{1/2} \sim 4.5$~TeV.

The key signal for GDM with a ${\tilde \tau}_1$ NLSP would in general be
the coincident appearance in adjacent or nearby bunch crossings of generic
high-$p_T$ triggers and subsequent `muon' triggers. Such coincidences 
would be very
rare for conventional trigger rates, which would give coincidence rates $< 
10^{-6}$ in adjacent bunch
crossings.

The ${\tilde \tau}_1$ NLSP is often sufficiently
non-relativistic, $\beta \gamma < 1$, that it would not exit an LHC 
detector such as ATLAS or
CMS before the next bunch crossing (25~ns after the event in which the
${\tilde \tau}_1$ was produced), in which case its tracking information 
might be lost. This issue could be addressed by reading
out of the detector all the tracking signals that occurred within several
crossing times following an `interesting' event. `Interest' would normally
be defined by a conventional high-$p_T$ lepton or calorimetric trigger. 
As seen in Figs.~\ref{fig:trigger5}, \ref{fig:trigger6} and
\ref{fig:trigger7}, most supersymmetric events would
indeed pass the normal ATLAS and CMS criteria for `interesting' events.

If a sample of interesting candidate events can be identified, one
possible ${\tilde \tau}_1$ search strategy would be to select out of the
usual high-$p_T$ lepton and calorimetric triggers a subsample of events
suspected of containing ${\tilde \tau}_1$ NLSPs. Even if the muon 
systems do not trigger on the ${\tilde \tau}_1$'s, the muon drift tubes of
both ATLAS and CMS integrate signals over a number of bunch crossings,
such that hits of particles which are out of the normal 25-ns time window
can still be recorded with the event. One would, however, need to adapt
the track reconstruction software so as to allow for such signal time
shifts and recuperate the full information. At the moment, in the absence
of any good reason to take seriously such scenarios with massive,
slow-moving charged particles, the only experimental strategy required 
from the ATLAS and
CMS Collaborations is to avoid precluding the possibility of such a
buffered readout, should it subsequently appear worthwhile for searches 
in GDM or other scenarios.

Alternatively, one could use the presence of a high-$p_T$ charged particle
in the muon system as a primary trigger, and then look back through
earlier bunch crossings for evidence of other high-$p_T$ jets and/or
leptons, that would already have triggered the detector and been recorded.

\subsection{Measuring the Stau Mass}

\begin{figure}
\begin{center}
\begin{tabular}{c c}
\mbox{\epsfig{file=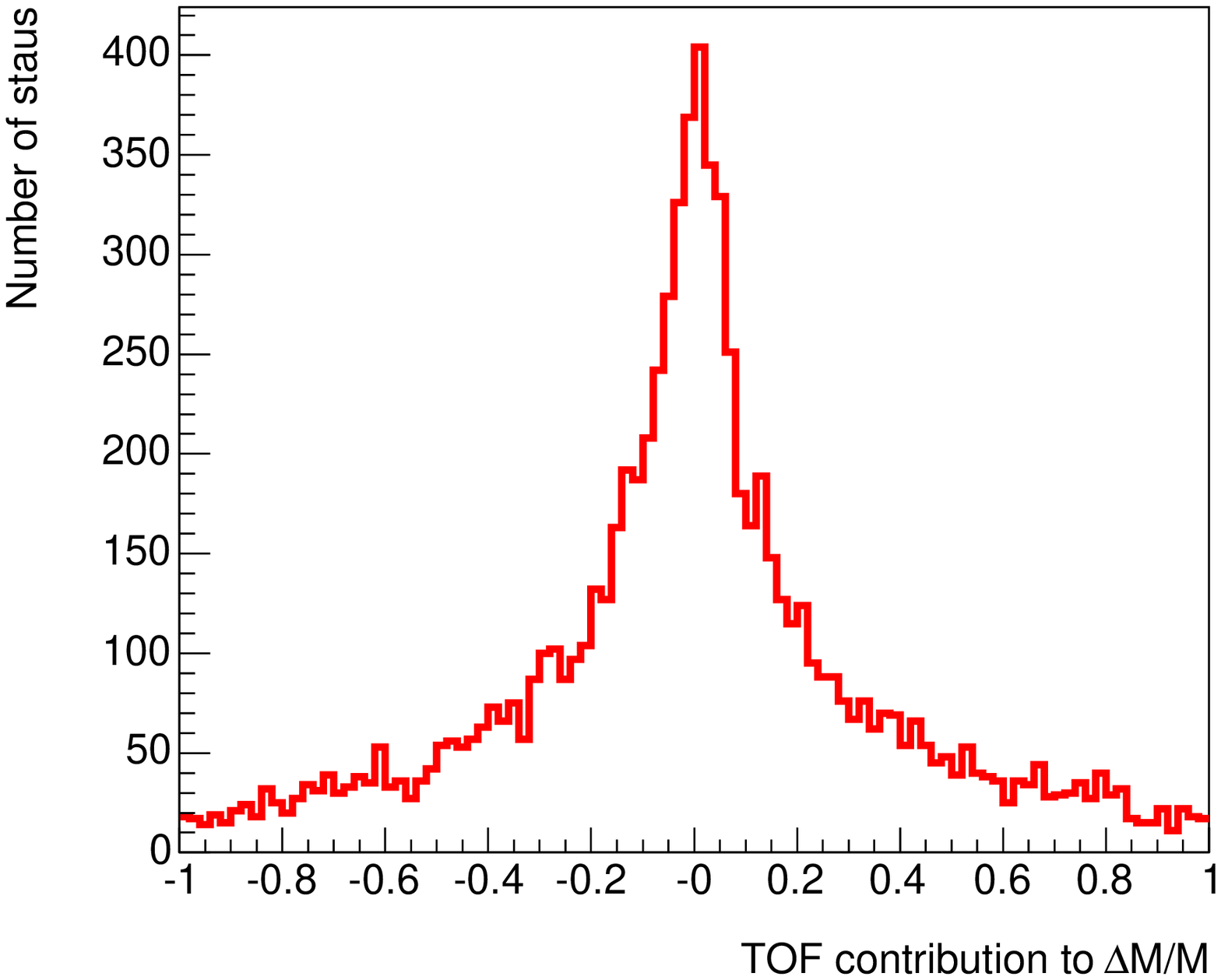,height=6cm}} &
\mbox{\epsfig{file=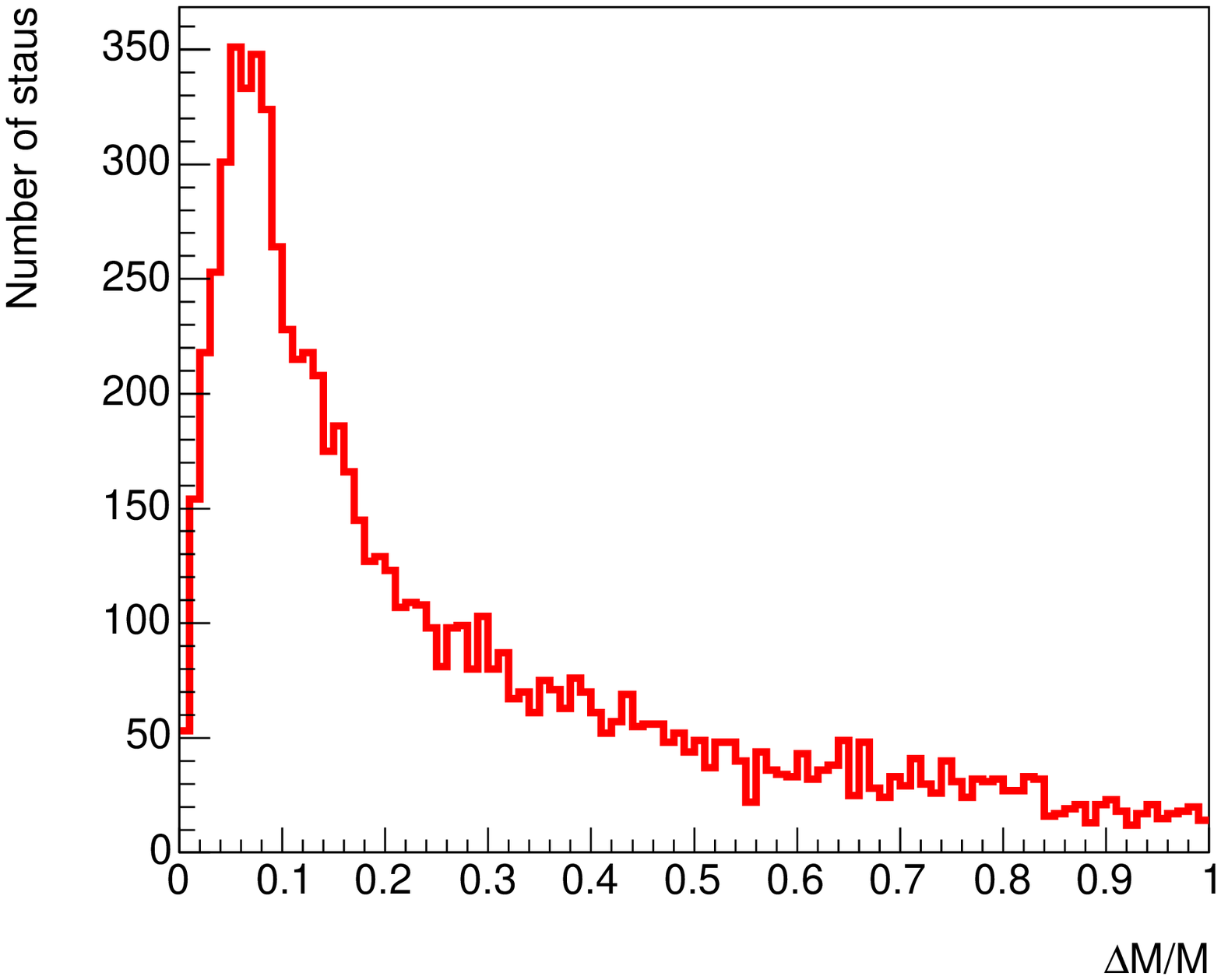,height=6cm}} \\
\end{tabular}
\end{center}   
\begin{center}
\begin{tabular}{c c}
\mbox{\epsfig{file=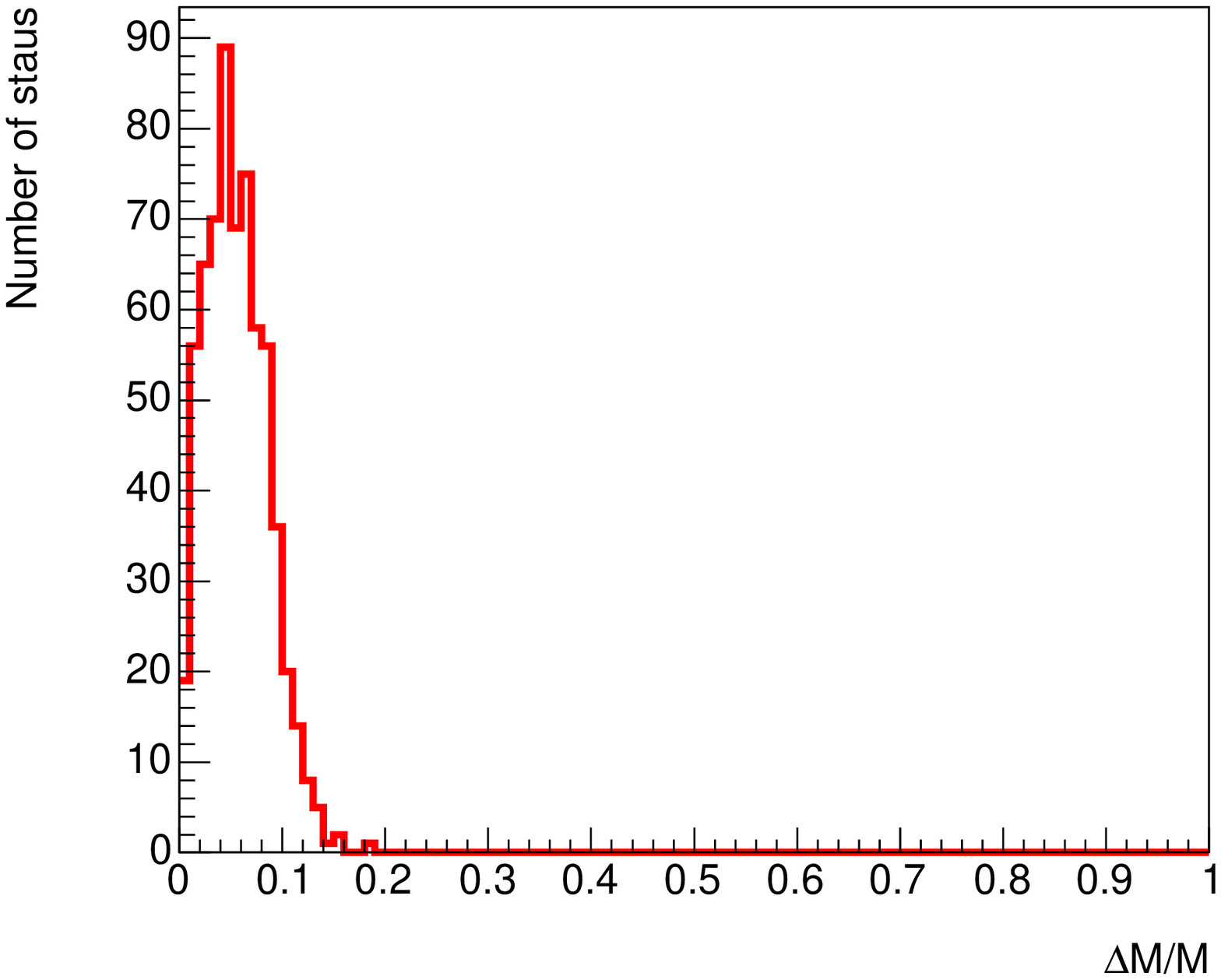,height=6cm}} &
\mbox{\epsfig{file=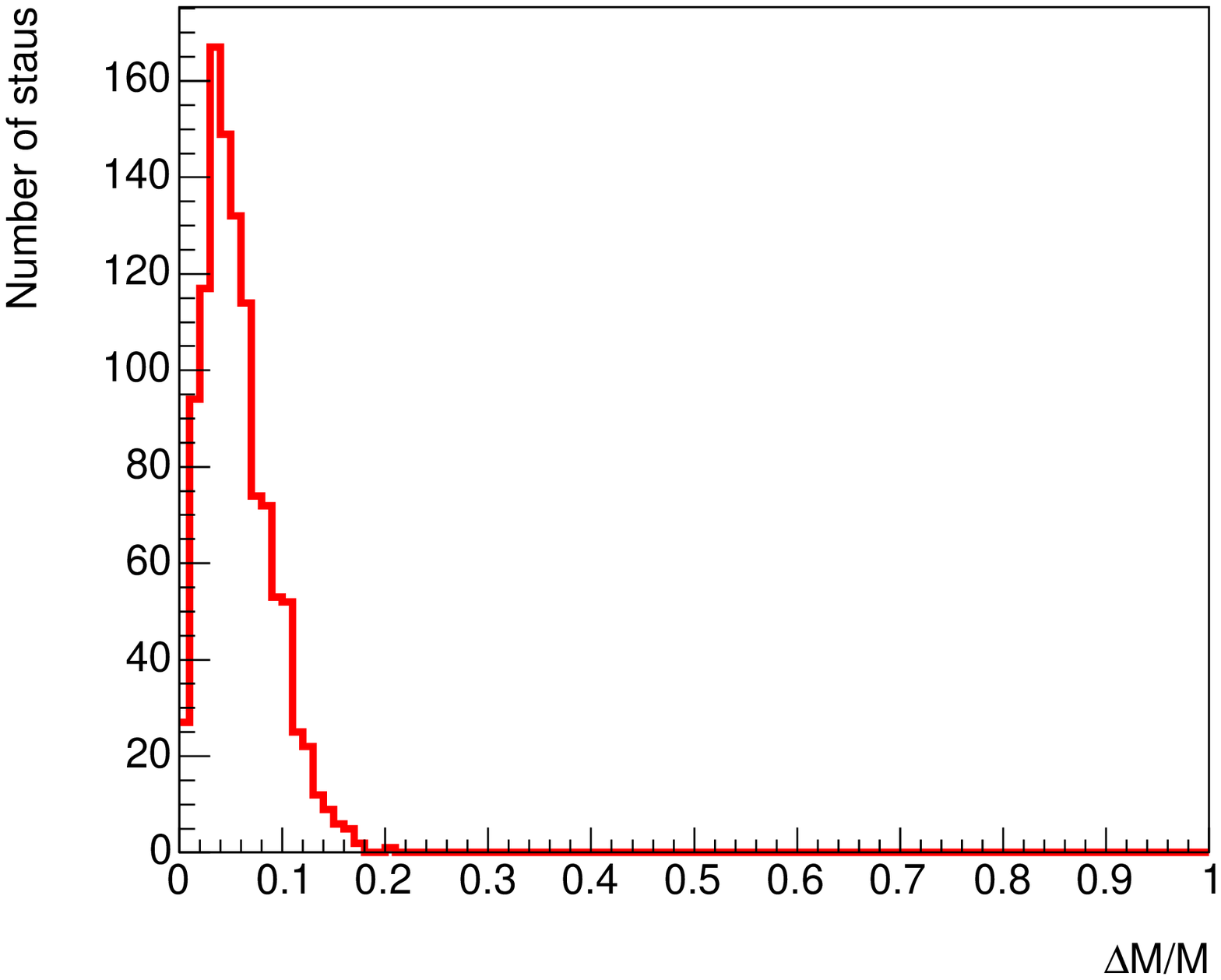,height=6cm}} \\
\end{tabular}
\end{center}   
\caption{\label{fig:deltam}\it
(a) The TOF contribution to the mass resolution $\Delta M/M$ 
(\protect\ref{deltam}) in the stau NLSP mass measurement, 
(b) the full uncertainty $\Delta M/M$ 
found for the complete event sample of
benchmark point $\epsilon$, (c) $\Delta M/M$ for the 10~\% of 
events at point $\epsilon$ with the lowest values of $\beta 
\gamma^2$, and (d) the same distribution for point $\zeta$ 
(the result for point $\eta$ is very similar).}
\end{figure}

A crude estimate of the obtainable $\tilde{\tau}_1$ mass resolution
can be derived by propagating the uncertainties in the momentum 
measurement $\Delta p$ and in the TOF resolution $\Delta 
t$, as determined in a detector at a distance $L$:
\begin{eqnarray}
\frac{\Delta M}{M} = \frac{\Delta p}{p} \bigoplus
    \beta \gamma^2 \frac{\Delta t}{L}
\label{deltam}
\end{eqnarray}
For ATLAS and CMS, the expectations are $\Delta p / p \simeq 1 - 10$~\%
and $\Delta t \simeq$ 1 ns at a distance of $\sim 
5$~m. Since the peak value of $\beta \gamma^2$, 
is $\sim 2$, as seen in Fig.~\ref{fig:betagamma}, 
we estimate that in each event
\begin{eqnarray}
\frac{\Delta M}{M} = (0.01 - 0.10) \bigoplus 0.12.
\label{deltam2}
\end{eqnarray}
We therefore estimate that the $\tilde{\tau}_1$ mass could be measured 
with an error of 10 - 20~\% in each event, which could be reduced by 
selecting low-momentum events as shown in Fig.~\ref{fig:deltam},
and further by combining measurements in 
many events. Panel (a) shows the distribution of the second
term in (\ref{deltam}) for $\tilde{\tau}_1$'s produced 
in a sample of events from benchmark point $\epsilon$, and panel (b) shows 
the distribution of the corresponding mass resolution $\Delta M$ 
obtainable event-by-event. Selecting now a sample of the 
10~\% of $\tilde{\tau}_1$'s with the lowest values of $\beta 
\gamma^2$, and hence, according to (\ref{deltam}), those with the 
smallest values of $\Delta M$~\footnote{In contrast, 
in the context of gauge-mediated supersymmetry-breaking models 
(GMSB)~\cite{GMSB} with masses similar to those in the GDM models 
considered here, 
Ref.~\cite{Aetal} considered the measurement of particles in the upper 
range of $\beta$ before the next bunch crossing. This would not be 
optimal for measuring the metastable particle mass, but it was 
nevertheless estimated that a precision $\Delta M /M < 1$~\% could be 
attained. Measuring the 
lifetime of the metastable particle inside the collider detector was also 
considered in such GMSB models~\cite{Aetal}, but in GDM models this would 
be feasible only after first stopping the NLSP, since its 
lifetime is much longer, as we discuss below.}, and assuming 
that $\Delta p / p$ has a Gaussian error of 5~\%, we obtain panel (c) of 
Fig.~\ref{fig:deltam}. We see that almost all these individual 
events have $\Delta M /M < 10$~\%. The same is true for an 
anologous sample of events produced in a simulation of benchmark 
point $\zeta$, as seen in panel (d) of Fig.~\ref{fig:deltam} and 
also for point $\eta$ (not shown). Therefore, if one could obtain 
a total sample of $\sim 1000$ $\tilde{\tau}_1$ events, and if 
systematic effects and correlations could be controlled, combining the 
best-measured 100 events could yield $\Delta M /M < 1$~\%.

\begin{figure}  
\begin{center}
\mbox{\epsfig{file=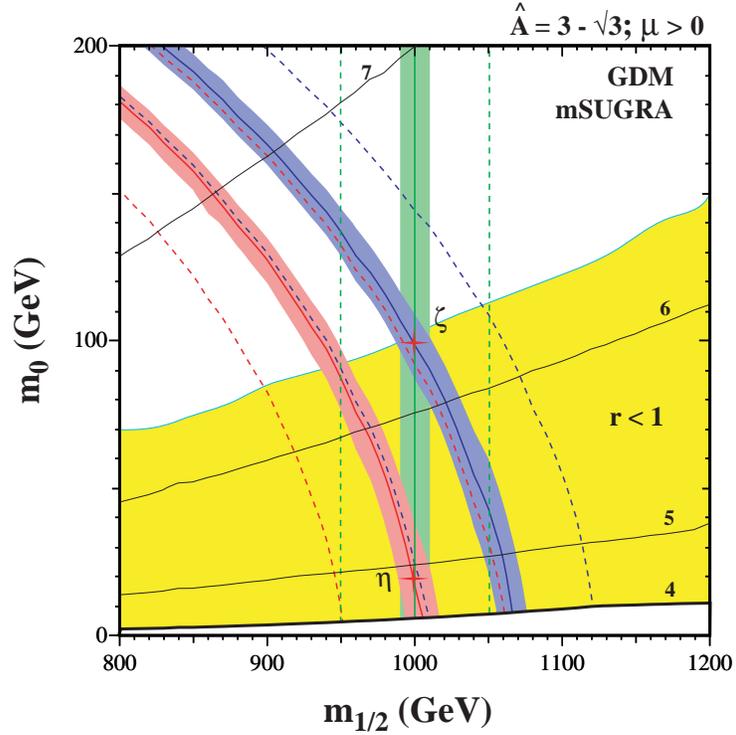,height=10cm}}
\end{center}
\caption{\it The potential impacts of prospective measurement errors 
of 1~\% and 5~\% for $m_{{\tilde \tau}_1}$ 
(diagonal bands and lines) and $m_{1/2}$ (vertical bands and lines) 
as constraints in the $(m_{1/2}, m_0)$ plane for GDM models in the 
mSUGRA framework. The smaller errors would enable the benchmark scenarios 
$\zeta$ and $\eta$ to be distinguished, and the possible NLSP lifetime 
to be estimated. The thin solid lines are labeled by the logarithm of the 
NLSP lifetime in seconds.}
\label{fig:combine} 
\end{figure}

To help assess the importance of such a measurement, we display in
Fig.~\ref{fig:combine} how a 1~\% ${\tilde \tau}_1$ mass measurement could
be combined with a determination of the supersymmetric mass scale
$m_{1/2}$ to determine the allowed range of parameters in the $(m_{1/2},
m_0)$ plane. If $m_{1/2}$ were
known exactly, the error in $m_{{\tilde \tau}_1}$ would correspond to
$\delta m_0 \sim 20$~GeV, which should be compounded with the error
induced by propagating the uncertainty in $m_{1/2}$. Such a determination
of $m_0$ would enable a ballpark estimate of the ${\tilde \tau}_1$ NLSP
lifetime to be made, within this mSUGRA GDM framework, enabling a strategy
to search for its decays to be better focused, as also seen in
Fig.~\ref{fig:combine}. 

The value of $m_{1/2}$ could perhaps be determined by measuring the gluino
mass, but here we discuss the use of the total supersymmetric cross
section. Comparing the total sparticle production cross section in
scenario $\epsilon$ with those for scenarios $\zeta, \eta$, we see that
$\sigma_{tot} \sim m_{1/2}^{-6}$, approximately. The statistical error in
measuring $\sigma_{tot}$ for either of the scenarios $\zeta, \eta$ would
be about 1.5~\%, but we expect that the systematic and theoretical errors
would be larger. Neglecting theoretical errors, an experimental error of
5~\% would enable $m_{1/2}$ to be estimated with an uncertainty $< 1$~\%
from the total cross section alone: see the vertical shaded band in 
Fig.~\ref{fig:combine}~\footnote{A more conservative error estimate of 
$\sim 25$~\% would
yield the wider band indicated by dashed vertical lines.}.  Assuming also
a measurement uncertainty of $\sim 1$~\% in $m_{{\tilde \tau}_1}$, we see
from Table~\ref{tab:points} and Fig.~\ref{fig:combine} (diagonal shaded
bands) that this should be sufficient to distinguish between scenarios
$\zeta, \eta$ at the 5-$\sigma$ level. However, this discrimination would
be lost if the error in either $m_{{\tilde \tau}_1}$ or $m_{1/2}$ rose to
$\sim 5$~\%.

One could in principle also look for slowly-moving massive charged
particles via an anomalous $dE/dx$ signal, for instance using
high-threshold signals in the ATLAS TRT detector, and maybe using the
electromagnetic calorimeter of either experiment~\footnote{It might also
be interesting to add to ATLAS or CMS a specialized time-of-flight
detector, either in the cavern itself or suspended in the access pit
outside it~\cite{ADONIS}.}. A good $dE/dx$ measurement is a design feature
that could be considered for the central trackers of future detectors at
the ILC and/or CLIC.

\subsection{Distinguishing GDM Benchmarks from GMSB Models}

We now consider the possibility of using spectroscopic measurements at the
LHC to distinguish the GDM benchmark scenarios considered here from
minimal gauge-mediated models of supersymmetry breaking
(mGMSB)~\footnote{LHC measurements of supersymmetric cascade decay
branching ratios might also help discriminate, but we do not consider them
here. Since the NLSP lifetimes are very different at our GDM benchmark
points from mGMSB models, the detection of decays into gravitinos, 
discussed in
the next subsection, would also help in the discrimination. The ILC would
be able to distinguish GDM from mGMSB very easily.}.  We first consider 
the
spectroscopic properties of the `easy' GDM scenario $\epsilon$, and
compare them with mGMSB models with the same value of $m_{1/2}$. The
principal discriminants we consider are the masses of the $\chi, \chi_2$,
the fact that the ${\tilde \ell}_R$ is significantly lighter than the
$\chi$, and the ${\tilde q}_R$ mass.

We recall that mGMSB mass spectra are characterized typically by the
messenger index $N$, a mass $\Lambda$ that sets the overall gaugino 
mass scale:
$M_a = (\alpha_a / 4 \pi) N \Lambda$ $(a = 1, 2, 3)$ and the scalar 
masses: $m_{0_i}^2 \propto N
\Lambda^2$, and an input scale $M$ from which the RGEs are used to evolve
the sparticle masses down to $Q_{EWSB} = 2$~TeV.

\begin{figure}
\begin{center}
\begin{tabular}{c c}
\mbox{\epsfig{file=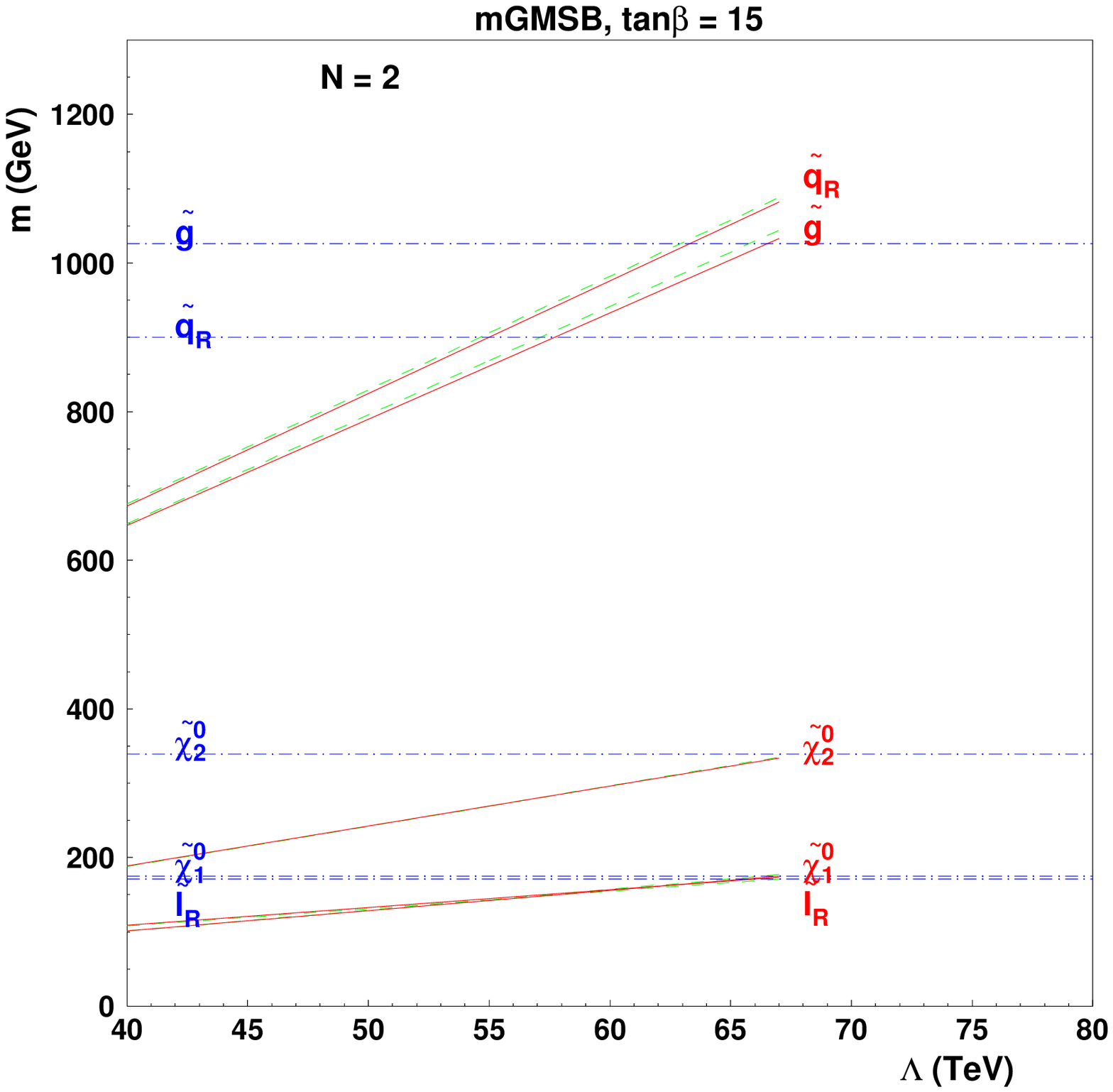,height=8cm}} &
\mbox{\epsfig{file=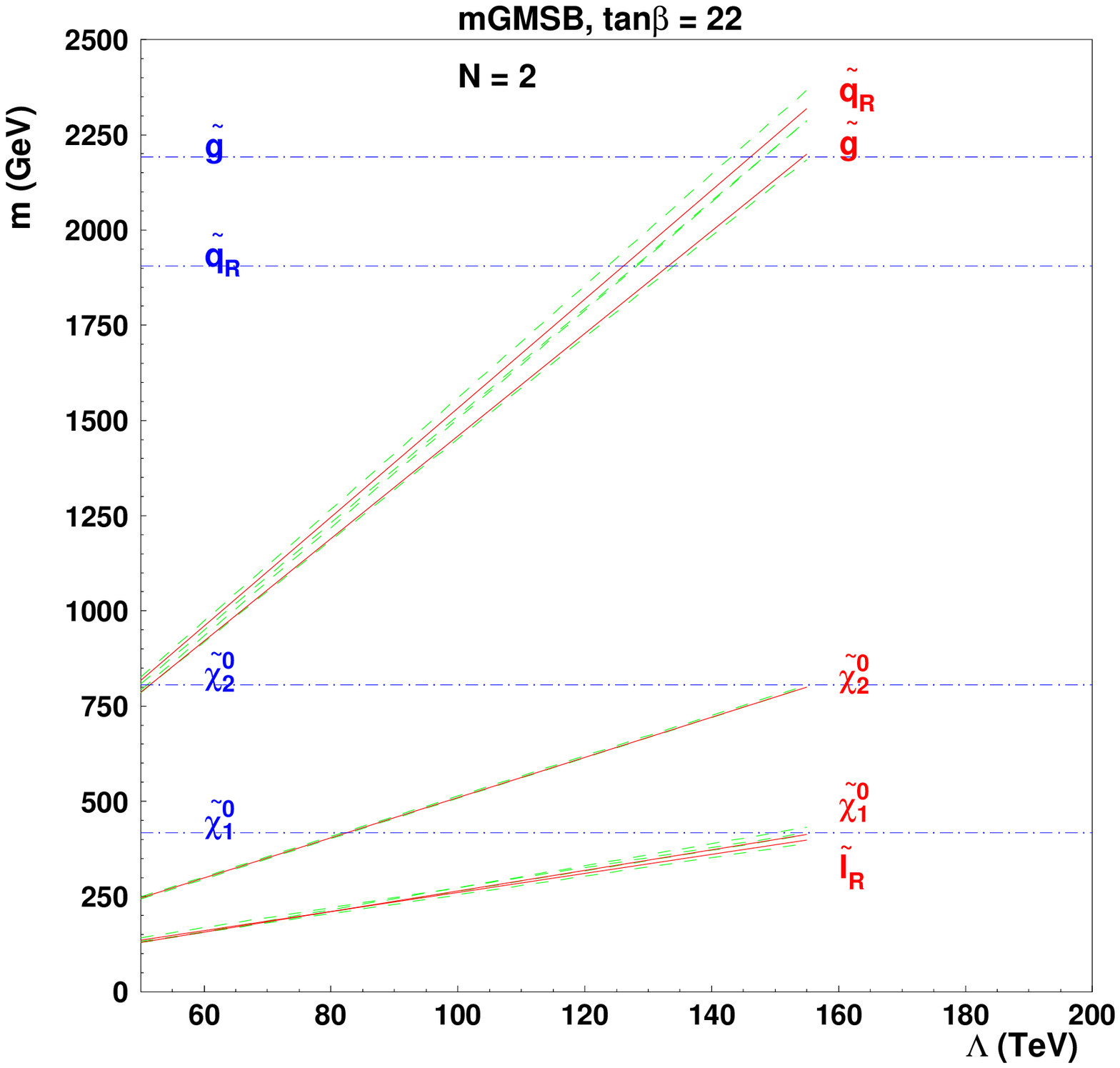,height=8cm}} \\
\end{tabular}
\end{center}
\caption{\label{fig:GMSB}\it
Sparticle masses in minimal GMSB models with $N = 2$ and (a) $m_{1/2} = 
440$~GeV, 
$\tan \beta = 15$ and (b) $m_{1/2} = 1000$~GeV and $\tan \beta = 22$, as 
functions of $\Lambda$ for (a) $M = 800, 300, 200$~TeV (diagonal dashed, 
solid and dashed lines, respectively), and (b) $M = 8000, 1000, 300$~TeV 
(diagonal dashed, solid 
and dashed lines, respectively), compared with GDM predictions (horizontal 
dot-dashed lines). The best agreement between the two models is found for 
(a) $\Lambda \simeq 67$~TeV and (b) 
$\Lambda \simeq 155$~TeV, respectively. 
}
\end{figure}

For $N = 1$, the ${\tilde \ell}_R$ is always heavier than the $\chi$, for
$N = 2$ the $\chi - {\tilde \ell}_R$ mass difference may be
approximately the same as at the GDM benchmarks if $M \sim 200 - 300$~TeV,
and for $N = 3$ the mass difference is significantly larger. Therefore, we
can discard $N = 1$ and concentrate on $N = 2$ while retaining $N = 3$ as
a second option. As seen in Fig.~\ref{fig:GMSB}(a), a GMSB model with $N =
2$ has approximately the same values of $m_\chi$ and $m_{\chi_2}$ as the
GDM benchmarks if $\Lambda \simeq 67$~TeV, whereas the best value of
$\Lambda$ would be somewhat lower for $N = 3$. In each case, $m_{\tilde
g}$ is very similar in the GDM and mGMSB model for the best value of
$\Lambda$. However, in the $N = 2$ case $m_{{\tilde q}_R} > m_{\tilde g}$,
whereas in the $N = 3$ case $m_{{\tilde q}_R} < m_{\tilde g}$. Thus, an
LHC measurement of $m_\chi - m_{{\tilde \ell}_R}$ has the potential to
exclude mGMSB with $N = 3$ and that of $m_{\tilde g} - m_{{\tilde q}_R}$
to exclude $N = 2$.

We have also considered the `average' spectroscopic properties of GDM
scenarios $\zeta$ and $\eta$, and compared them with mGMSB models with the
same value of $m_{1/2} = 1000$~GeV and $\tan \beta = 22$. The principal
discriminants we consider are again the masses of the $\chi, \chi_2$,
${\tilde \ell}_R$ and ${\tilde q}_R$. As in the case $m_{1/2} = 440$~GeV,
we again find that $N = 1$ gives a ${\tilde \ell}_R$ heavier than the
$\chi$, and hence can be discarded, whereas $N = 3$ gives too large a mass
difference $m_\chi - m_{{\tilde \ell}_R}$. On the other hand, as seen in
panel (b) of Fig.~\ref{fig:GMSB}, $N = 2$ gives a ${\tilde q}_R$ heavier
than the ${\tilde g}$ whereas $N = 3$ again gives a lighter ${\tilde
q}_R$. As in the previous case, an LHC measurement of $m_\chi - m_{{\tilde
\ell}_R}$ has the potential to exclude mGMSB with $N = 3$ and that of
$m_{\tilde g} - m_{{\tilde q}_R}$ to exclude $N = 2$.

We conclude that a combination of these mass measurements and the 
different lifetimes of NLSP decays would be sufficient to distinguish 
the GDM models of the type considered here from mGMSB models.

\subsection{Stau Trapping and the Detection of Decays}

If produced ${\tilde \tau}_1$'s are sufficiently slow-moving, they may be
stopped inside the detector or its neighbourhood. We consider three
possibilities: that the ${\tilde \tau}_1$ may be trapped inside the
detector itself, or in adjacent water tank or calorimetric detector, or in
the walls of the experimental cavern. In the case of the LHC, the trapping
rate can be calculated using the spectra shown in Fig.~\ref{fig:betagamma}
and the known rates of energy loss by charged particles passing though
different types of matter~\cite{PDG}. As representative examples, we 
consider Iron -
as in an experimental calorimeter - and Carbon - which has similar
stopping power to water and less than any other plausible detector and/or
surrounding material.

We display in Table~\ref{tab:stop} the numbers of ${\tilde \tau}_1$'s
expected to be produced in GDM benchmark scenarios $\epsilon, \zeta, \eta$
at the LHC with 100~fb$^{-1}$ of integrated luminosity, using the cross 
sections
shown in Table~\ref{tab:LHCxsec} calculated with {\tt 
PROSPINO}~\cite{PROSPINO}. We
consider numbers of particles produced with $\beta \gamma < 0.25
(0.5)$ and show in each case the corresponding ranges in Carbon and Iron.

\begin{table}[htb]
\centering
\begin{tabular}{|c||r|r|r|}
\hline
Model          & $\epsilon$ & $\zeta$ & $\eta$ \\
\hline
Number of particles with & 850 & 7 & 7 \\
$\beta \gamma < 0.25$ &     &   &   \\  
\hline
Range in C (cm) & 60 & 136 & 129 \\
Range in Fe (cm) & 29 & ~65 & ~61 \\
\hline
\hline
Number of particles with & 7700 & 100 & 90 \\
$\beta \gamma < 0.5$  &      &     &    \\  
\hline
Range in C (cm) & 600 & 1360 & 1290 \\
Range in Fe (cm) & 290 & ~650 & ~610 \\
\hline
\end{tabular}
\caption{\it
Numbers of slow-moving ${\tilde \tau}_1$'s produced with 100/fb at the LHC 
in 
GDM benchmark scenarios $\epsilon, \zeta, \eta$, and the corresponding 
ranges in Carbon and Iron.}
\label{tab:stop}
\end{table}

Benchmark scenario $\epsilon$ is in a different class from
scenarios $\zeta, \eta$. The numbers of slow-moving
particles are orders of magnitude larger, and secondly the ranges are
shorter typically by a factor of two. Both these features are directly
related to the sparticle mass scale, which is set essentially by
$m_{1/2}$. In the case of benchmark $\epsilon$, hundreds of ${\tilde
\tau}_1$'s would be trapped within either the ATLAS or CMS calorimeter,
and thousands more would be trapped within a few metres of surrounding
material. On the other hand, in the cases of scenarios $\zeta, \eta$, only
a handful of particles would be trapped within either detector, and only a
few dozen events would be trapped within $\sim 10$~m of surrounding
material.

The situation at the ILC would be very favourable for any of the scenarios
considered. The above reference values of $\beta \gamma$ correspond to
$\sqrt{s} = 2 m_{{\tilde \tau}_1} \times 1.0625 (1.25)$, and there should 
be
no problem tuning the beam energy very close to the ${\tilde \tau}_1$ mass
and obtaining large samples of ${\tilde \tau}_1$'s stopped within the
calorimeter. The same would be true at CLIC for mSUGRA GDM models with 
$m_{1/2} < 4.5$~TeV.

In the case of the LHC, unfortunately there is very little room left in
the ATLAS cavern for a trapping water tank or calorimeter, and it would be
difficult to envisage inserting any detector over a metre thick between
the barrel and the cavern walls. On the other hand, in the case of CMS
there may be some more room after restructuring the infrastructure
(balconies and services), permitting the installation of an ${\cal
O}$(kton) trapping detector, as discussed in~\cite{nojiri}. There would be
more room in the forward direction at CMS, but this possibility would have
limited angular acceptance. Moreover, the ${\tilde \tau}_1$
pseudorapidity distributions are generally very central, within the 
ATLAS or CMS acceptance, as shown in
Figs.~\ref{fig:trigger5}, \ref{fig:trigger6}, \ref{fig:trigger7}, so a
forward trap would not be very efficient. In the case of the ILC or CLIC,
if the experiments at the LHC reveal interest, it would be possible to
design the experimental areas ahead of time so as to allow for a trapping
detector.

In the interim, we speculate on the alternative
possibility of looking for the decays of ${\tilde
\tau}_1$'s that are trapped in the walls of the ATLAS and CMS caverns. One
possible strategy would be to use the tracking information from CMS or
ATLAS to determine the ${\tilde \tau}_1$'s impact point and angle, then
bore a hole into the wall and extract a core with an optimal chance of
containing a trapped ${\tilde \tau}_1$. The tracking systems of CMS 
and ATLAS should each yield an
experimental uncertainty in the impact point that is about half a cm, and 
a corresponding angular error $\sim 10^{-3}$ radians. Using the standard 
formula~\cite{PDG}
\begin{equation}
\theta_0 = \frac{13.6~{\rm MeV}}{\beta p} \sqrt{\frac{x}{X_0}} \{ 1 + 
0.038 ~ {\rm ln}(x/X_0) \}
\label{scangle}
\end{equation}
for the 98~\% C.L. width of the projected distribution of the multiple 
scattering angle, where $\beta, p$ are the ${\tilde 
\tau}_1$ velocity and momentum and $x$ the penetration depth relative to 
the scattering length $X_0$, we find typical values $\theta_0 <
10^{-3}$, within the experimental angular error. As we can see 
from Table~\ref{tab:stop}, one might want to extract ${\cal O}(100)$ to
${\cal O}(10000)$ `interesting' cores with dimensions $\sim 1$~cm 
$\times 1$~cm $\times 10$~m each year. This technique might be appropriate 
for
the upper part of the mSUGRA wedge shown in Fig.~\ref{fig:planes}, where
the ${\tilde \tau}_1$ lifetime is measured in weeks, such as scenarios
$\epsilon, \zeta$. However, this is unlikely to be feasible in the lower
part of the mSUGRA wedge, e.g., at point $\eta$, because radiation levels
in the LHC caverns would preclude access on the necessary time-scale
$\tau_{{\tilde \tau}_1} \sim 5$~hours.

The baseline operating plan for the LHC foresees one multi-month stop each
winter, and half-a-dozen two-day technical stops at regular intervals
during the rest of the year. Each of these would provide an opportunity to
extract a limited number of cores from the cavern walls. This would be
interesting if the ${\tilde \tau}_1$ lifetime is several weeks or more, as
in benchmarks $\epsilon, \zeta$, but not point $\eta$.

We have also considered the possibility of measuring directly the mass of
a stopped ${\tilde \tau}_1$ in a mass spectrometer. A typical extracted
core of size 1~cm $\times$ 1~cm $\times$ 10~m would contain $\sim 1 \times
10^{28}$ protons. On the basis of estimates of the mass of the ${\tilde
\tau}_1$ and the velocity of the specific particle being sought in the
core, we estimate that a `high-interest' sample of about 10~\% of the 
length of each
core might be selected for exploration in more detail using a mass 
spectrometer. For comparison, we note that the
best available upper limit on the relative abundance in water of a
positively-charged stable relic particle with mass between 40 and 400~GeV 
is $1
\times 10^{-29}$~\cite{Smith82b}. It might therefore be feasible to pass
the high-interest samples of each core through a mass spectrometer and
measure the ${\tilde \tau}_1$ mass very precisely. The issue would be how
quickly this study could be completed, in comparison with the ${\tilde
\tau}_1$ lifetime, as discussed above.

Another possibility might be to look for upward- or sideways-going muons
coming out of the wall, produced by ${\tilde \tau}_1$ decays in the
neighbouring rock. We estimate that typical muon momenta would be tens of
GeV, in which case they should be able to traverse tens of metres of rock.  
However, the acceptance for decays back into the cavern would not be large
unless the ${\tilde \tau}_1$ decays within a few metres of the cavern
wall. In the benchmarks studied, detecting these might be feasible for the
thousands of ${\tilde \tau}_1$'s produced with $\beta \gamma < 0.5$ in
scenario $\epsilon$, but looks very marginal for the few dozen ${\tilde
\tau}_1$'s produced with $\beta \gamma < 0.5$ in scenarios $\zeta, \eta$,
which would also have longer ranges, diminishing the angular acceptance
for the decays.

We have compared the possible measurement of the `albedo' due to ${\tilde
\tau}_1 \to \tau \to \mu$ decay, which has a branching ratio of $\sim 
16$~\%,
with the irreducible background due to the known atmospheric $\nu \to \mu$
flux. Assuming that the gravitino mass is small compared with $m_{{\tilde
\tau}_1}$, the characteristic $\mu$ energy will be $\sim m_{{\tilde
\tau}_1}/6$, corresponding to $\sim 25$~GeV for point $\epsilon$ and 
$\sim 50$~GeV for
points $\epsilon, \zeta$. If we consider a representative LHC or ILC
detector with linear dimensions 20~m $\times$ 20~m and ${\tilde \tau}_1$
decay at a characteristic distance $\sim 10$~m, the detector subtends 1/6
of the total solid angle, namely ${2 \pi \over 3}$ steradians. However,
${\tilde \tau}_1$ decays in the upper hemisphere surrounding the detector
will surely be drowned in cosmic-ray $\mu$ background, so we consider
only ${\tilde \tau}_1$ decays in the lower hemisphere. Therefore, only
1/12 of the decay muons are in principle observable, corresponding to just
1.3~\% of the stopped ${\tilde \tau}_1$ decays. As seen from
Table~\ref{tab:stop}, at the LHC this would give ${\cal O}(100)$
events in benchmark scenario $\epsilon$, but only ${\cal O}(1)$ event in
either of scenarios $\zeta, \eta$. For comparison, the MACRO 
experiment~\cite{MACRO}
has reported a sample of about 900 upward-going atmospheric $\nu \to \mu$
events passing through a detector of area 76~m $\times$ 12~m in five years
of operation, corresponding to $\sim 80$ events/year through our nominal
20~m $\times$ 20~m collider detector. The energy spectrum of the MACRO
`up-through' $\mu$ sample has a broad peak around 50~GeV, so there is no
clear separation in energy between the ${\tilde \tau}_1 \to \tau \to \mu$
signal and the atmospheric $\nu \to \mu$ background. On the other hand, 
the tracking system defines the direction in which a 
candidate ${\tilde \tau}_1$ exited the detector, and the momentum 
measurement constrains the distance at which it is likely to have 
stopped. Together, these measurements define the direction 
from which a candidate `albedo' decay muon might emerge from the wall 
into any part of the detector. 
Using this information, it might be possible to detect the `albedo' decay 
muons at the LHC 
in scenario $\epsilon$, though not in scenarios 
$\zeta, \eta$. On the
other hand, detection at the ILC could be optimized with a dedicated 
detector.

Comparing the mSUGRA spectra for benchmark scenarios $\epsilon, \zeta,
\eta$ with the sample of GMSB models in~\cite{GMSB}, we see that there
exist GMSB models with identical values of $(m_{{\tilde \tau}_1}, \langle
m_{\tilde q} \rangle )$. We are therefore pessimistic that these
spectroscopic measurements at the LHC will be enough alone to distinguish 
mSUGRA
from GMSB.

We have also considered the possibility of using ${\tilde \tau}_1 \to \tau
{\tilde G}$ decay kinematics to constrain directly the mass of the 
gravitino
${\tilde G}$, and hence perhaps also distinguish between mSUGRA and GMSB
scenarios. In the three ${\tilde \tau}_1$ NLSP scenarios $\epsilon, \zeta,
\eta$ considered here, the mean $\tau$ energies are $\langle E_{\tau}
\rangle = 74, 155, 160$~GeV, respectively. On the other hand, in GMSB
scenarios with the same values of $m_{{\tilde \tau}_1}$, one would have
$\langle E_{\tau} \rangle = 75, 170, 161$~GeV, respectively. We recall
that $m_{{\tilde \tau}_1}$ would be measurable with an accuracy that is
probably not much better than 1~\%, which already removes any chance of
measuring the percentage difference in $\langle E_{\tau} \rangle$ in
scenarios $\epsilon, \eta$, where it is ${\cal O}(1)$~\%. On the other
hand, the percentage difference in scenario $\zeta$ is about 10~\%, which
should be measurable in principle with enough events and accurate energy
measurements. However, in practice, as we have discussed above, a sample 
size of more than 100 ${\tilde
\tau}_1$ decays, as would be required for a 10~\% measurement of $\langle
E_{\tau} \rangle$ in scenario $\zeta$, even if $E_{\tau}$ could be
measured perfectly event-by-event, could not be obtained inside ATLAS
or CMS with the 100~fb$^{-1}$ integrated luminosity assumed here.

On the other hand, at the ILC $m_{{\tilde \tau}_1}$ should be measurable
with an accuracy better than 0.1~\%, and it should be possible to obtain a
large enough sample to measure $\langle E_{\tau} \rangle$ with an accuracy
of a few \%. We therefore think that distinguishing scenario $\zeta$ from
a GMSB model with the same $m_{{\tilde \tau}_1}$ should be possible, but
making the same distinction for scenarios $\zeta, \eta$ would be much more
challenging, in view of the very accurate $\langle E_{\tau} \rangle$
measurement required.

\section{Conclusions}

We have discussed in this paper various alternatives to the constrained
MSSM (CMSSM) framework with universal soft supersymmetry breaking, which
has been used in most previous benchmark studies of supersymmetric
signatures at the LHC, ILC and CLIC. Specifically, we have considered both
a less constrained framework with non-universal Higgs masses (NUHM) and a
more constrained gravitino dark matter (GDM) framework inspired by minimal 
supergravity (mSUGRA). As
we have shown, both of these scenarios offer distinctive phenomenological
signatures that were absent in the previous CMSSM studies, at least 
when
cosmological constraints from WMAP and other experiments were taken into
account. For example, Gravitino Dark Matter (GDM) becomes a generic
possibility.

In the case of NUHM models, the freedom to vary the scalar masses relative
to the gaugino mass $m_{1/2}$ opens up the possibility of different
characteristic heavier neutralino decays such as $\chi_2 \to \chi h, 
\chi Z$,
which were disfavoured along the WMAP lines allowed in the CMSSM. Three of
the new benchmark scenarios we have proposed and examined here are of this
type. One has a dominant $\chi_2 \to \chi Z$ decay mode (point $\alpha$),
one has a dominant $\chi_2 \to \chi h$ decay mode (point $\beta$), and one
has non-resonant $\chi_2 \to \chi \ell^+ \ell^-$ decays (point $\gamma$).
The $h$
signal at point $\beta$ should easily be detected at the LHC, and {\it a
fortiori} at the ILC or CLIC. These features open up new possibilities for
reconstructing the cascade decays of heavier sparticles and measuring
better the mass of the lightest neutralino $\chi$.

In the case of GDM models, there are two generic possibilities for the
next-to-lightest supersymmetric particle (NLSP). Either it is neutral and
presumably the lightest neutralino $\chi$, as exemplified by benchmark
scenario $\delta$, or it is charged and presumably the lighter stau
slepton ${\tilde \tau}_1$, as exemplified by benchmark scenarios
$\epsilon, \zeta$ and $\eta$.

In the former case, which is a CMSSM scenario with the gravitino
specifically chosen to be light, the collider signatures are very similar
to those for scenarios with a stable $\chi$, since generic decays of the
unstable $\chi$ occur far outside the collider detector. However, the
sparticle spectrum is somewhat different from what it would be along a
WMAP line, providing an opportunity to distinguish this possibility, for 
example in ${\tilde g} \to {\tilde b} b$ decays. 

The three benchmark scenarios $\epsilon, \zeta$ and $\eta$ with a ${\tilde
\tau}_1$ NLSP span the range of possibilities that are likely to be
detectable at the LHC or ILC within the mSUGRA framework. Point $\epsilon$
was chosen to have sparticles as light as possible, with heavier spectra 
at points $\zeta, \eta$. At point $\zeta$ the
${\tilde \tau}_1$ lifetime is close to its theoretical maximum, and at
point $\eta$ the ${\tilde \tau}_1$ lifetime is close to the minimum we 
consider.

In all three cases, the ${\tilde \tau}_1$ should be detectable as a
slowly-moving metastable particle, and TOF measurements at the LHC might
enable its mass to be measured with an accuracy at the \% level. We have
discussed how these mSUGRA GDM scenarios could be distinguished from mGMSB
models.

As pointed out in~\cite{Nojiri,Fengslep}, a small fraction of the produced
${\tilde \tau}_1$'s may be stopped inside the collider detector and/or in
surrounding material, either cavern walls or a specialized stopping
detector. At the LHC, there is limited space in the constructed caverns
for such a specialized detector~\cite{nojiri}, but this could be envisaged 
in designing
intersection regions for the ILC or CLIC, if needed. We have discussed a
couple of strategies for detecting ${\tilde \tau}_1$'s that might have
stopped in the cavern walls surrounding ATLAS or CMS, either by extracting
cores from the surrounding wall material or by looking for muons from the
decays ${\tilde \tau}_1 \to \tau +$ gravitino followed by $\tau \to \mu +
\nu_e {\bar \nu}_\mu$. The validities of these strategies depend on 
practical feasibility as well as the
${\tilde \tau}_1$ lifetime, which is uncertain by two orders of magnitude
even in the restricted mSUGRA framework discussed here.

Our preliminary investigations indicate that the LHC has good capabilities 
to discover supersymmetry
in each of the benchmark scenarios proposed, and may be able to uncover
several different sparticles in each case. However, in general the LHC
experiments would not be able to make many very accurate spectroscopic
measurements. In several of the benchmark scenarios, an ILC with $\sqrt{s} 
=
0.5$~TeV would already be able to make several precise measurements of
weakly-interacting sparticles, and a 1~TeV ILC would extend this
capability to all the benchmarks studied here. In all the scenarios
studied, the additional measurements possible with CLIC at $\sqrt{s} = 3$ 
or
5~TeV would contribute significant added value to the exploration of the
new physics previously uncovered by the LHC and followed up by the ILC.

These new benchmark scenarios merely scratch the surface of the myriad
possibilities open in supersymmetric phenomenology once one explores the
high-dimensional parameter space of soft supersymmetry breaking. These
examples should offer general encouragement to think outside the CMSSM
box, and suggest the likelihood that even more exotic possibilities might
be waiting out there, perhaps including one chosen by Nature.

\vskip 0.5in
\vbox{
\noindent{ {\bf Acknowledgments} } \\
\noindent
The work of K.A.O. was supported partly by DOE grant
DE--FG02--94ER--40823. We thank S.~Abdullin, E.~Tsesmelis, 
D.~Forkel-Wirth, R.~Voss and M.~White for useful discussions.}

\end{document}